\documentclass[preprint,11pt]{elsarticle}

\usepackage{amsthm}
\usepackage{amsmath}
\usepackage{subfig}
\usepackage{caption}
\usepackage{geometry}
\usepackage{gensymb}
\usepackage{multirow}
\usepackage{algorithmic}
\usepackage[linesnumbered,ruled,vlined]{algorithm2e}

\usepackage{setspace}

\usepackage{gensymb}
\usepackage[utf8]{inputenc}
\usepackage{array}

\usepackage{graphicx}
\usepackage{tikz}
\usepackage{soul}

\usepackage{amssymb}

\usepackage{lineno}

\journal{Journal of Computational Physics}

\begin{document}

\begin{frontmatter}

\title{Subdomain POD-TPWL with Local Parameterization for Large-Scale Reservoir History Matching Problems}

\author{Cong Xiao$^{1}$, Olwijn Leeuwenburgh$^{2,3}$, Hai Xiang Lin$^{1}$, Arnold Heemink$^{1}$}

\address{$^{1}$Delft Institute of Applied Mathematics, Delft University of Technology, Mekelweg 4, 2628 CD Delft, the Netherlands}
\address{$^{2}$Civil Engineering and Geosciences, Delft University of Technology, Mekelweg 4, 2628 CD Delft, the Netherlands}
\address{$^{3}$TNO Princetonlaan 6 PO Box 80015 3508 TA Utrecht, the Netherlands}

\begin{abstract}
We have recently proposed an efficient subdomain POD-TPWL (Xiao et al, 2018) for history matching without the need of model intrusion. 
From a computational point of view, a local parameterization where the parameters are separately defined in each subdomain is very attractive. 
In this study we present one such local parameterization through integrating principle component analysis with domain decomposition to independently 
represent the spatial parameter field within low-order parameter subspaces in each subdomain. 
This local parameterization allows us to perturb parameters in each subdomain simultaneously and the effects 
of all these perturbations can be computed with only a few full-order model simulations. 
To avoid non-smoothness at the boundaries of neighboring subdomains, the optimal local parameters are projected onto a global parameterization.
The performance of subdomain POD-TPWL with local parameterization has been assessed through two cases tested on a modified version of SAIGUP model.  
This local parameterization has high scalability, since the number of training models depends primarily on 
the number of the local parameters in each subdomain and not on the dimension of the underlying full-order model.
Activating more subdomains results in less local parameter patterns and enables us to run less model simulations.
For a large-scale case-study in this work, to optimize 282 global parameters, LP-SD POD-TPWL only needs 90 fidelity simulations.
In comparison to the finite-difference based history matching, and subdomain POD-TPWL with global parameterization where the parameters are defined over the entire domain,  
the CPU cost is reduced by a factor of several orders of the magnitude while reasonable accuracy remains.
\end{abstract}


\end{frontmatter}
 
\section*{Abbreviation}
POD, proper orthogonal decomposition;  PCA, principal component analysis; RBF, radial basis function;  TPWL, trajectory piecewise linearizaton; DD, domain decomposition; FOM, full-order model

\section{Introduction}
History matching problems have been investigated for several decades, and many algorithms have been developed, among them, the gradient-based minimization algorithm is an attractive option, 
where the gradient of the cost function with respect to the parameters is computed using an adjoint model \cite{courant1962methods}. 
It has been widely acknowledged that the adjoint approach is one of the most efficient approaches nowadays to handle large-scale history matching problems.

Many efforts have been taken to make the implementation of the adjoint model more feasible. 
Courtier et al. \cite{courtier1994strategy} proposed an incremental approach by replacing a high resolution nonlinear model with an approximated linear model so that the adjoint model can be easily obtained. 
Liu et al. \cite{liu2008ensemble,liu2009ensemble} developed an ensemble-based four-dimensional variational (En4DVar) data assimilation scheme where the approximated 
linear model is constructed by use of an ensemble of model forecast. Recently, Frolov and Bishop et al. \cite{frolov2016localized,bishop2017local} incorporated a 
local ensemble tangent linear model (LETLM) into a variational scheme. The LETLM is capable of capturing localized physical features of dynamic models with a relatively small ensemble size. 
Reduced-order modeling technique is a promising approach in decreasing the dimensionality of the original model. 
The POD approach has been applied to various areas with success in efficiently speeding up the model simulations \cite{heijn2003generation,markovinovic2006accelerating,cardoso2009development}.

The combination of model linearization and reduced-order modeling technique provides the possibilities to further ease the implementation of adjoint model for high-dimensional
complex dynamic system. Vermeulen and Heemink \cite{vermeulen2006model} combined POD and a non-intrusive perturbation-based linearization method to build a reduced-order 
linear approximation of the original high-dimensional non-linear model, subsequently, the adjoint of this reduced-order linear model 
can be easily constructed and therefore the minimization of the objective function can be handled efficiently.
Altaf et al. \cite{altaf2009inverse} and Kaleta et al \cite{kaleta2011model} applied this method to a coastal engineering and reservoir history matching problem, respectively, 
and showed it is very efficient. This algorithm was implemented in a low-order linear space and considered any simulators as black boxes without modifying the source code. 
This construction of a reduced-order linear model is non-intrusive \cite{vermeulen2006model,altaf2009inverse,kaleta2011model}, 
however, the required derivative information is estimated using a global perturbation-based finite difference method, 
which needs a large number of full-order model simulations for large-scale problem and is therefore computationally demanding. 

Xiao et al. \cite{Xiao2018} recently introduced a variational data assimilation method where domain decomposition (DD) and radial basis function (RBF) 
interpolation are integrated with trajectory piecewise linearization (TPWL) resulting in a new subdomain POD-TPWL. The adjoint model of the original high-dimensional non-linear model is replaced 
by a subdomain reduced-order linear model, which is easily incorporated into an adjoint-based parameter estimation procedure. It has been shown to be a feasible 
approach to assist the adjoint-based history matching without intruding the legacy source code. 
In the original subdomain POD-TPWL, a global parameterization of the log-permeability field is considered where the PCA patterns are defined over the entire domain. 
The total number of full-order model simulations required for the subdomain POD-TPWL is roughly 2-4 times the number of global parameter patterns,
which makes it almost intractable for large-scale history matching with a large amount of parameters, although this subdomain POD-TPWL algorithm is still more efficient than the finite-difference method.

From a computational point of view, a local parameterization where the parameters are separately defined in each subdomain is very attractive \cite{DaVeiga2012}. 
Since in this case we can mainly focus on the independent target zones, and parameters in each subdomain can be perturbed simultaneously of each other and the effects 
of all these perturbations can be computed with only a few full-order model simulations \cite{Ding2011Development,Ding2013Study}. 
Currently, some classical parameterization techniques.e.g, PCA \cite{Chen2014Integration}, 
KPCA \cite{Sarma2008}, DCT \cite{Jafarpour2008} and DWT \cite{Chen2012Multiscale}
have been extensively applied into petroleum engineering to effectively reconstruct spatial parameter field. 
When implementing these parameterizations in the global domain, a smooth spatial parameter field can be reconstructed.
However, when separately conducting parameterization in each subdomain, the physical smoothness will be lost at the boundary of neighboring subdomains, 
which severely violates the characteristics of the original spatial parameters.
In this paper we explore a new local parameterization by combining principle component analysis (PCA) and domain decomposition (DD) to represent the spatial parameter field 
with a low-order parameter subspace in each subdomain. 
To avoid non-smoothness at the boundaries of neighboring subdomains, the optimal local parameters are projected onto a global parameterization 
based on a global PCA. This local parameterization simultaneously retains the advantages of GPCA (e.g., smooth and differentiability) and LPCA (relatively low-order subspace) to 
improve its ability to reconstruct smooth spatial parameter fields with a relatively low-order subspace. 
In addition, the degree of freedom for global parameterization is preserved via a low-order local parameterization. 

We extend our original subdomain POD-TPWL to large-scale history matching problems by incorporating a local parameterization. 
To simplify the notation, subdomain POD-TPWL with global parameterization is denoted as GP-SD POD-TPWL, 
while Subdomain POD-TPWL with Local Parameterization is referred to LP-SD POD-TPWL.
Firstly, an ensemble of parameters is statistically 
generated based on the prior statistic characteristics,
an optimization-based local smooth PCA, is proposed to construct a low-order parameter subspace in each subdomain, respectively. Secondly, 
the parameters from the local subdomain are perturbed to construct reduced-order linear model in each subdomain through modifying our previous GP-SD POD-TPWL. 
After constructing the reduced-order linear model, the implementation of adjoint model is easily realized and incorporated into an adjoint-based reservoir history matching procedure. 
The performance of this new local parameterization has been assessed through separately assimilating sparse production data 
and dense seismic data, i.e, water saturation in each gridblock, on a modified version of the SAIGUP model.
Comparisons between the LP-SD POD-TPWL, GP-SD POD-TPWL and the finite-difference (FD) method have been carried out to evaluate this new history matching algorithm. 

The paper is organized as follows. The formulation of the history
matching problem is described in Section 2. 
The traditional global parameterization and the step by step procedure of local parameterization is described in Section 3.
The procedure of the modified subdomain POD-TPWL and adjoint-based history matching
formulation with local parameterization is presented in Section 4. 
Section 5 discusses and analyzes two ‘twin’ experiments results using a modified version of SAIGUP model. 
Finally, section 6 summarizes our contributions and discusses future work.

\section{Formulation of History Matching Problem}
We now provide an overview of the mathematical formulation for history matching, along the
lines of the descriptions given in our previous work \cite{Xiao2018}. The dynamic equation for a two-phase oil water system is described as follows, 
\begin{equation}
\label{eq1}
\textbf{x}^{n} = \textbf{f}^{n}(\textbf{x}^{n-1},\boldsymbol{\beta}), \quad n=1,\cdot\cdot\cdot,N
\end{equation}
where, the dynamic operator ${\textbf{f}}^{n}$:$R^{2N_{\beta}}{\rightarrow}R^{2N_{\beta}}$ represents the nonlinear time-dependent model evolution. 
$ \textbf{x}^{n}{\in}R^{2N_{\beta}} $ represents the state vector (pressure and saturation in every gridblock), $N_{\beta}$ is the total number of gridblock,
$n$ and $n$-1 indicate the timestep, and $N$ denotes the total number of simulation steps. 
$\boldsymbol{\beta}$ denotes the vector of uncertain parameters, which is the spatial permeability field in our case. 

The relationship between simulated data and state variables can be described by a nonlinear operator.
The specific formula of this nonlinear operator depends on the type of measured data. In this study we consider two types of data:
\begin{itemize}
\item well injection/production rate data $\textbf{y}_{w}^{m}$ with some noise can be described by 

\begin{equation}
\label{eq2}
\textbf{y}_{w}^{m_{w}}=\textbf{h}_{w}^{m_{w}}(\textbf{x}^{m_{w}},\boldsymbol{\beta})+\textbf{r}_{w}^{m_{w}}, \quad m_{w}=1,\cdot\cdot\cdot,N_{w}
\end{equation}

\item time-lapse seismic data $\textbf{y}_{s}^{m_{s}}$, e.g., phase saturation, given by a seismic inversion or a direct collection from the forward simulation
\begin{equation}
\label{eq3}
\textbf{y}_{s}^{m_{s}}=\textbf{h}_{s}^{m_{s}}(\textbf{x}^{m_{s}},\boldsymbol{\beta})+\textbf{r}_{s}^{m_{s}}, \quad m_{s}=1,\cdot\cdot\cdot,N_{s}
\end{equation}

\end{itemize}
where, ${N}_{w}$ or ${N}_{s}$ is the number of timesteps where the measurements are taken. $\textbf{r}_{w}^{m_{w}}$ or $\textbf{r}_{s}^{m_{s}}$ is the time-dependent vector of observation error, 
which is uncorrelated over time and generally assumed to satisfy the Gaussian distribution $G (\textbf{0},\textbf{R}_{s}^{m_{w}})$ or $G (\textbf{0},\textbf{R}_{s}^{m_{s}})$, 
where $\textbf{R}_{s}^{m_{w}}$ and $\textbf{R}_{s}^{m_{s}}$ are observation error covariance matrix.

History matching process calibrates the uncertain parameters by maximizing a likelihood (MAP) or alternatively minimizing a cost function. 
In general, the cost function can be formulated within the Bayesian framework, where 
the posterior probability density function (PDF) of uncertain model parameters $\boldsymbol{\beta}$ is conditioned to measurements $\textbf{d}_{obs}$.
If we assume that the prior PDF of parameters $\boldsymbol{\beta}$ is Gaussian with a mean $\boldsymbol{\beta}_{b}$ 
and covariance matrix $\textbf{R}_{b}$, then an objective function $\textit{J}$ $(\boldsymbol{\beta})$ can be derived correspondingly. 
The particular formulation of this objective function also depends on the type of measured data.

\begin{itemize}
\item assimilate well injection/production data $\textbf{y}_{w}^{m_{w}}$

\begin{align}
\label{eq4}
J(\boldsymbol{\beta})_{w} = \frac{1}{2}(\boldsymbol{\beta}-\boldsymbol{\beta}_{b})^{T}{\textbf{R}_{b}}^{-1}(\boldsymbol{\beta}-\boldsymbol{\beta}_{b})
+\frac{1}{2}\sum_{m_{w}=1}^{N_{w}}[\textbf{d}_{obs}^{m_{w}}-\textbf{h}_{w}^{m_{w}}(\textbf{x}^{m_{w}},\boldsymbol{\beta})]^{T}{\textbf{R}_{w}^{m_{w}}}^{-1}[\textbf{d}_{obs}^{m_{w}}-\textbf{h}_{w}^{m_{w}}(\textbf{x}^{m_{w}},\boldsymbol{\beta})]
\end{align}

\item assimilate time-lapse seismic data $\textbf{y}_{s}^{m_{s}}$
\begin{align}
\label{eq5}
J(\boldsymbol{\beta})_{s} = \frac{1}{2}(\boldsymbol{\beta}-\boldsymbol{\beta}_{b})^{T}{\textbf{R}_{b}}^{-1}(\boldsymbol{\beta}-\boldsymbol{\beta}_{b})
+\frac{1}{2}\sum_{m_{s}=1}^{N_{s}}[\textbf{d}_{obs}^{m_{s}}-\textbf{h}_{s}^{m_{s}}(\textbf{x}^{m_{s}},\boldsymbol{\beta})]^{T}{\textbf{R}_{s}^{m_{s}}}^{-1}[\textbf{d}_{obs}^{m_{s}}-\textbf{h}_{s}^{m_{s}}(\textbf{x}^{m_{s}},\boldsymbol{\beta})]
\end{align}

\item simultaneously assimilate well data $\textbf{y}_{w}^{m_{w}}$ and time-lapse seismic data $\textbf{y}_{s}^{m_{s}}$
\begin{align}
\label{eq6}
& J(\boldsymbol{\beta})_{w,s} = J(\boldsymbol{\beta})_{w} + J(\boldsymbol{\beta})_{s}
\end{align}

\end{itemize}

In general, gradient-based algorithms can be used to minimize the cost function, see Eq.\ref{eq5}.
The key step of gradient-based minimization algorithm is to determine the gradient of the objective function with respect to the parameters. Through the constraints of the dynamic
model Eq.\ref{eq1}, the gradient is formulated by introducing the adjoint model
\cite{brouwer2002dynamic,sarma2006efficient,vlemmix2009adjoint}. However, the implementation of the adjoint model requires an overwhelming programming effort 
by rearranging the source code. We recently have developed a subdomain POD-TPWL which allows us to easily construct the adjoint model in a reduced-order linear subspace \cite{Xiao2018}.
In this paper, we propose a subdomain POD-TPWL with local parameterization for large-scale problem, which will be briefly described in the following section.

\section{Smooth Local Parameterization}
The main idea of local parameterization is to project the optimal local parameters onto a global parameterization 
based on an integration of local PCA and global PCA. 
In this section, we first discuss the construction of PCA-based procedures for representing spatial parameter field in global domain and local subdomain, respectively.
The drawbacks of local PCA and global PCA are illustrated as well. And then a procedure for local parameterization is described in detail. 

\subsection{PCA-based Local Representation of Spatial Parameter}
The PCA, also referred to as Karhunen-Loeve expansion (KLE), reduces the dimensionality of the parameter by projecting the high-dimensional parameter into an optimal lower-
dimensional subspace \cite{fukunaga1970application}. The basis of this subspace is obtained by performing an
eigenvalue decomposition of a covariance matrix of the parameter, represented by an
ensemble of prior parameter field and the corresponding ensemble mean. The PCA-based parameterization in global domain and subdomain will be illustrated separately.

Given a spatial domain $\Omega$, containing $N_{\beta}$ gridblocks, that is fully characterized in terms of the value of parameters, e.g, permeability, in each gridblock. 
The parameter field is assumed to statistically satisfy Gaussian distribution. 
Our specific procedure for the construction of the PCA mode is as follows. First, a set of $N_{r}$ realizations each as a column of the data matrix $\textbf{X}_{c}$, is denoted as
\begin{equation}
\label{eq8}
\textbf{X}_{c} = [ \boldsymbol{\beta}_{1}-\boldsymbol{\beta}_{m}, \boldsymbol{\beta}_{2}-\boldsymbol{\beta}_{m}, \cdot\cdot\cdot, \boldsymbol{\beta}_{i}-\boldsymbol{\beta}_{m}, 
\cdot\cdot\cdot, \boldsymbol{\beta}_{N_{r}}-\boldsymbol{\beta}_{m} ], \quad \boldsymbol{\beta}_{m} =\frac{1}{N_{r}}\sum_{i=1}^{N_{r}}\boldsymbol{\beta}_{i}
\end{equation}
where, $\beta_{i}$ denotes a vector of parameter realization and $\boldsymbol{\beta}_{m}$ represents the average of these $N_{r}$ realizations. To implement the PCA parameterization
in global domain, we can directly compute the covariance matrix $\textbf{C}$ by $\tfrac{\textbf{X}_{c}\textbf{X}_{c}^{T}}{N_{r}-1}$. To implement the PCA parameterization
in each subdomain, first, the global domain $\Omega$ is assumed to be decomposed into $\textit{S}$ non-overlapping subdomains $\Omega^{d}$ , $d \in \{1,2,\cdot\cdot\cdot,S\}$
(such as $\Omega=\bigcup_{d=1}^{S}\Omega^{d} $ and $ \Omega^{i} \cap \Omega^{j}=0 $ for $ i\neq j$) and each subdomain is assigned a set of local realizations from the global realizations.
\begin{equation}
\label{eq9}
\textbf{X}_{c}^{d} =[ \boldsymbol{\beta}_{1}^{d}-\boldsymbol{\beta}_{m}^{d}, \boldsymbol{\beta}_{2}^{d}-\boldsymbol{\beta}_{m}^{d}, \cdot\cdot\cdot, 
\boldsymbol{\beta}_{i}^{d}-\boldsymbol{\beta}_{m}^{d}, \cdot\cdot\cdot, \boldsymbol{\beta}_{N_{r}}^{d}-\boldsymbol{\beta}_{m}^{d}], \quad \boldsymbol{\beta}_{m}^{d} =\frac{1}{N_{r}}\sum_{i=1}^{N_{r}}\boldsymbol{\beta}_{i}^{d}
\end{equation}
where, $\boldsymbol{\beta}_{1}^{d}$ and $\boldsymbol{\beta}_{m}^{d}$ separately denotes a vector of local parameter realization and the average of these $N_{r}$ local realizations in subdomain $\Omega^{d}$. 
Thus, we can separately compute the local covariance matrix $\textbf{C}^{d}$ by $\tfrac{\textbf{X}_{c}^{d}{\textbf{X}_{c}^{d}}^{T}}{N_{r}-1}$ for each subdomain.

The basis of this subspace is obtained by performing a singular value decomposition (SVD) of a covariance matrix, e.g. global covariance matrix $\textbf{C}$, or local covariance matrix 
$\textbf{C}^{d}$ at subdomain $\Omega^{d}$. The global parameter vector $\boldsymbol{\beta}$ can be approximated by
\begin{equation}
\label{eq10}
\boldsymbol{\beta} = \boldsymbol{\beta}_{m}+ \boldsymbol{\Phi}_{\boldsymbol{\beta}} \boldsymbol{\xi}_{G}, \boldsymbol{\Phi}_{\boldsymbol{\beta}}= \textbf{U}_{\boldsymbol{\beta}} \sqrt{\boldsymbol{\Sigma}_{\boldsymbol{\beta}}}
\end{equation}

Similarly, the local parameter vector $\boldsymbol{\beta}^{d}$ in subdomain $\Omega^{d}$ is approximated by
\begin{equation}
\label{eq11}
\boldsymbol{\beta}^{d} = \boldsymbol{\beta}_{m}^{d}+ \boldsymbol{\Phi}_{\boldsymbol{\beta}}^{d} \boldsymbol{\xi}^{d}, \quad
\boldsymbol{\Phi}_{\boldsymbol{\beta}}^{d}= \textbf{U}_{\boldsymbol{\beta}}^{d} \sqrt{\boldsymbol{\Sigma}_{\boldsymbol{\beta}}^{d}}, \quad d=1,\cdot\cdot\cdot,S
\end{equation}
where, $\boldsymbol{\Phi}_{\boldsymbol{\beta}}$ or $\boldsymbol{\Phi}_{\boldsymbol{\beta}}^{d}$ denotes the basis matrix to project the full-order parameter space into the reduced subspace; 
$\textbf{U}_{\boldsymbol{\beta}}$, $\boldsymbol{\Sigma}_{\boldsymbol{\beta}}$, $\textbf{U}_{\boldsymbol{\beta}}^{d}$ and $\boldsymbol{\Sigma}_{\boldsymbol{\beta}}^{d}$ denote
the left eigenvector matrix and eigenvalue matrix of the SVDs of the parameter covariance matrix, respectively; $\boldsymbol{\xi}_{G}$ and $\boldsymbol{\xi}^{d}$ denote the independent Gaussian 
random variables with zeros mean and unit variance. 

The number of columns of the left singular matrix to be retained (denoted as $N_{G}$ and $l_{d}$ for global domain and each subdomain, respectively, $N_{L}=\sum_{d=1}^{S}l_{d}$ denotes 
the total number of local PCA patterns) 
is determined by either the energy criterion or through a basis optimization procedure \cite{he2015constraint}. We take the energy criterion as an example. Considering $\boldsymbol{\Phi}_{\boldsymbol{\beta}}$ 
and using the energy criterion, we first compute the total energy $E_{t}$, which is defined as $E_{t}=\sum_{i=1}^{L}\kappa_{i}^{2}$, 
where $\kappa_{i}$ denotes a singular value of covariance matrix. The energy associated with the first $N_{G}$ singular vectors is given by $E_{N_{G}}=\sum_{i=1}^{N_{G}}{\kappa_{i}}^{2}$. 
Then $N_{G}$ is determined such that $E_{N_{G}}$exceeds a specific fraction of $E_{t}$. The same process can be assigned to determine $l_{d}$.

\begin{figure}[!h]
\centering\includegraphics[width=0.5\linewidth]{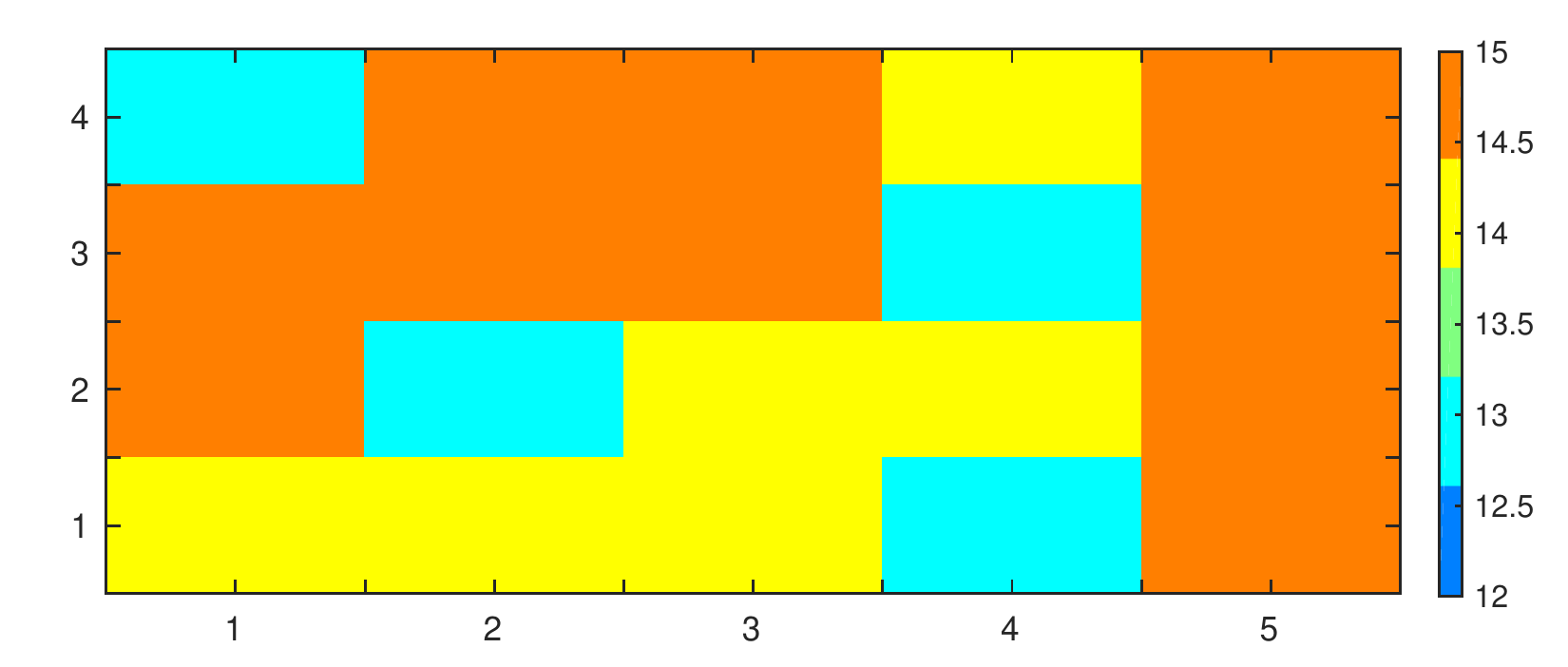}
\caption{Illustration of the number of PCA patterns in each subdomain after LPCA}\label{fig1}
\end{figure}

\begin{figure}[!h]
\centering
\subfloat[]%
  {\includegraphics[width=0.45\linewidth]{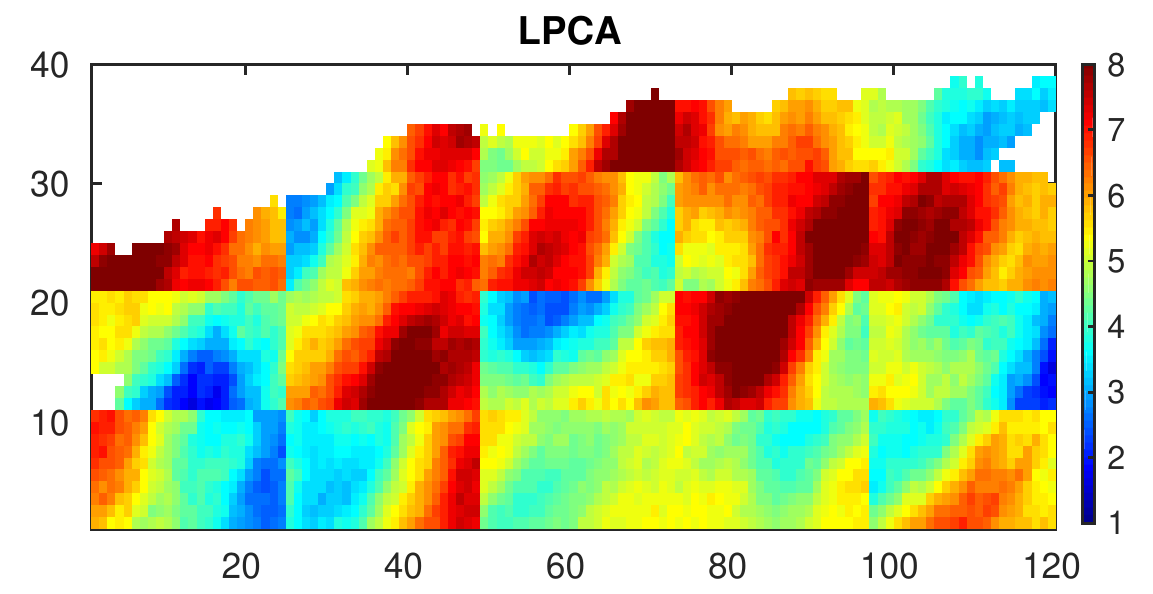}} 
\subfloat[]%
  {\includegraphics[width=0.45\linewidth]{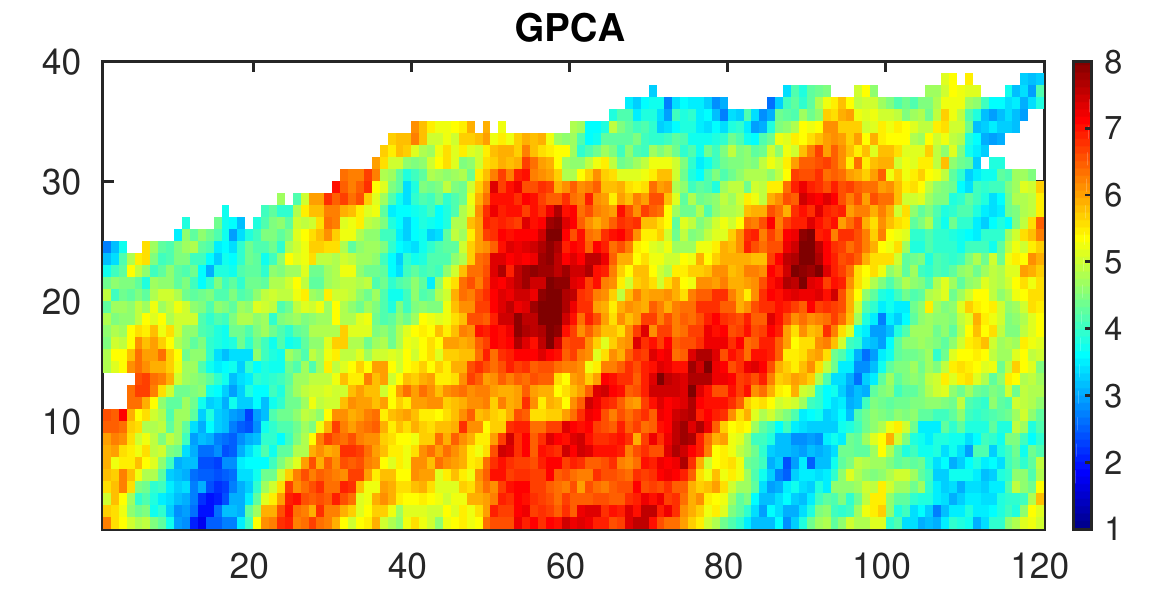}}
\caption{Illustration of parameter reconstruction using LPCA and GPCA}\label{fig2}
\end{figure}

We now illustrate an application of global PCA (GPCA) and local PCA (LPCA) procedure described above to reconstruct a parameter field. 
We generate $N_{r}$=1000 Gaussian log-permeability realizations for SAIGUP model \cite{Matthews35} which will be used in the following numerical experiments. 
The dimensionality of this 2-D parameter field is 40$\times$120, which is decomposed into 20 non-overlapping subdomains. 
After the SVDs of global and local parameter covariance matrix with retaining 95\% energy, $N_{G}$ = 48 and $l_{d}$ for each subdomain is summarized in Fig.\ref{fig1}. 
Given $\boldsymbol{\xi}_{G}$ and $\boldsymbol{\xi}^{d}$ for global domain 
and each subdomain, the corresponding parameter fields are reconstructed.
Fig.\ref{fig2} shows two representations of the parameter fields using GPCA and LPCA for random realization 
of $\boldsymbol{\xi}_{G}$ and $\boldsymbol{\xi}^{d}$ coefficients. 
It demonstrates that although LPCA represents the parameter field with relative low-order parameter subspace, 
it does not reconstruct a smooth parameter realization, which severely offends the spatial properties of original parameters. 
While GPCA indeed represents a smooth parameter field but with relative high-order parameter subspace. 
These facts motivate us to develop a new Optimization-based Local Smooth PCA (O-LS-PCA) through merging 
the advantages of GPCA (smooth representation) and LPCA (low-order representation).

\subsection{O-LS-PCA Representation of Spatial Parameter} 
The main idea of O-LS-PCA is described as follows. Instead of directly projecting LPCA coefficients back to local parameter field for each subdomain and 
then stitching all subdomains together to form a global
parameter field, which leads to a non-smooth parameter field as illustrated in Fig.\ref{fig2}(a), we firstly transform the 
LPCA coefficients to GPCA coefficients using an optimization algorithm, and then project the GPCA coefficients back to global parameter.

Here we use Bayesian theorem to derive O-LS-PCA. The global parameter field satisfies Gaussian distribution, $N\sim(\boldsymbol{\beta}_{g},\textbf{R}_{g})$, which is approximated
using the generated ensemble of parameter realizations, $\boldsymbol{\beta}_{g}\approx\boldsymbol{\beta}_{m}$, and $\textbf{R}_{g}\approx\textbf{C}$. Given LPCA coefficients $\boldsymbol{\xi}^{d}$ for 
each subdomains, the reconstructed parameter field using LPCA is considered to 
be "observed data", which should satisfy Gaussian distribution $N\sim(\boldsymbol{\beta}_{g},\textbf{R}_{g})$ to preserve the statistical properties of original parameter field. 
The corresponding GPCA coefficients $\boldsymbol{\xi}_{G}$
are our objective variables, and the reconstructed parameter field using GPCA with these objective variables is considered to be "simulated data".
At the same time, GPCA coefficients $\boldsymbol{\xi}_{G}$ also satisfy Gaussian normal distribution $N\sim(0,1)$. Based on Bayesian theorem, 
the objective function for O-LS-PCA is defined as a sum of weighted squared differences between "observed" and "simulated" data. One possible approach to make this optimization problem well-posed is 
to incorporate the prior information into the cost function as a regularization term to further
constrain the minimization procedure. Eventually, the cost function is described by a sum of two terms.
\begin{align}
\label{eq12}
J(\boldsymbol{\xi}_{G}) = \frac{1}{2}\boldsymbol{\xi}_{G}^{T}\boldsymbol{\xi}_{G} 
+\frac{1}{2}[\sum_{d=1}^{S}\textbf{T}^{d}(\boldsymbol{\beta}_{m}^{d}+\boldsymbol{\Phi}_{\boldsymbol{\beta}}^{d}\boldsymbol{\xi}^{d})-\boldsymbol{\beta}_{m}-\boldsymbol{\Phi}_{\boldsymbol{\beta}}\boldsymbol{\xi}_{G}]^{T}{\textbf{R}_{g}}^{-1}
[\sum_{d=1}^{S}\textbf{T}^{d}(\boldsymbol{\beta}_{m}^{d}+\boldsymbol{\Phi}_{\boldsymbol{\beta}}^{d}\boldsymbol{\xi}^{d})-\boldsymbol{\beta}_{m}-\boldsymbol{\Phi}_{\boldsymbol{\beta}}\boldsymbol{\xi}_{G}]
\end{align}
where, $\textbf{T}^{d} {\in} R^{N_{\beta} \times l_{d}} $ 
is a transformation matrix for mapping the grid position of local parameters at subdomain $\Omega^{d}$ to the corresponding grid position of global parameters at global domain. 
The elements of matrix $\textbf{T}^{d}$ are either 0 or 1.

Given LPCA coefficients $\boldsymbol{\xi}^{d}$, $d \in \{1,2,\cdot\cdot\cdot,S\}$, the corresponding GPCA coefficients $\boldsymbol{\xi}_{G}$ are obtained by minimizing cost function Eq.\ref{eq12}. 
The "simulated data" operator, e.g, $\boldsymbol{\Phi}_{\boldsymbol{\beta}}$ is linear,
we can analytically minimize the above objective function as follows
\begin{align}
\label{eq13}
[\frac{d J(\boldsymbol{\xi}_{G})}{d \boldsymbol{\xi}_{G}}]^{T} = \boldsymbol{\xi}_{G}-\boldsymbol{\Phi}_{\boldsymbol{\beta}}^{T}{\textbf{R}_{g}}^{-1}
[\sum_{d=1}^{S}\textbf{T}^{d}(\boldsymbol{\beta}_{m}^{d}+\boldsymbol{\Phi}_{\boldsymbol{\beta}}^{d}\boldsymbol{\xi}^{d})-\boldsymbol{\beta}_{m}-\boldsymbol{\Phi}_{\boldsymbol{\beta}}\boldsymbol{\xi}_{G}]=0
\end{align}
and the optimal GPCA coefficients are obtained
\begin{align}
\label{eq14}
\boldsymbol{\xi}_{G} =\frac{1}{2}\boldsymbol{\Phi}_{\boldsymbol{\beta}}^{T}{\textbf{R}_{g}}^{-1}\sum_{d=1}^{S}\textbf{T}^{d}\boldsymbol{\Phi}_{\boldsymbol{\beta}}^{d}\boldsymbol{\xi}^{d}
\end{align}

And then, the global parameter field is reconstructed using the global basis matrix $\boldsymbol{\Phi}_{\boldsymbol{\beta}}$ as follows,
\begin{equation}
\label{eq15}
\boldsymbol{\beta} = \boldsymbol{\beta}_{m}+ \frac{1}{2}\boldsymbol{\Phi}_{\boldsymbol{\beta}}\boldsymbol{\Phi}_{\boldsymbol{\beta}}^{T}{\textbf{R}_{g}}^{-1}\sum_{d=1}^{S}\textbf{T}^{d}\boldsymbol{\Phi}_{\boldsymbol{\beta}}^{d}\boldsymbol{\xi}^{d}
\end{equation}

\begin{figure}[!h]
\centering
\subfloat[]%
  {\includegraphics[width=0.45\linewidth]{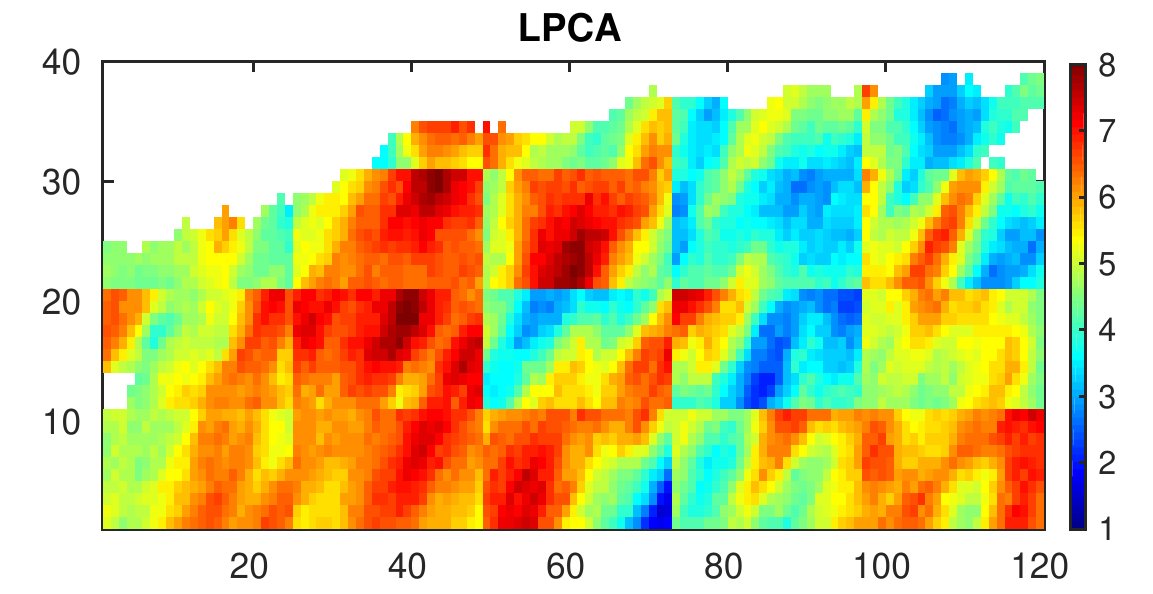}}
\subfloat[]%
  {\includegraphics[width=0.45\linewidth]{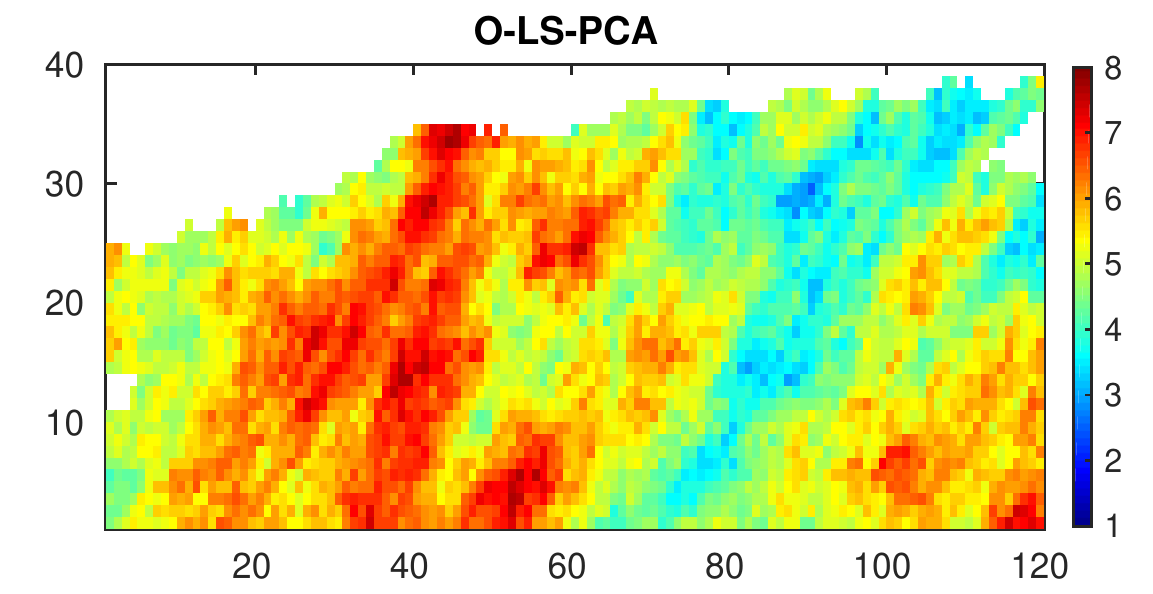}}
\caption{Illustration of parameter reconstruction, the left one is LPCA, the right one is O-LS-PCA}\label{fig3}
\end{figure}

Fig.\ref{fig3} shows the reconstructed parameter field using LPCA and O-LS-PCA
which demonstrates that O-LS-PCA procedure reconstructs a smooth parameter field in a low-order parameter subspace in each subdomain. 
From Eq.\ref{eq14} and Eq.\ref{eq15}, we can see that the O-LS-PCA algorithm is a linear transformation with smoothness and differentiability, 
which makes it particularly compatible with the gradient-based optimization problems.
In addition, we also show some results for different domain decomposition strategies, e.g., different number of subdomains. 
Table \ref{tab1} lists the retained number of PCA coefficients in each subdomain using O-LS-PCA. 
The more the number of subdomains, the smaller the number of LPCA coefficients in each subdomain while the more the total number of PCA coefficients. 
For the same number of subdomains, different domain decomposition strategies also yield different number of PCA coefficients.

\begin{table}[!h]
\footnotesize
\centering
\caption{Summary of total number of PCA coefficients for GPCA and O-LS-PCA}\label{tab1}
\begin{spacing}{1.25}
\begin{tabular}{|c|c|c|c|}
\hline
  & \multicolumn{2}{|c|}{O-LS-PCA} & GPCA \\
\cline{2-4}
  &  $N_{L}=\sum_{d=1}^{S}l_{d}$ & $\max$\{$l_{1}$,...,$l_{S}$\} & $N_{G}$ \\
\hline
3 $\times$ 4 & 236 & 18 & \multirow{4}{*}{48} \\
4 $\times$ 5 & 275 & 15 & \multirow{4}{*}{}  \\
2 $\times$ 10 & 278 & 14 & \multirow{4}{*}{}  \\
5 $\times$ 6 & 312 & 12 & \multirow{4}{*}{} \\
\hline
\end{tabular}
\end{spacing}
\end{table}

\subsection{The minimum number of local PCA patterns}
Theoretically speaking, to fully cover the original global PCA patterns in this local parameterization, we should retain the number of local PCA patterns as much as possible,
however, retaining a large number of local PCA patterns requires numerous FOM simulations for implementing subdomain POD-TPWL, which has been demonstrated in our previous work \cite{Xiao2018}. 
Thus, it is essential to determine an appropriate number of local PCA patterns to trade-off both sides.

Given LPCA coefficients $\boldsymbol{\xi}_{L}$=[$\boldsymbol{\xi}^{1}$,...,$\boldsymbol{\xi}^{d}$,...$\boldsymbol{\xi}^{S}$]$^{T}$, $d \in \{1,2,\cdot\cdot\cdot,S\}$, 
the corresponding GPCA coefficients $\boldsymbol{\xi}_{G}$ are obtained using local parameterization as follows
\begin{align}
\label{eq16}
\boldsymbol{\xi}_{G} = \textbf{T}_{GL} \boldsymbol{\xi}_{L}
\end{align}
where $\textbf{T}_{GL} {\in} R^{N_{G} \times N_{L}} $ is a transformation matrix for converting LPCA coefficients to GPCA coefficients which can be determined using Eq.\ref{eq15}.
A compact expression of $\textbf{T}_{GL}$ can be presented as follows
\begin{align}
\label{eq14}
\textbf{T}_{GL} =\frac{1}{2}\boldsymbol{\Phi}_{\boldsymbol{\beta}}^{T}{\textbf{R}_{g}}^{-1} [\textbf{T}^{1}\boldsymbol{\Phi}_{\boldsymbol{\beta}}^{1},
\textbf{T}^{2}\boldsymbol{\Phi}_{\boldsymbol{\beta}}^{2}, ...,\textbf{T}^{d}\boldsymbol{\Phi}_{\boldsymbol{\beta}}^{d},
...,\textbf{T}^{S}\boldsymbol{\Phi}_{\boldsymbol{\beta}}^{S}]
\end{align}

The global projection basis matrix $\boldsymbol{\Phi}_{\boldsymbol{\beta}}$ can be unfolded as follows
\begin{equation}
\label{eq17}
\boldsymbol{\Phi}_{\boldsymbol{\beta}} = [\boldsymbol{\nu}_{1},  \boldsymbol{\nu}_{2}, ...,\boldsymbol{\nu}_{i},..., \boldsymbol{\nu}_{N_{G}}]
\end{equation}
where, $ \boldsymbol{\nu}_{i} {\in} R^{N_{\beta}} $, $i \in \{1,2,\cdot\cdot\cdot,N_{G}\}$ is the global basis vector. 
The global PCA coefficients corresponding to each basis vector $\boldsymbol{\nu}_{i}$ are as follows
\begin{align}
\label{eq18}
\boldsymbol{\nu}_{i} = [\boldsymbol{\nu}_{1},  \boldsymbol{\nu}_{2}, ...,\boldsymbol{\nu}_{i},..., \boldsymbol{\nu}_{N_{G}}] \times \textbf{e}_{i}^{T}, i=1,2,...,N_{G}
\end{align}
where $\textbf{e}_{i}$ is a unit vector with only a 1 at position $i$. To fully cover the original global PCA patterns in this local parameterization procedure, the following equation should have solution.
\begin{align}
\label{eq19}
\textbf{T}_{GL} \boldsymbol{\xi}_{L}^{i} = \textbf{e}_{i}^{T}, i=1,2,...,N_{G}
\end{align}

In theory, the transformation matrix $\textbf{T}_{GL}$ should be full-row rank, that is to say, the total number of local PCA patterns $N_{L}$ is at least equal to or larger than the number of global 
PCA patterns $N_{G}$. We should note that this is a necessary condition, but not a sufficient condition. The rank of transformation matrix $\textbf{T}_{GL}$ cannot be explicitly derived from Eq.\ref{eq14}
, some numerical tests will be used to determine $N_{L}$. 
Here we assume that $\boldsymbol{\xi}_{L}^{i}$ is the solution of the $i$th equation, and the reconstructed $i$th global PCA
basis vector is $\boldsymbol{\nu}_{i}^{*}$. $RMSE_{i}$ is a root mean square error to characterize the accuracy of the reconstructed global PCA
basis vector $\boldsymbol{\nu}_{i}$. Finally, the total RMSE for all the global PCA basis vectors is defined as follows
\begin{align}
\label{eq20}
& RMSE_{i} = \|\boldsymbol{\nu}_{i}-\boldsymbol{\nu}_{i}^{*}\|_{2} = \|\boldsymbol{\nu}_{i}-[\boldsymbol{\nu}_{1},  \boldsymbol{\nu}_{2}, ...,\boldsymbol{\nu}_{i},..., \boldsymbol{\nu}_{N_{G}}] \times \textbf{T}_{GL} \boldsymbol{\xi}_{L}^{i}\|_{2}  \notag \\
& RMSE = \sum_{i=1}^{N_{G}} RMSE_{i} 
\end{align}

\begin{figure*}[!h]
\centering
\subfloat[]{%
  \includegraphics[width=0.45\linewidth]{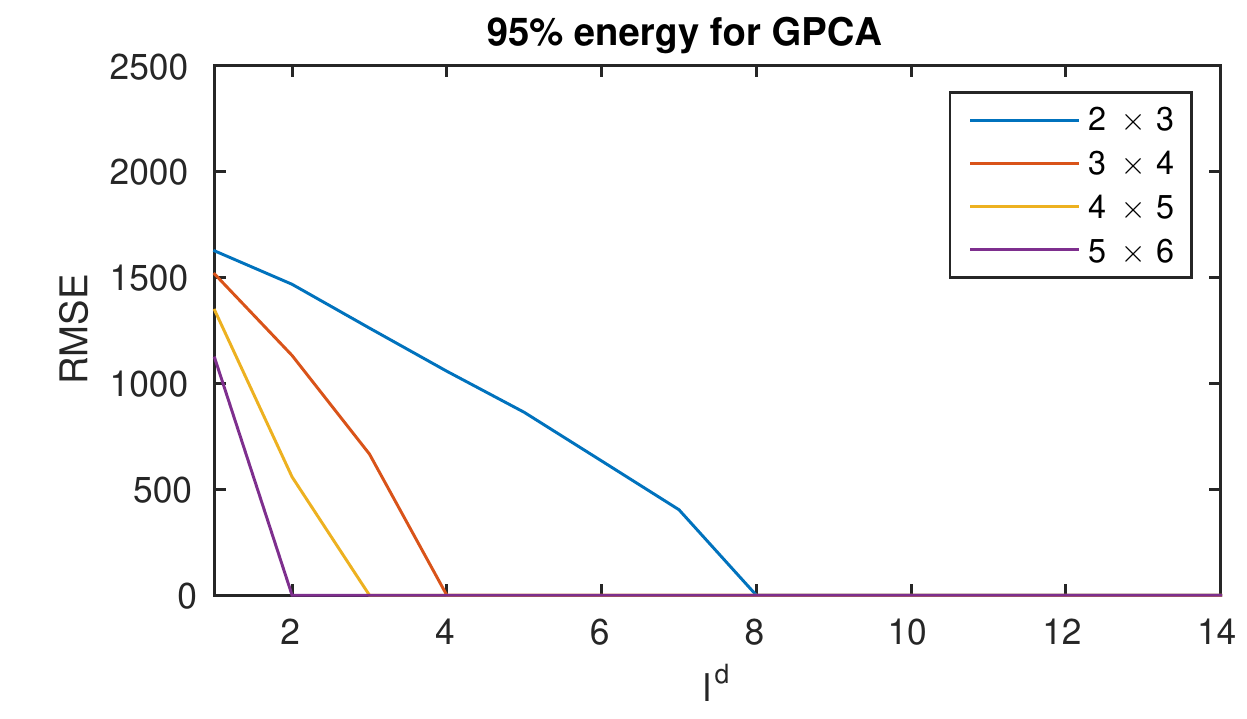}}
\subfloat[]{%
\includegraphics[width=0.45\linewidth]{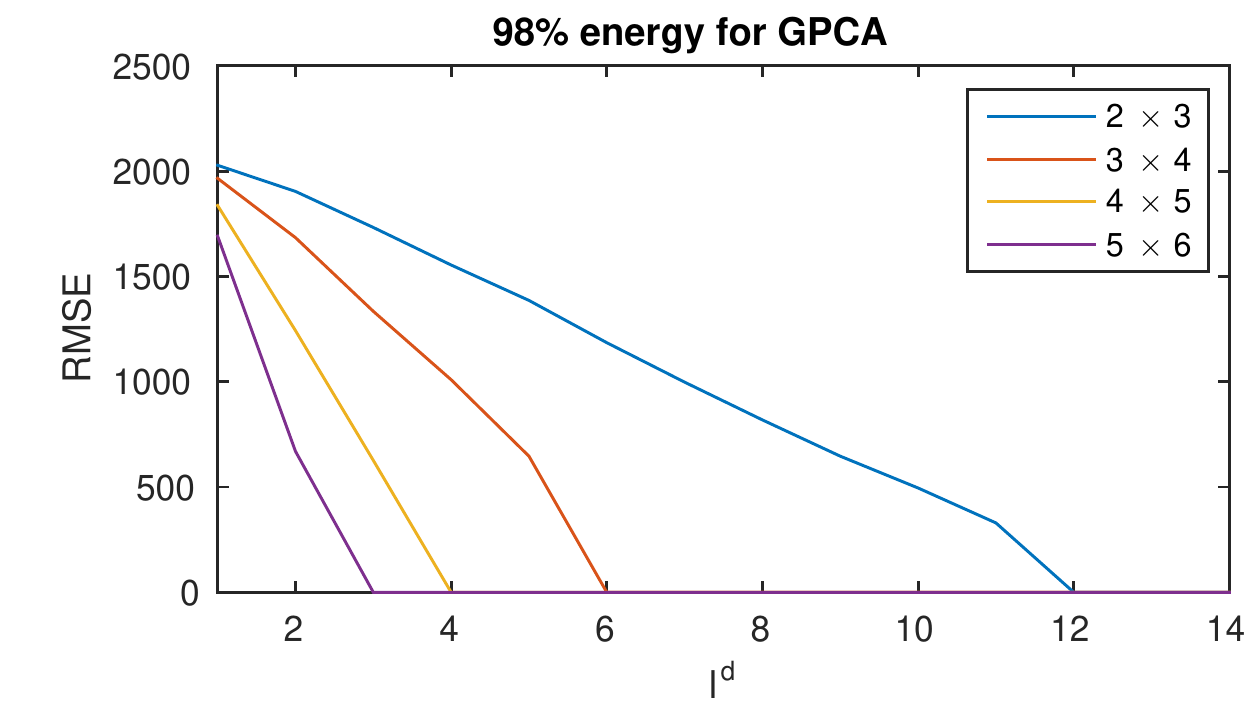}}\\
\caption{The RMSE Eq.\ref{eq20} for different number of local PCA patterns in each subdomain}\label{fig5}
\end{figure*}

\begin{figure*}[!h]
\centering
\subfloat[1st Basis vector $\boldsymbol{\nu}_{1}$]{%
  \includegraphics[width=0.32\linewidth]{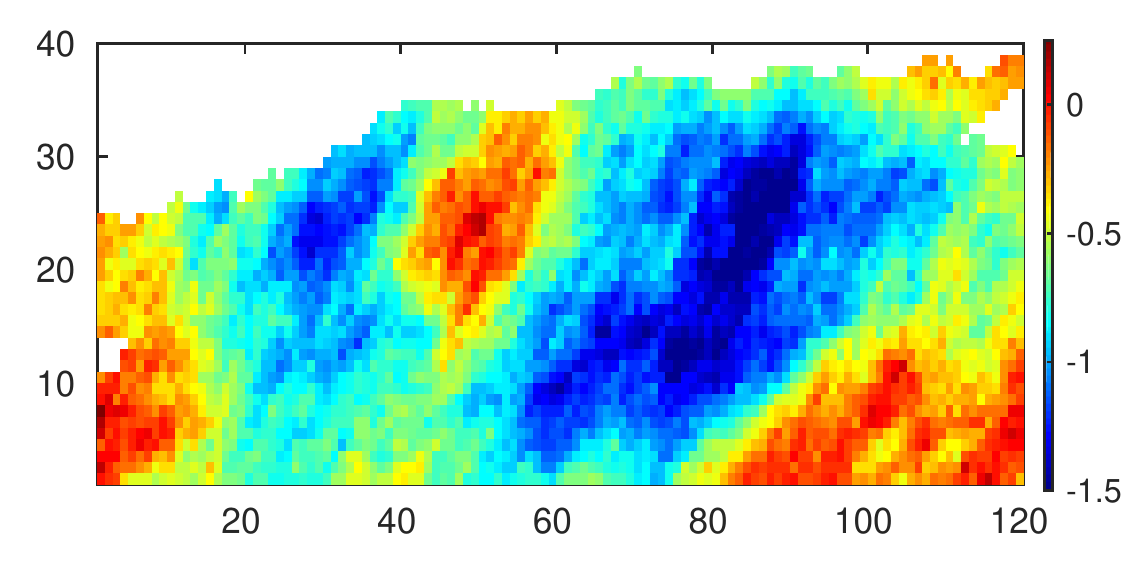}} \\
\subfloat[2 $\times$ 3]{%
  \includegraphics[width=0.32\linewidth]{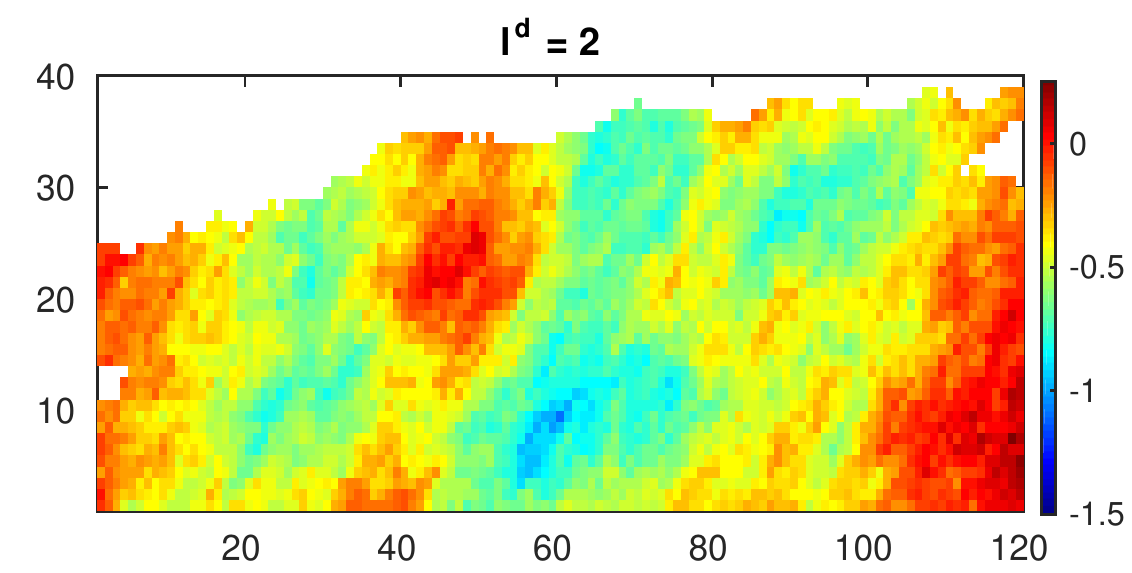}} 
\subfloat[2 $\times$ 3]{%
\includegraphics[width=0.32\linewidth]{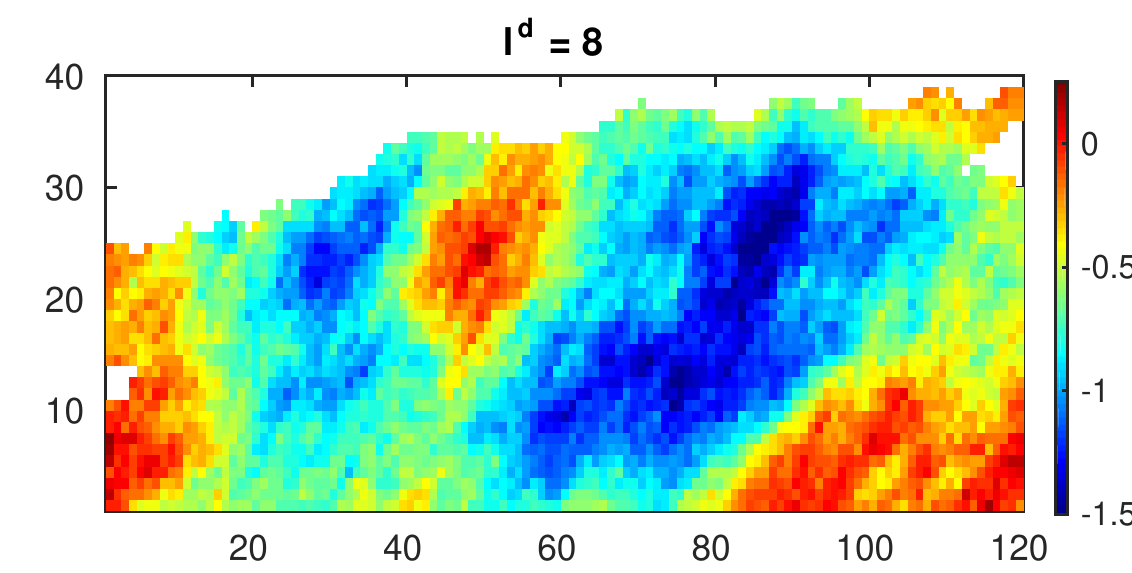}}
\subfloat[2 $\times$ 3]{%
\includegraphics[width=0.32\linewidth]{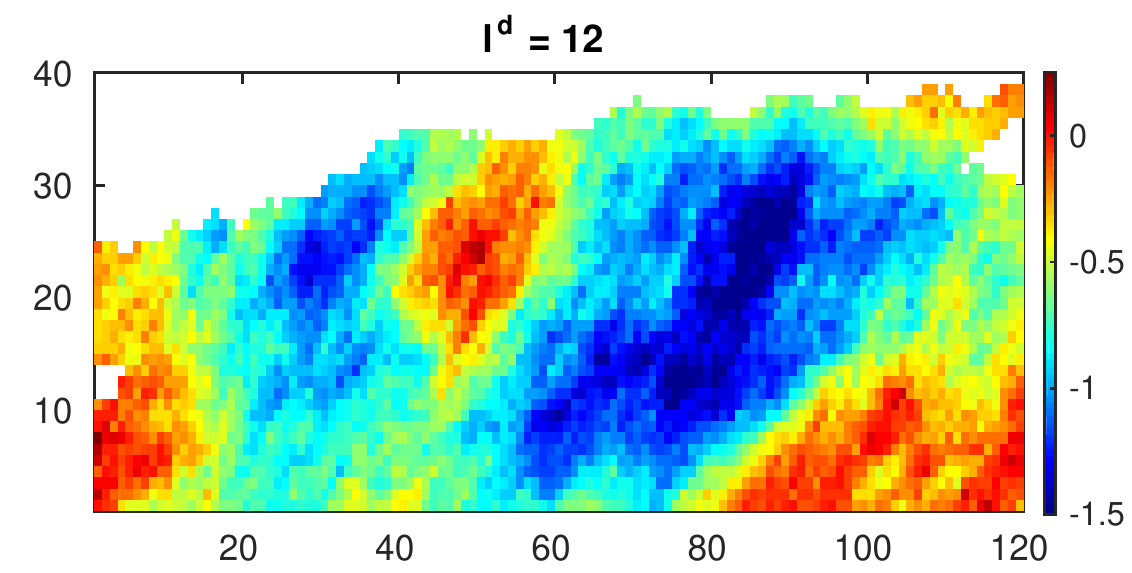}} \\

\subfloat[3 $\times$ 4]{%
  \includegraphics[width=0.32\linewidth]{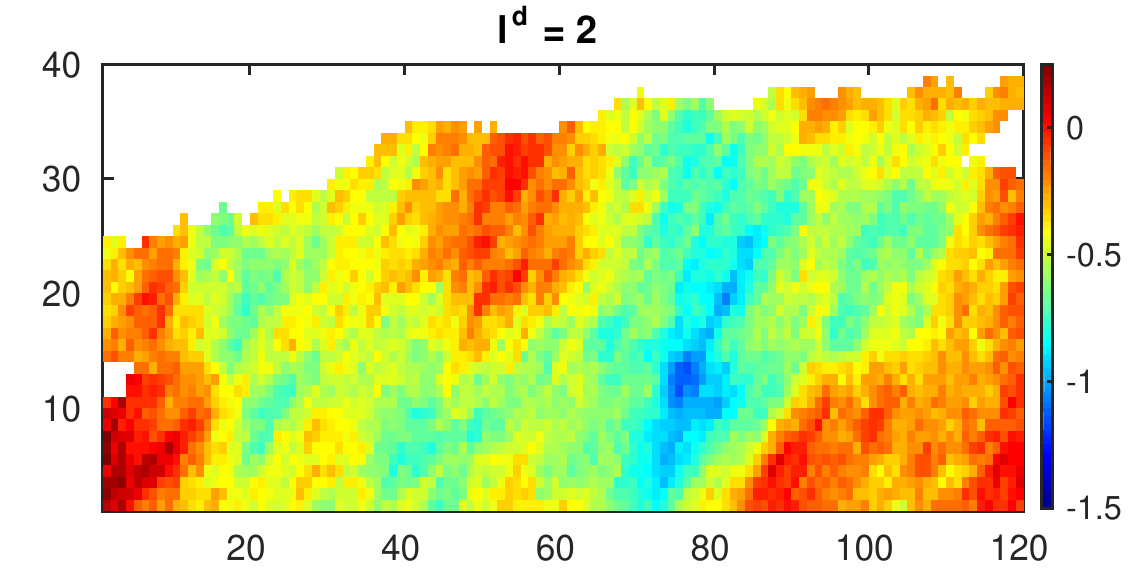}} 
\subfloat[3 $\times$ 4]{%
\includegraphics[width=0.32\linewidth]{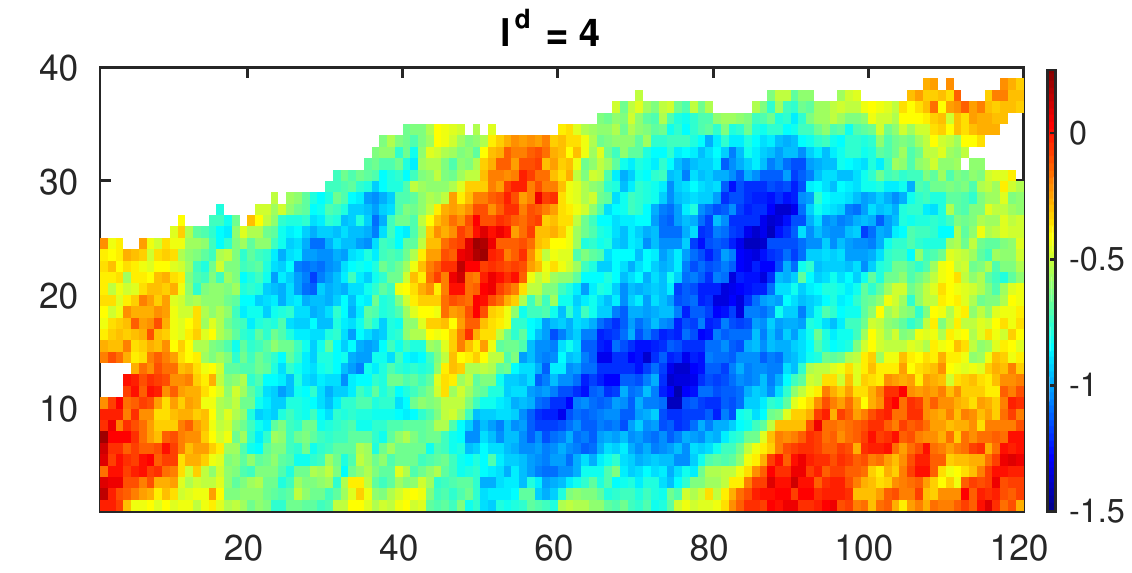}}
\subfloat[3 $\times$ 4]{%
\includegraphics[width=0.32\linewidth]{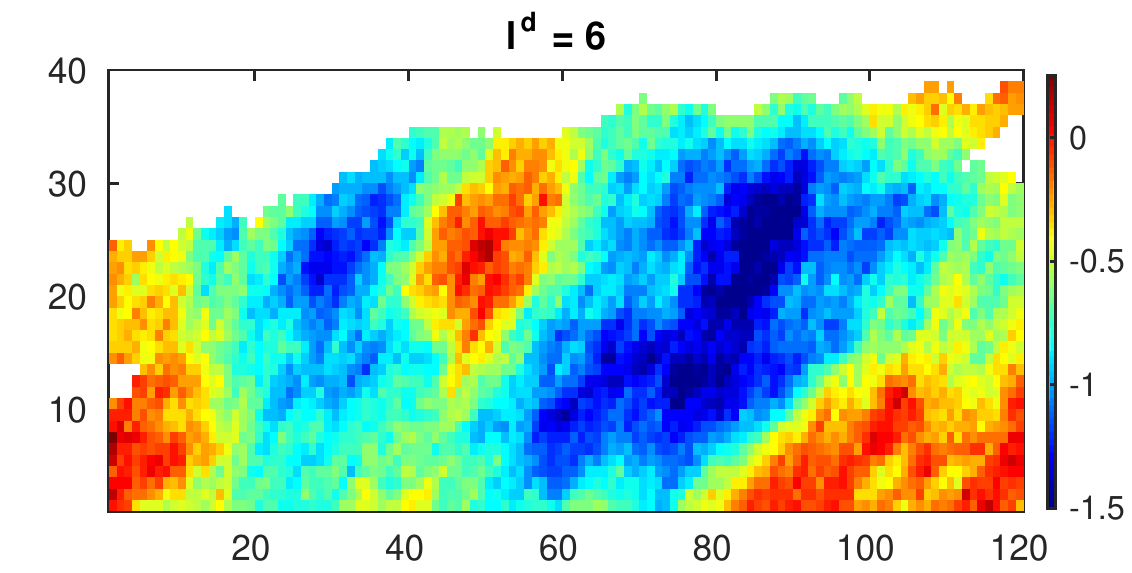}} \\

\subfloat[4 $\times$ 5]{%
  \includegraphics[width=0.32\linewidth]{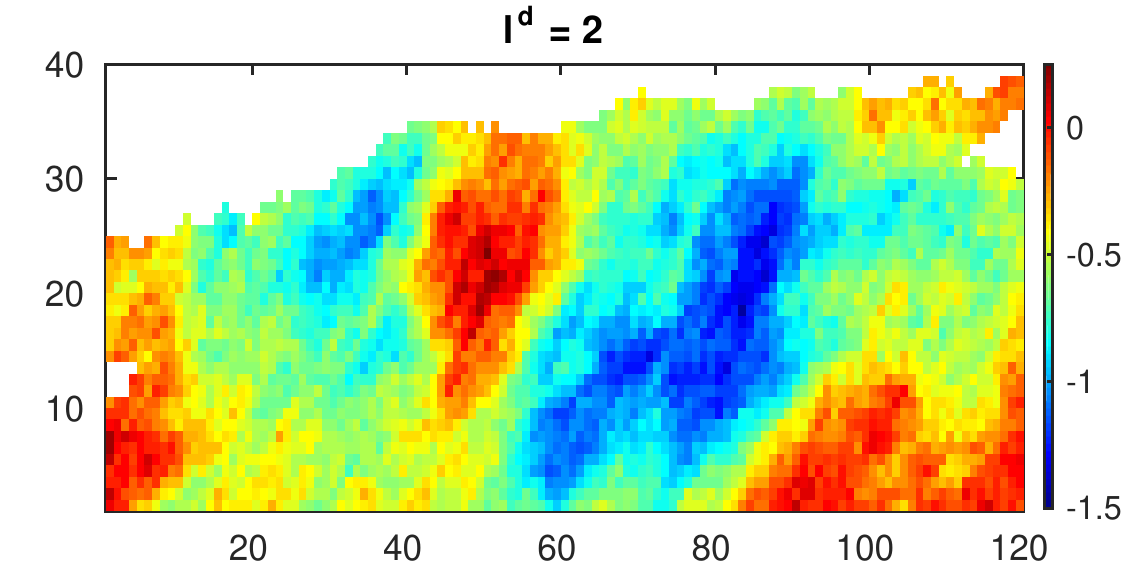}} 
\subfloat[4 $\times$ 5]{%
\includegraphics[width=0.32\linewidth]{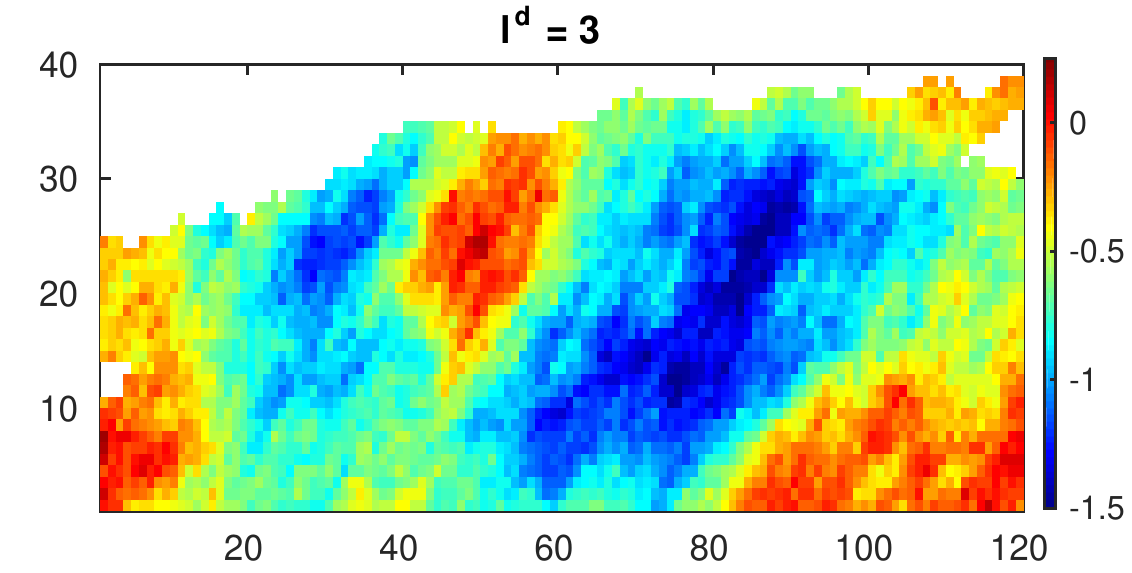}}
\subfloat[4 $\times$ 5]{%
\includegraphics[width=0.32\linewidth]{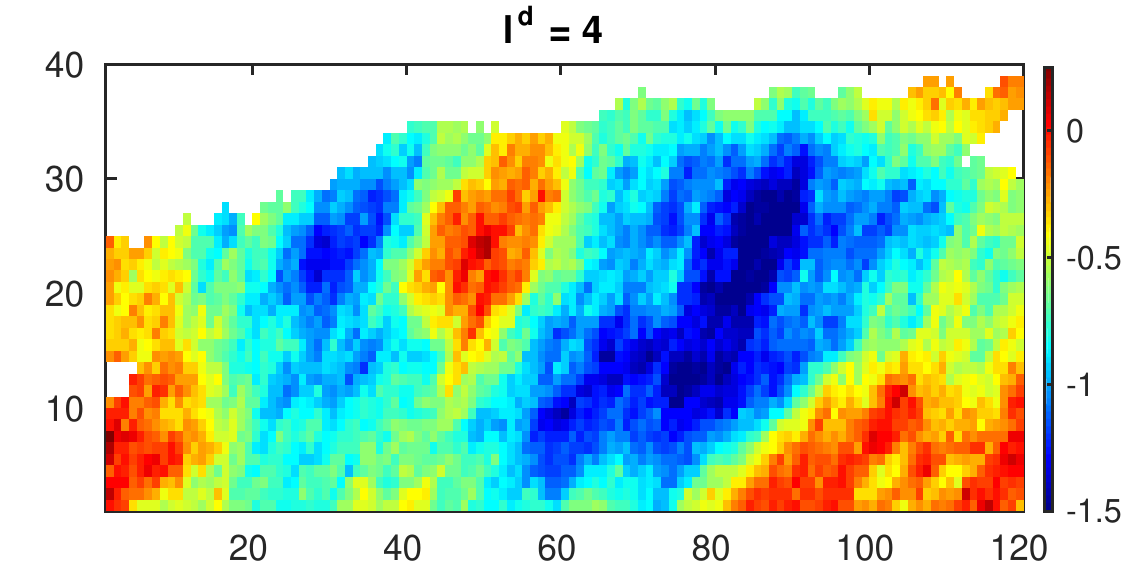}} \\

\caption{The reconstructed 1st global basis vector using a different number of local PCA patterns in each subdomain}\label{fig6}
\end{figure*}

\begin{figure*}[!h]
\centering
\subfloat[72nd Basis vector $\boldsymbol{\nu}_{72}$]{%
  \includegraphics[width=0.32\linewidth]{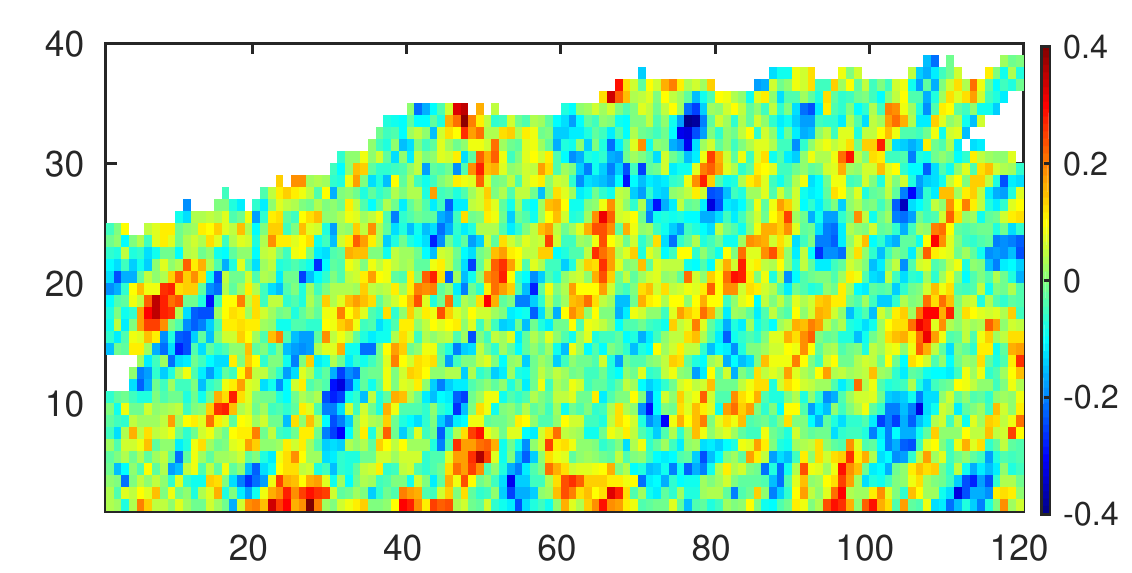}} \\
\subfloat[2 $\times$ 3]{%
  \includegraphics[width=0.32\linewidth]{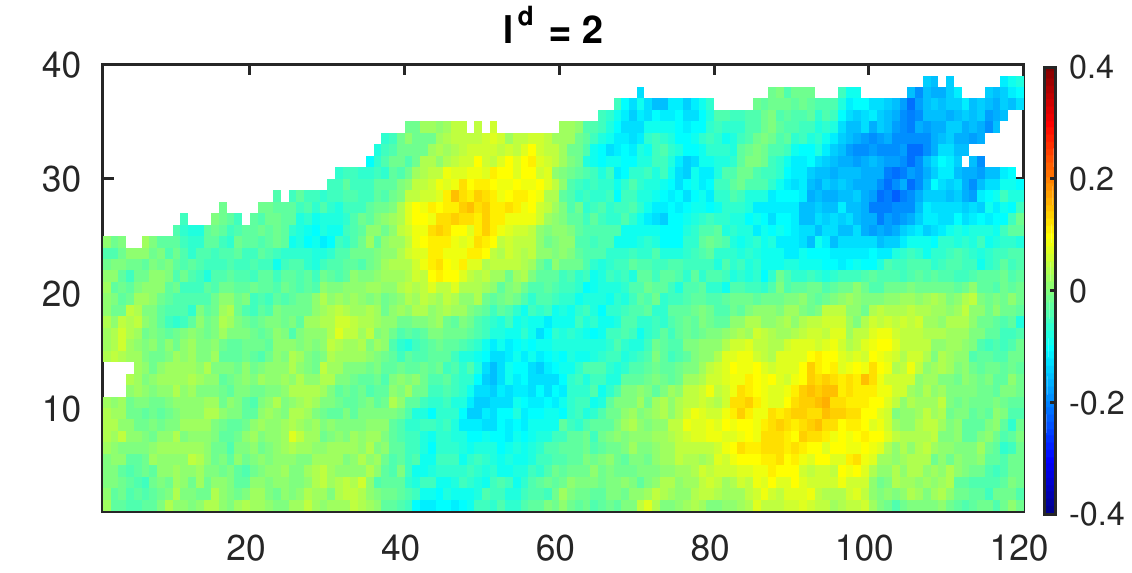}} 
\subfloat[2 $\times$ 3]{%
\includegraphics[width=0.32\linewidth]{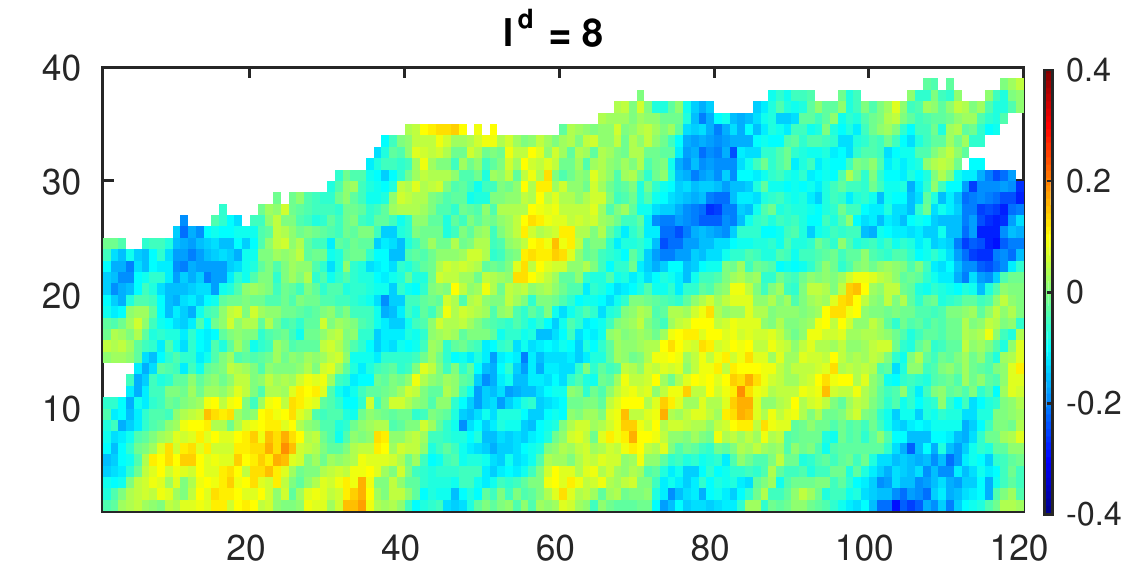}}
\subfloat[2 $\times$ 3]{%
\includegraphics[width=0.32\linewidth]{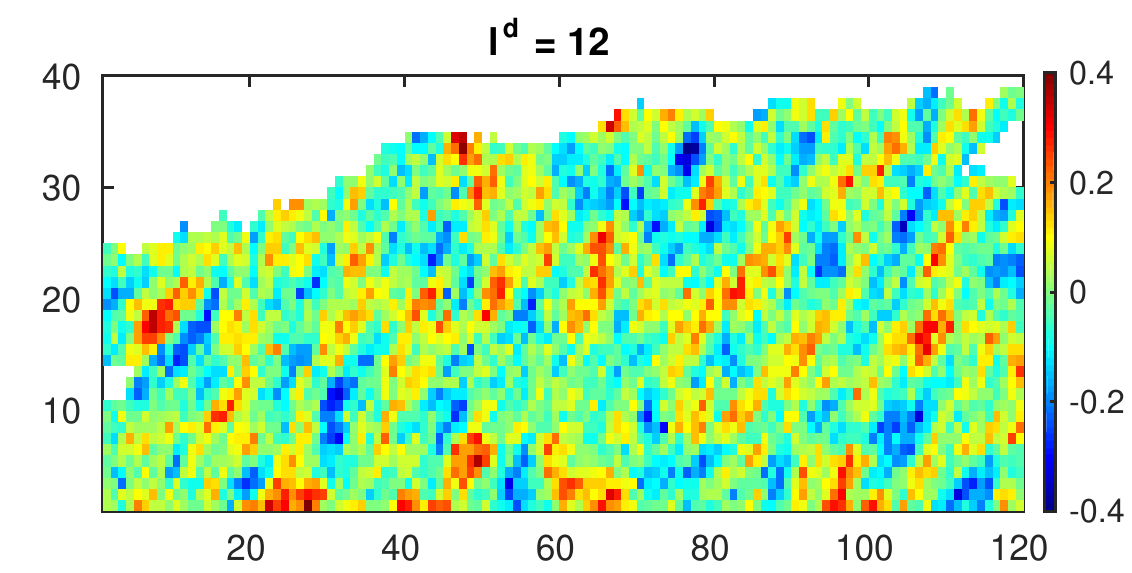}} \\

\subfloat[3 $\times$ 4]{%
  \includegraphics[width=0.32\linewidth]{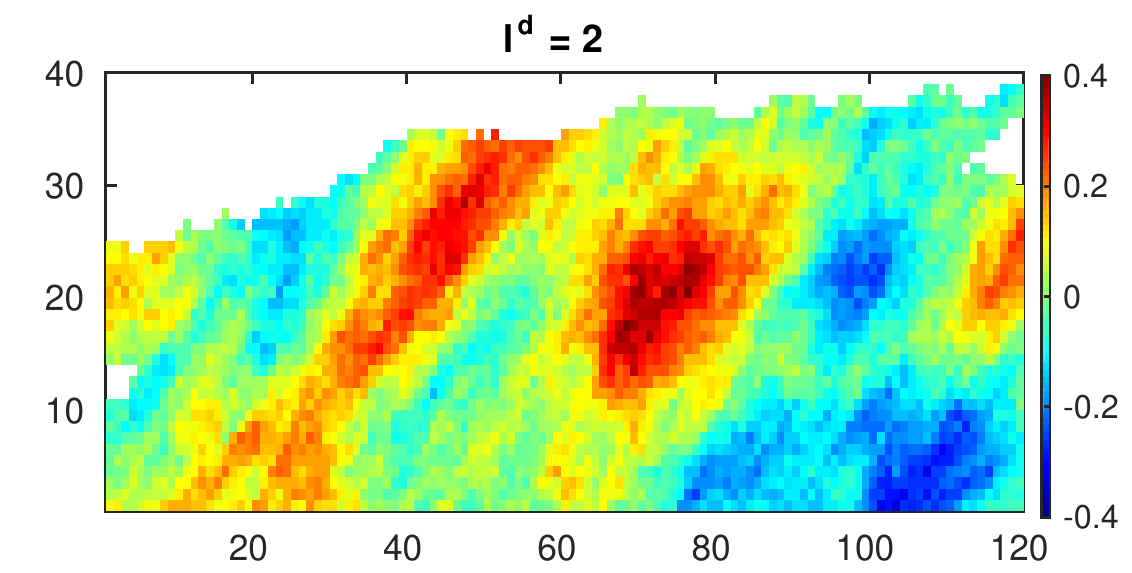}} 
\subfloat[3 $\times$ 4]{%
\includegraphics[width=0.32\linewidth]{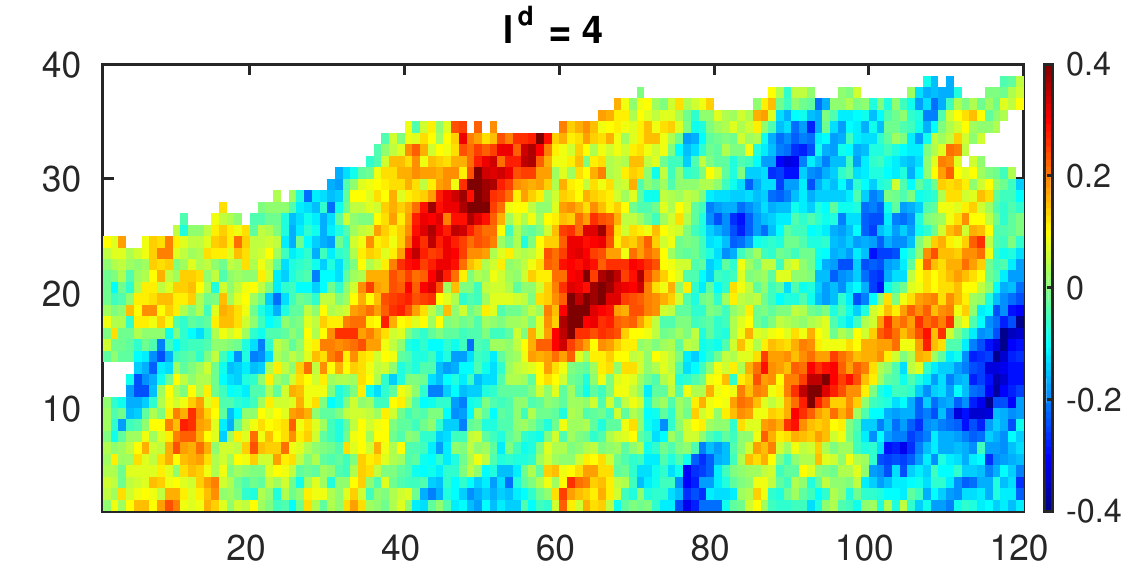}}
\subfloat[3 $\times$ 4]{%
\includegraphics[width=0.32\linewidth]{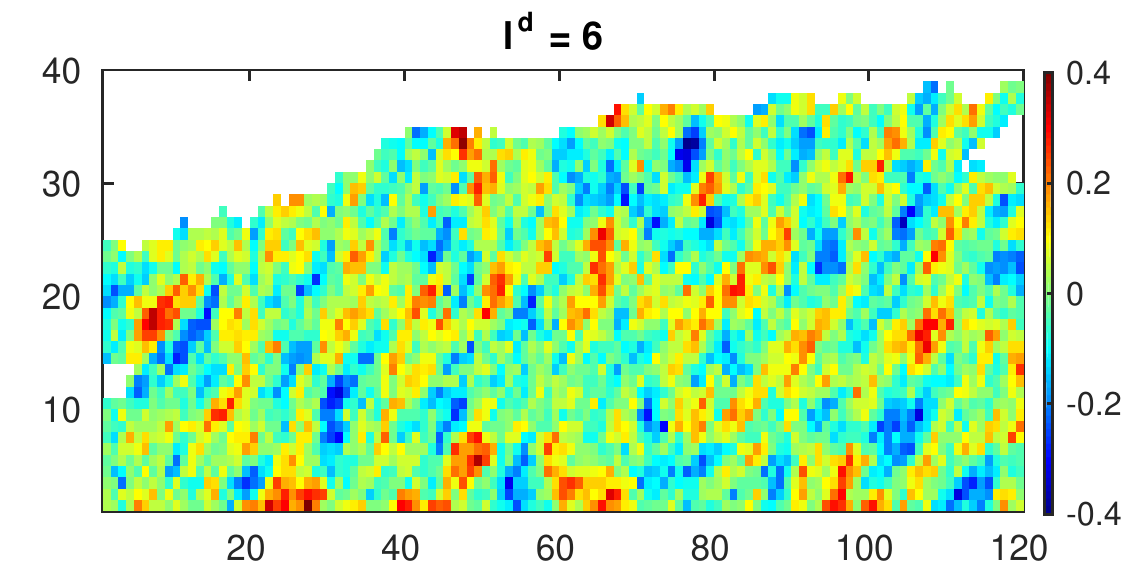}} \\

\subfloat[4 $\times$ 5]{%
  \includegraphics[width=0.32\linewidth]{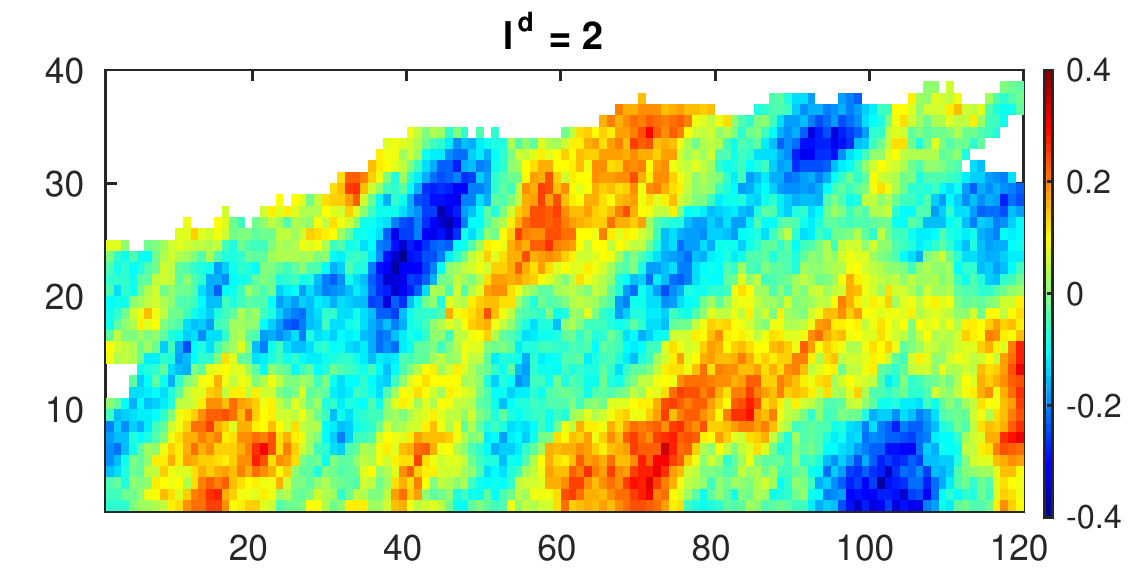}} 
\subfloat[4 $\times$ 5]{%
\includegraphics[width=0.32\linewidth]{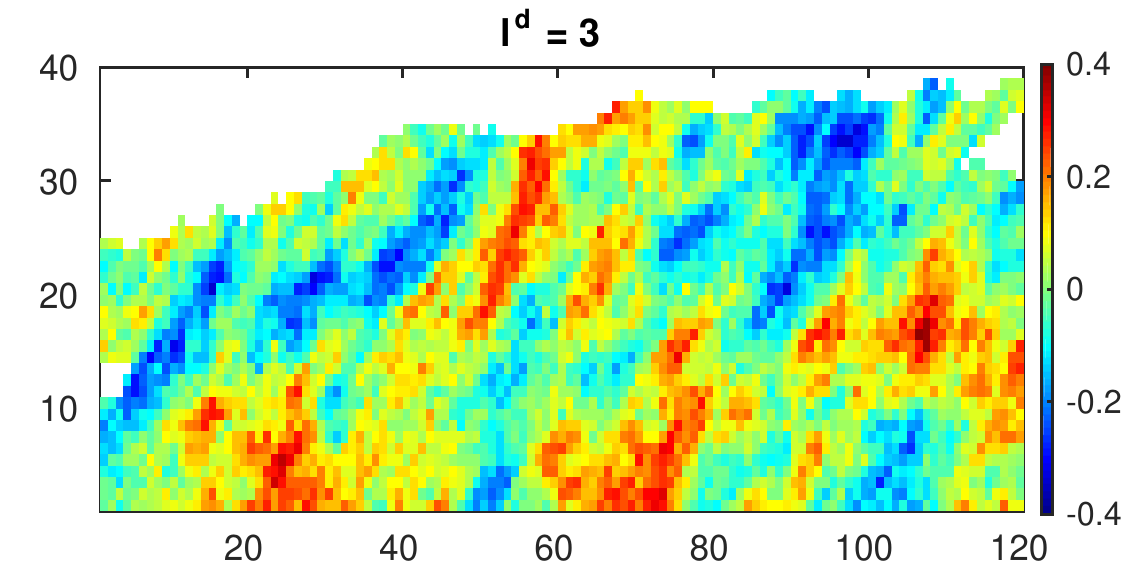}}
\subfloat[4 $\times$ 5]{%
\includegraphics[width=0.32\linewidth]{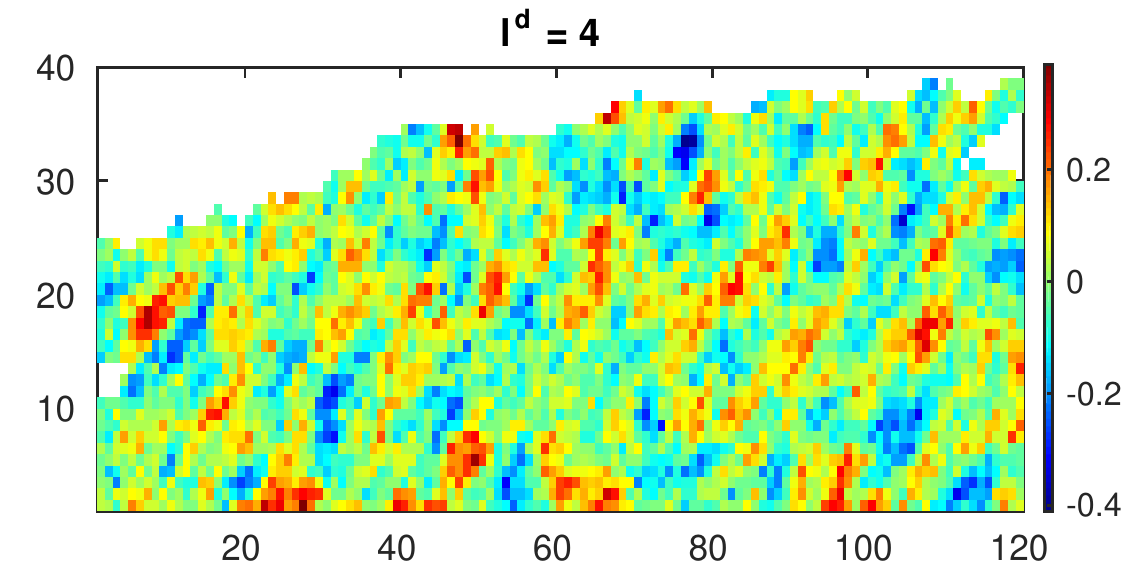}} \\

\caption{The reconstructed 72nd global basis vector using a different number of local PCA patterns in each subdomain}\label{fig7}
\end{figure*}

We continuously increase the number of local PCA patterns to make the total RMSE as small as possible. Fig.\ref{fig5} shows the change of RMSE with respect to the number of local PCA patterns.
It is demonstrated that RMSE indeed approaches to zero when we retain a certain number of local PCA patterns and therefore 
we can specify this certain value to the minimum number of local PCA patterns for fully covering the global PCA patterns.
Table \ref{tab2} summarizes the minimum number of local PCA patterns corresponding to different domain decomposition. 
As long as the total number of local PCA patterns is equal to or slightly larger than the number of global PCA patterns, an optimal local parameterization has been obtained. 
Fig.\ref{fig6} and Fig.\ref{fig7} separately shows the 1st global basis vector $\boldsymbol{\nu}_{1}$, 72nd global basis vector $\boldsymbol{\nu}_{72}$ and their corresponding reconstructed 
$\boldsymbol{\nu}_{1}^{*}$ and $\boldsymbol{\nu}_{72}^{*}$ by retaining different number of local PCA patterns. 
The true basis vectors can be perfectly reconstructed when the retained number of local PCA patterns is larger than the minimum value.
We should note that although this holds for there, this methodology is very general and therefore can be applied to any other parameter fields.

\begin{table}[!h]
\footnotesize
\centering
\caption{The minimum total number of local PCA patterns when RMSE = 0}\label{tab2}
\begin{spacing}{1.25}
\begin{tabular}{|c|c|c|c|c|}
\hline
 Energy for GPCA  & \multicolumn{2}{|c|}{95\%} & \multicolumn{2}{|c|}{98\%} \\
\hline
$N_{G}$  & \multicolumn{2}{|c|}{48} & \multicolumn{2}{|c|}{72} \\
\hline
Domain Decomposition & $l^{d}$ & $N_{L}$ & $l^{d}$ & $N_{L}$ \\
\hline
2$\times$3  & 8 & 48 & 12 & 72  \\
3$\times$4  & 4 & 48 & 6 & 72  \\
4$\times$5  & 3 & 60 & 4 & 80  \\
5$\times$6  & 2 & 60 & 3 & 90  \\
\hline
\end{tabular}
\end{spacing}
\end{table}

\section{Subdomain Adjoint-based History Matching Using Local Parameterization}
In the following, we will briefly introduce a procedure on how to implement the subdomain adjoint-based history matching using local parameterization. 
Detailed information about subdomain POD-TPWL algorithm can be found in \cite{Xiao2018}. Due to the introduction of local parameterization in each subdomain, 
the original subdomain POD-TPWL needs some modifications.

\subsection{Modifications of Subdomain POD-TPWL}
According to the traditional POD-TPWL algorithm \cite{he2015constraint}, the nonlinear reservoir model as Eq.\ref{eq1} is linearized around a "closest" trajectory 
specified through certain criteria \cite{he2015constraint}. In addition to the model linearization, the global parameters $\boldsymbol{\beta}$ and model state $\textbf{x}$ 
are simultaneously reduced using global PCA and POD, respectively.  
Finally, the reduced-order linear model of Eq.\ref{eq1} is reformulated as 
\begin{equation}
\label{eq21}
\boldsymbol{\psi}^{n+1} {\approx} \boldsymbol{\psi}_{tr}^{n+1}+\textbf{E}_{\boldsymbol{\psi}_{tr}}^{n+1}(\boldsymbol{\psi}^{n}-\boldsymbol{\psi}_{tr}^{n})+\textbf{G}_{\boldsymbol{\xi}_{tr}}^{n+1}(\boldsymbol{\xi}-\boldsymbol{\xi}_{tr})
\end{equation}

Similarly, the well model defined as Eq.\ref{eq2} or Eq.\ref{eq3} (here we compress the subscript for the type of data to simplify the notation) is also linearized as
\begin{equation}
\label{eq22}
\textbf{y}^{m+1} {\approx} \textbf{y}_{tr}^{m+1}+\textbf{A}_{\boldsymbol{\psi}_{tr}}^{m+1}(\boldsymbol{\psi}^{m+1}-\boldsymbol{\psi}_{tr}^{m+1})+\textbf{B}_{\boldsymbol{\xi}_{tr}}^{m+1}(\boldsymbol{\xi}-\boldsymbol{\xi}_{tr})
\end{equation}
In Eq.\ref{eq21} $\sim$ Eq.\ref{eq22}, the original state vector $\textbf{x}$ is represented using a low-order POD coefficient
vector $\boldsymbol{\psi}$. $\boldsymbol{\xi}$ denotes the vector of reduced parameter coefficients in global domain. 
The matrix, $\textbf{E}_{\boldsymbol{\psi}}$, $\textbf{G}_{\boldsymbol{\xi}}$, $\textbf{A}_{\boldsymbol{\psi}}$, $\textbf{B}_{\boldsymbol{\xi}}$ are the required derivative matrices. 

In subdomain POD-TPWL \cite{Xiao2018}, the radial basis function (RBF) interpolation and domain decomposition (DD) are applied
to estimate these derivative matrices using locally low-dimensional 
RBF interpolation with a small number of interpolation points. 
The entire domain $\Omega$ is assumed to be decomposed into $\textit{S}$ non-overlapping subdomains $\Omega^{d}$ , $d \in \{1,2,\cdot\cdot\cdot,S\}$
(such that $\Omega=\bigcup_{d=1}^{S}\Omega^{d} $ and $ \Omega^{i} \cap \Omega^{j}=0 $ for $ i\neq j$). 
The reduced-order linear model Eq.\ref{eq21} is further modified to represent the underlying dynamic associated with subdomain $\Omega^{d}$ and its neighboring subdomains $\Omega^{sd}$.

Finally, the coefficient $\boldsymbol{\psi}^{d,n+1}$ for the subdomain $\Omega^{d}$ is presented as follows
\begin{align}
\label{eq23}
\boldsymbol{\psi}^{d,n+1} \approx \boldsymbol{\psi}_{tr}^{d,n+1}+\textbf{E}_{\boldsymbol{\psi}_{tr}}^{d,n+1}(\boldsymbol{\psi}^{d,n}-\boldsymbol{\psi}_{tr}^{d,n}) 
+\textbf{E}_{\boldsymbol{\psi}_{tr}}^{sd,n+1}(\boldsymbol{\psi}^{sd,n+1}-\boldsymbol{\psi}_{tr}^{sd,n+1})+\textbf{G}_{\boldsymbol{\xi}_{tr}^{d}}^{n+1}(\boldsymbol{\xi}^{d}-\boldsymbol{\xi}_{tr}^{d})
\end{align}
\begin{equation}
\label{eq24}
\textbf{E}_{\boldsymbol{\psi}_{tr}}^{d,n+1} \approx \frac{\partial \boldsymbol{\pounds}^{d,n+1}}{\partial \boldsymbol{\psi}_{tr}^{d,n}}, \quad \textbf{E}_{\boldsymbol{\psi}_{tr}}^{sd,n+1} \approx \frac{\partial \boldsymbol{\pounds}^{d,n+1}}{\partial \boldsymbol{\psi}_{tr}^{sd,n+1}}, 
\quad \textbf{G}_{\boldsymbol{\xi}_{tr}^{d}}^{n+1} \approx \frac{\partial \boldsymbol{\pounds}^{d.n+1}}{\partial \boldsymbol{\xi}_{tr}^{d}}
\end{equation}

Similarly, the simulated measurements $\textbf{y}^{d,m+1}$ of the subdomain $\Omega^{d}$ is reformulated as
\begin{equation}
\label{eq25}
\textbf{y}^{d,m+1} \approx \textbf{y}_{tr}^{d,m+1}+\textbf{A}_{\boldsymbol{\psi}_{tr}}^{d,m+1}(\boldsymbol{\psi}^{d,m+1}-\boldsymbol{\psi}_{tr}^{d,m+1})+\textbf{B}_{\boldsymbol{\xi}_{tr}^{d}}^{m+1}(\boldsymbol{\xi}^{d}-\boldsymbol{\xi}_{tr}^{d})
\end{equation}
\begin{equation}
\label{eq26}
\textbf{A}_{\boldsymbol{\psi}_{tr}}^{d,m+1} \approx \frac{\partial \boldsymbol{\hbar}^{d,n+1}}{\partial \boldsymbol{\psi}_{tr}^{d,m+1}}, \textbf{B}_{\boldsymbol{\xi}_{tr}^{d}}^{m+1} \approx \frac{\partial \boldsymbol{\hbar}^{d,n+1}}{\partial \boldsymbol{\xi}_{tr}^{d}}
\end{equation}
where, $\boldsymbol{\pounds}^{d,n+1}(\boldsymbol{\psi}^{d,n},\boldsymbol{\psi}^{sd,n+1},\boldsymbol{\xi}^{d})$ denotes a RBF interpolation function for the POD coefficient $\boldsymbol{\psi}^{d,n+1}$ 
at timestep $\textit{n}$+1 for subdomain $\Omega^{d}$
. $\boldsymbol{\hbar}^{d,n+1}(\boldsymbol{\psi}^{d,m+1},\boldsymbol{\xi}^{d})$ denotes a RBF interpolation function for the simulated measurements $\textbf{y}^{d,m+1}$ at 
 timestep $\textit{m}$+1 for subdomain $\Omega^{d}$.

The RBF interpolation for $\boldsymbol{\pounds}^{d,n+1}$ and $\boldsymbol{\hbar}^{d,n+1}$ are represented as a linear combination of $\textit{M}$ radial basis functions in the form of,
\begin{align}
\label{eq27}
\boldsymbol{\pounds}^{d,n+1}(\boldsymbol{\psi}^{d,n},\boldsymbol{\psi}^{sd,n+1},\boldsymbol{\xi}^{d}) 
= \sum_{j=1}^{M}\boldsymbol{\omega}_{j}^{d,n+1}\times \theta (||(\boldsymbol{\psi}^{d,n},\boldsymbol{\psi}^{sd,n+1},\boldsymbol{\xi}^{d})-(\boldsymbol{\psi}_{j}^{d,n},\boldsymbol{\psi}_{j}^{sd,n+1},\boldsymbol{\xi}_{j}^{d})||)
\end{align}

\begin{align}
\label{eq28}
\boldsymbol{\hbar}^{d.n+1}(\boldsymbol{\psi}^{d,m+1},\boldsymbol{\xi}^{d})= \sum_{j=1}^{M}\boldsymbol{\varepsilon}_{j}^{d,m+1}\times \theta (\|(\boldsymbol{\psi}^{d,m+1},\boldsymbol{\xi}^{d})-(\boldsymbol{\psi}_{j}^{d,m+1},\boldsymbol{\xi}_{j}^{d})\|)
\end{align}

This LP-SD POD-TPWL algorithm also consists of an offline stage and an online stage. 
(1) The offline stage describes a local parameterization and a computational procedure on how to construct a set of local 
RBF and estimate the derivative information for each subdomain. 
(2) The online stage describes how to iteratively implement LP-SD POD-TPWL where the dynamic interactions between a subdomain and its surrounding subdomains are considered. 
We represent the dynamic interactions among neighboring subdomains using an implicit formula, thus, additional iterations are required.
More information on the determination of radial basis function $\theta$, its corresponding weighting coefficients $\boldsymbol{\omega}$ 
and the iterative implementation of LP-SD POD-TPWL can be found in \cite{Xiao2018}.

\subsection{Mathematical formulation of Adjoint-based parameter estimation}
It is convenient to incorporate the reduced-order linear model as Eq.\ref{eq23} and Eq.\ref{eq25} into an adjoint-based reservoir history matching.
the objective function $\hat{\jmath}(\boldsymbol{\xi})$ as Eq.\ref{eq4} or Eq.\ref{eq5} can be computed using reduced-order linear models as follows 
(here we also compress the subscript for the type of data), 
\begin{align}
\label{eq29}
& \jmath(\boldsymbol{\xi}_{L}) = \frac{1}{2}[\sum_{d=1}^{S}\textbf{T}^{d}(\boldsymbol{\beta}_{m}^{d}+\boldsymbol{\Phi}_{\boldsymbol{\beta}}^{d}\boldsymbol{\xi}^{d})-\boldsymbol{\beta}_{b}]^{T}{\textbf{R}_{b}}^{-1}[\sum_{d=1}^{S}\textbf{T}^{d}(\boldsymbol{\beta}_{m}^{d}+\boldsymbol{\Phi}_{\boldsymbol{\beta}}^{d}\boldsymbol{\xi}^{d})-\boldsymbol{\beta}_{b}]  \notag \\
& +\frac{1}{2}\sum_{d=1}^{S}\sum_{m=1}^{N_{0}}[\textbf{d}_{o}^{d,m}-\textbf{y}_{tr}^{d,m}-\textbf{A}_{\boldsymbol{\psi}_{tr}}^{d_{l},m}(\boldsymbol{\psi}^{d_{l},m}-\boldsymbol{\psi}_{tr}^{d_{l},m})-\textbf{B}_{\boldsymbol{\xi}^{d}_{tr}}^{m}(\boldsymbol{\xi}^{d}-\boldsymbol{\xi}^{d}_{tr})]^{T} \notag \\
& {\textbf{R}_{m}}^{-1}[\textbf{d}_{o}^{d,m}-\textbf{y}_{tr}^{d,m}-\textbf{A}_{\boldsymbol{\psi}_{tr}}^{d_{l},m}(\boldsymbol{\psi}^{d_{l},m}-\boldsymbol{\psi}_{tr}^{d_{l},m})-\textbf{B}_{\boldsymbol{\xi}^{d}_{tr}}^{m}(\boldsymbol{\xi}^{d}-\boldsymbol{\xi}^{d}_{tr})]
\end{align}
Where LPCA coefficients $\boldsymbol{\xi}_{L}$=[$\boldsymbol{\xi}^{1}$,...,$\boldsymbol{\xi}^{d}$,...$\boldsymbol{\xi}^{S}$], $d \in \{1,2,\cdot\cdot\cdot,S\}$, 
The key step of a gradient-based minimization algorithm is to determine the gradient of the cost function with respect to the parameters. 
In the adjoint approach, a modified function $\hat{\jmath}(\boldsymbol{\xi}_{L})$ is obtained by adjoining the reduced model equation Eq.\ref{eq23} and Eq.\ref{eq25} to $\jmath(\boldsymbol{\xi}_{L})$
\begin{align}
\label{eq30}
\hat{\jmath}(\boldsymbol{\xi}_{L}) = \jmath(\boldsymbol{\xi}_{L})+ 
\sum_{d=1}^{S}\sum_{n=1}^{N}[\boldsymbol{\psi}^{d,n+1} - \boldsymbol{\psi}_{tr}^{d,n+1}-\textbf{E}_{\boldsymbol{\psi}_{tr}}^{d,n+1}(\boldsymbol{\psi}^{d,n}-\boldsymbol{\psi}_{tr}^{d,n}) 
-\textbf{E}_{\boldsymbol{\psi}_{tr}}^{sd,n+1}(\boldsymbol{\psi}^{sd,n+1}-\boldsymbol{\psi}_{tr}^{sd,n+1})-\textbf{G}_{\boldsymbol{\xi}_{tr}^{d}}^{n+1}(\boldsymbol{\xi}^{d}-\boldsymbol{\xi}_{tr}^{d})]^{T}\boldsymbol{\lambda}^{d,n}
\end{align}

And then, the gradient of the cost function with respect to $\boldsymbol{\xi}^{d}$ for subdomain $\Omega^{d}$ is derived 
after introducing the adjoint model as follows (more details about the mathematical derivation can be found in \cite{jansen2011adjoint}),
\begin{align}
\label{eq31}
& [\frac{d \hat{\jmath}}{d \boldsymbol{\xi}^{d}}]^{T} =(\boldsymbol{\Phi}_{\boldsymbol{\beta}}^{d}\textbf{T}^{d})^{T}{\textbf{R}_{b}}^{-1}[\textbf{T}^{d}(\boldsymbol{\beta}_{m}^{d}+\boldsymbol{\Phi}_{\boldsymbol{\beta}}^{d}\boldsymbol{\xi}^{d})-\boldsymbol{\beta}_{b}]
-\sum_{n=1}^{N}[\textbf{G}_{\boldsymbol{\xi}_{tr}^{d}}^{n}]^{T}\boldsymbol{\lambda}^{d,n}  \notag \\
& -\sum_{m=1}^{N_{0}}[\textbf{B}_{\boldsymbol{\xi}_{tr}}^{m}]^{T} 
{\textbf{R}_{m}}^{-1}[\textbf{d}_{o}^{d,m}-\textbf{y}_{tr}^{d,m}-\textbf{A}_{\boldsymbol{\psi}_{tr}}^{d,m}(\boldsymbol{\psi}^{d,m}-\boldsymbol{\psi}_{tr}^{d,m})
-\textbf{B}_{\boldsymbol{\xi}_{tr}}^{m}(\boldsymbol{\xi}^{d,sd}-\boldsymbol{\xi}_{tr}^{d,sd})] 
\end{align}

where the adjoint model in terms of the Lagrange multipliers $\boldsymbol{\lambda}^{d,n}$ for the subdomain $\Omega^{d}$ is given by
\begin{align}
\label{eq32}
[\textbf{I}-(\textbf{E}_{\boldsymbol{\psi}_{tr}}^{d,n})^{T}]\boldsymbol{\lambda}^{d,n}  = [\textbf{A}_{\boldsymbol{\psi}_{tr}}^{d,n}]^{T} 
{\textbf{R}_{n}}^{-1}[\textbf{d}_{o}^{d,n}-\textbf{y}_{tr}^{d,n} 
-\textbf{A}_{\boldsymbol{\psi}_{tr}}^{d,n}(\boldsymbol{\psi}^{d,n}-\boldsymbol{\psi}_{tr}^{d,n})-\textbf{B}_{\boldsymbol{\xi}_{tr}}^{n}(\boldsymbol{\xi}^{d,sd}-\boldsymbol{\xi}_{tr}^{d,sd})]+[\textbf{E}_{\boldsymbol{\psi}_{tr}}^{sd,n}]^{T}\boldsymbol{\lambda}^{d,n+1}   
\end{align}
for $\textit{n}$ = $\textit{N},\cdot\cdot\cdot ,1 $ with an ending condition $\boldsymbol{\lambda}^{d,N+1}$ = 0. 
The gradient of the cost function with respective to all local PCA coefficients is $\bigtriangledown \hat{\jmath}$=[$\frac{d \hat{\jmath}}{d \boldsymbol{\xi}^{1}}$,...,$\frac{d \hat{\jmath}}
{d \boldsymbol{\xi}^{d}}$,...,$\frac{d \hat{\jmath}}{d \boldsymbol{\xi}^{S}}$]$^{T}$. 
Once the gradient for the $k^{th}$ iteration step is available, the next estimate of the optimal variables that minimize the cost function is given by
\begin{equation}
\label{eq33}
\boldsymbol{\xi}_{k+1}^{L}=\boldsymbol{\xi}_{k}^{L}-\alpha_{k} \frac{\bigtriangledown \hat{\jmath}_{k}}{\lVert \bigtriangledown \hat{\jmath}_{k} \rVert_{\infty}} 
\end{equation}
Where $\alpha_{k}$ is the iteration length at the $k^{th}$ iteration step. Some possible methods can also be used to minimize the objective function $\jmath$ because the simulation of reduced-order model is not computational expensive. 
One can approximate the Hessian calculation using first-order gradient, 
and use a Newton's like update, e.g., quasi-Newton and Gaussian Newton. The accuracy of Hessian matrix is mainly determined by the quality of gradient computed from the reduced-order linear model, thus, 
the convergence of Newton's like optimization methods cannot be ensured very well. In this work, the minimization procedure is directly performed using a steepest descent algorithm \cite{Nocedal1999Numerical}. 
The minimization algorithm terminates when either one of the following three stopping criteria is satisfied,
\begin{itemize}
\item The objective function Eq.\ref{eq29} hardly changes, i.e.,

\begin{equation}
\label{eq34}
\frac{|\hat{\jmath}(\boldsymbol{\xi}_{k+1})-\hat{\jmath}(\boldsymbol{\xi}_{k})|}{max\{|\hat{\jmath}(\boldsymbol{\xi}_{k+1})|,1\}}< \eta_{\hat{\jmath}}
\end{equation}

\item The estimate of parameters almost do not change, i.e.,

\begin{equation}
\label{eq35}
\frac{|\boldsymbol{\xi}_{k+1}-\boldsymbol{\xi}_{k}|}{max\{|\boldsymbol{\xi}_{k+1}|,1\}}< \eta_{\boldsymbol{\xi}}
\end{equation}

\item The maximum iterative steps in iterative loop has been reached, i.e.,

\begin{equation}
\label{eq36}
k <= N_{max}
\end{equation}

\end{itemize}

where $\eta_{\hat{\jmath}}$, $\eta_{\boldsymbol{\xi}}$ and $N_{max}$ denote the predefined error constraints and maximum iterative steps respectively.

Oliver \cite{oliver2008inverse} discussed the expected range of the final mismatch value. If the relationship
between simulated data and parameters is linear, with a tolerance of five standard deviations from
the mean as used in \cite{oliver2008inverse}, the final mismatch $\hat{\jmath}$($\boldsymbol{\xi}$) should follow the inequality
\begin{equation}
\label{eq37}
N_{d}N_{o}-5\sqrt{2N_{d}N_{o}} \leqslant 2 \jmath(\boldsymbol{\xi}_{L}) \leqslant N_{d}N_{o}+5\sqrt{2N_{d}N_{o}}
\end{equation}
However, since we construct an approximate linear model in a reduced-order subspace, we apply a less strict criterion. Throughout this study, we use the following criterion
\begin{equation}
\label{eq38}
\jmath(\boldsymbol{\xi}_{L}) \leqslant 5N_{d}N_{o}
\end{equation}
where, $\textit{N}_{o}$ is the number of timesteps where the measurements are taken, $ N_{d} $ is the number of measurements at each timestep. $\jmath$ is the cost function computed as Eq.\ref{eq4}.  

As mentioned in \cite{Xiao2018}, the solution for one specific reduced-order linear system as Eq.\ref{eq23} is not the exact one for the original cost function as Eq.\ref{eq4} or Eq.\ref{eq5}. 
Additional outer-loops are required to reconstruct new reduced-order linear models with the previously updated parameters and then the iterative inner-loop is performed again. After finishing each outer-iteration, 
the criterion computed using Eq.\ref{eq37} is used to determine the termination of the outer-loop. 

\section{Application to a simplified version of SAIGUP Model}
We demonstrate the performance of LP-SD POD-TPWL, through a modified version of the SAIGUP model containing 13 wells \cite{Matthews35}. 
Our model grid is based on the top layer of one of the structural models generated in the SAIGUP project, 
but that we generate our own permeability realizations, well positions and operating strategy.
In the following, we consider two history matching scenarios that involve: sparse well production data, e.g., fluid rate, watercut and BHP, is assimilated; 
and both well production data and seismic data, e.g., water saturation in each grid block, are simultaneously assimilated 
for calibrating the SAIGUP model with a large amount of uncertain parameters. 
In the numerical experiments, MRST, a free open-source software for reservoir modeling and simulation \cite{lie2012open}, is used to run the full-order model simulations.
 
\subsection{Description of model settings}
The first layer of SAIGUP model containing a total of 3895 active and 905 inactive grid cells is chosen for our test case. 
The realistic geological properties, e.g, faults, are preserved. 
The reservoir model describes a waterflooding system. Six producers and seven injectors, labeled from $P_{1}$ to $P_{6}$, and $I_{1}$ to $I_{7}$ respectively, 
are almost uniformly located in this reservoir, see Fig.\ref{fig10}. Some detailed
information about the reservoir geometry, rock properties, fluid properties, and well controls are shown in Table \ref{tab3}. 

\begin{table}[h]
\footnotesize
\centering
\caption{Experiment settings using MRST}\label{tab3}
\begin{spacing}{1.25}
\begin{tabular}{l l}
\hline
\textbf{Description} & \textbf{Value} \\
\hline
Dimension & 40$\times$120$\times$1 \\
Number of wells & 6 producers, 7 injectors \\
Fluid density & 1014 kg/$m^{3}$, 859 kg/$m^{3}$ \\
Fluid viscosity & 0.4 mP$\cdot$s, 2 mP$\cdot$s \\
Initial pressure & 25 MPa  \\
Initial saturation & S$_{o}$=0.80,  S$_{w}$=0.20 \\
Connate water saturation & $S_{wc}$=0.20 \\
Residual oil saturation & $S_{or}$=0.20  \\
Corey exponent, oil & 4.0   \\
Corey exponent, water & 4.0  \\
Injection rate & 200$m^{3}$/d  \\
BHP & 20MPa  \\
History production time & 10 year  \\
Prediction time & 15 year  \\
Timestep & 0.1 year  \\
Measurement timestep & 0.2 year  \\
\hline
\end{tabular}
\end{spacing}
\end{table}

\subsection{Description of reduced model procedure}
\begin{figure}[!h]
\centering
\subfloat['True' model]%
  {\includegraphics[width=0.3\linewidth]{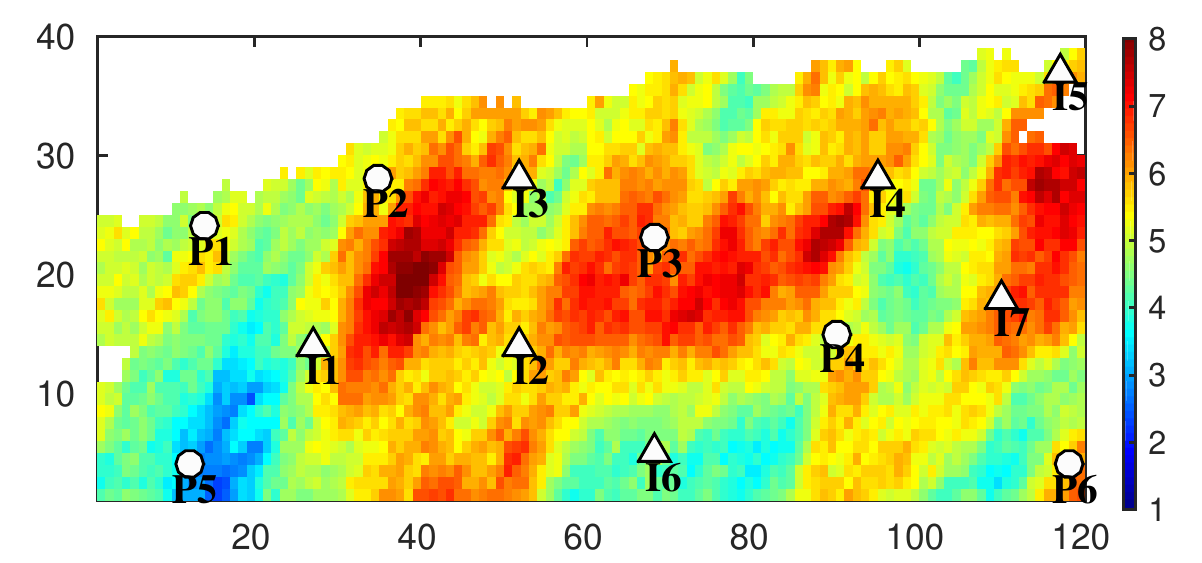}} 
\subfloat[Projected 'true' model using LPCA]%
  {\includegraphics[width=0.3\linewidth]{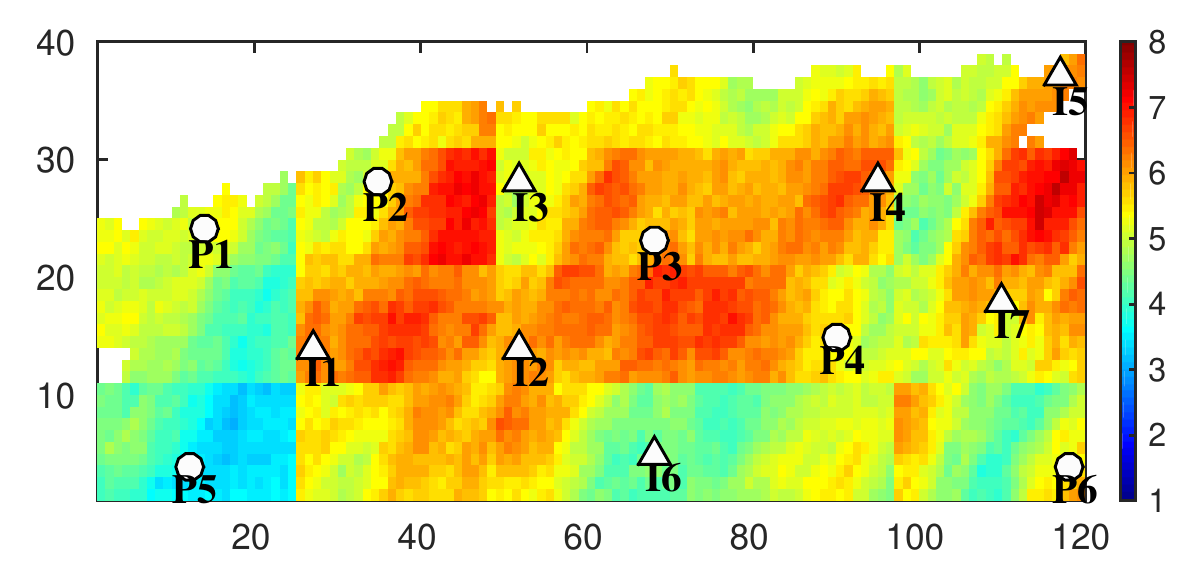}} 
\subfloat[Projected 'true' model where the LPCA results are projected onto the global PCA's]%
  {\includegraphics[width=0.3\linewidth]{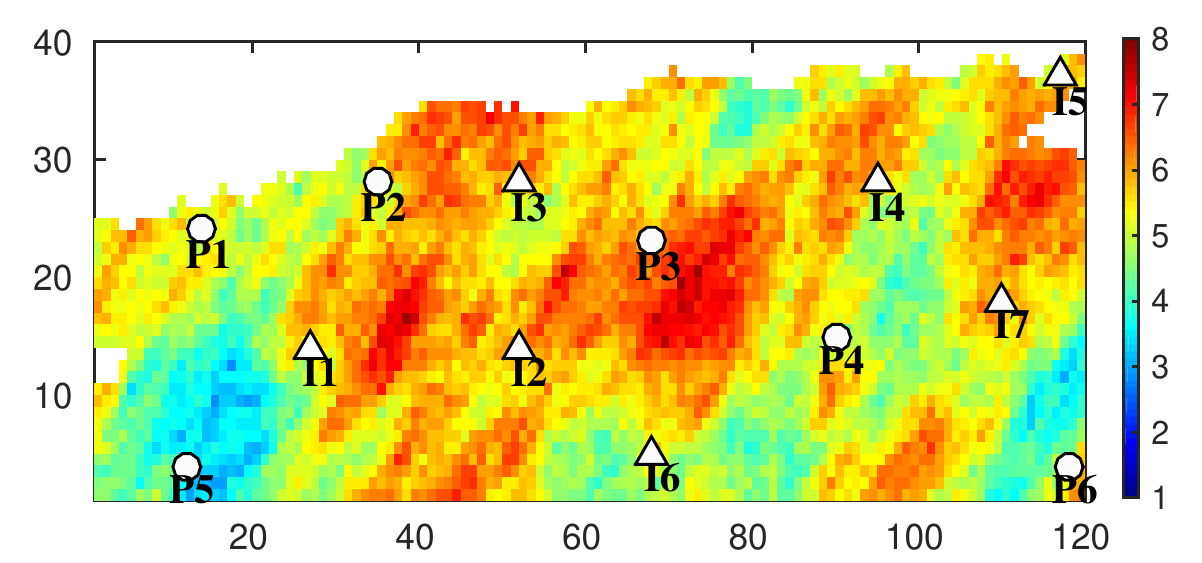}}
\caption{Comparison of the 'true' reservoir model in full-order space and reduced-order space for scenario S1}\label{fig9}
\end{figure}

We manually generate 1000 Gaussian-distributed log-permeability. 
These 1000 log-permeability fields are used to form the global parameter covariance matrix and local parameter covariance matrix for each subdomain. 
One of these realizations is considered to be the truth, see Fig.\ref{fig10}. 
In this numerical experiments, 95\% energy of the global PCA patterns are retained for global parameterization which corresponds to $N_{G}$=48 global PCA patterns.
The log-permeability fields are also locally parameterized with retaining 95\% energy of the local PCA patterns in each subdomain. 
The resulting local PCA coefficients for each subdomain are shown in Fig.\ref{fig11}. The total number of local PCA coefficients is $N_{L}=\sum_{d=1}^{S}l_{d}$=275. 
The maximum number of local PCA coefficients among all subdomains is 15. 
Fig.\ref{fig9}(a) and Fig.\ref{fig9}(b) separately represents the projected 'true' permeability field using O-LS-PCA and local PCA. 
We divide the entire domain into 20 rectangle subdomains (4 subdomains in $\textit{x}$ direction and 5 subdomains in $\textit{y}$ direction) as in Fig.\ref{fig10} 
which is also considered to be a base-case.

As has been explained in our previous work \cite{Xiao2018}, since the required number of training runs for generating snapshots is not known a priori we follow the following procedure: 
(1) generate a sample PCA coefficient vector by sampling from the set $\{-1,1\}$, 
(2) run a full-order model simulation with these parameters, (3) extract snapshots and form the snapshot matrix, 
(4) compute the singular value decomposition of the snapshot matrix (5) repeat steps (1) to 4) until changes in the singular values are insignificant. 
Experiments showed that the changes in singular value spectrum of the snapshot matrix are insignificant 
when the snapshots at every timestep are selected from 22 or more FOM simulations.
Finally, we collect 2200 snapshots for pressures and saturations separately. 
For each subdomain, two separate eigenvalue problems for pressure and saturation are solved using POD, 
where 95\% energy is preserved, and the number of POD patterns for each subdomain is shown in Fig.\ref{fig11}.

\begin{figure}[!h]
\centering\includegraphics[width=0.4\linewidth]{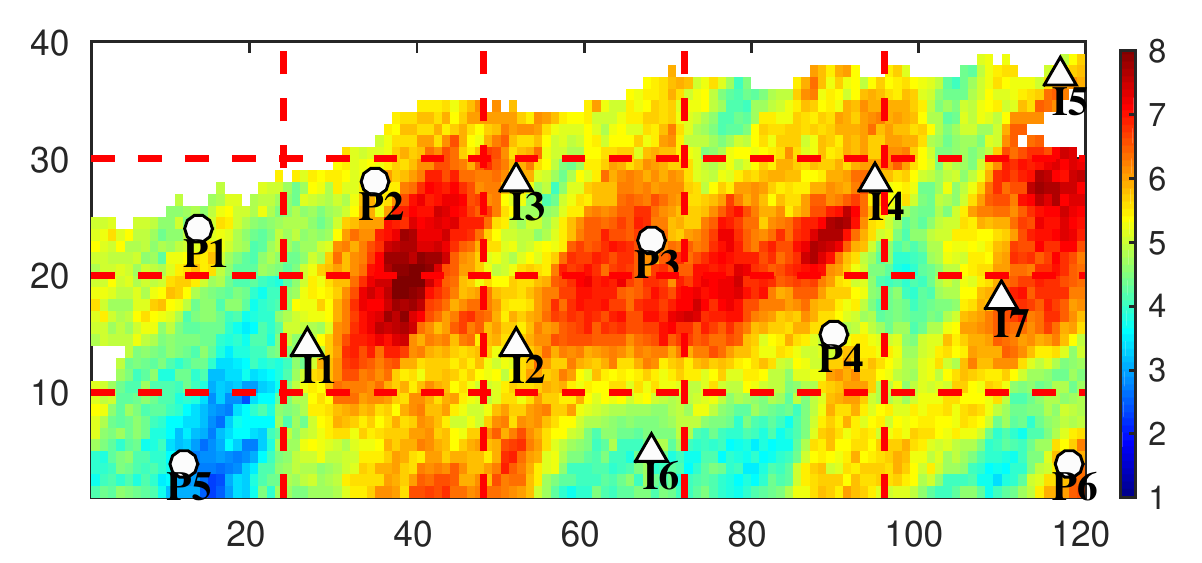}\\
\caption{Illustration of domain decomposition in the 2-D synthetic model for scenario S1}\label{fig10}
\end{figure}

\begin{figure}[!h]
\centering\includegraphics[width=0.8\linewidth]{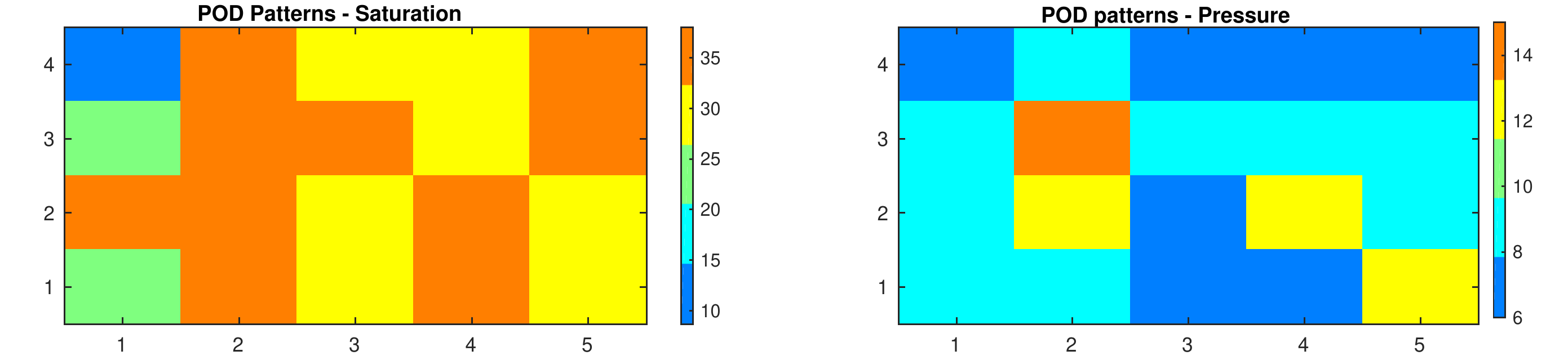}\
\caption{The number of reduced pressure and saturation patterns in each subdomain for scenario S1}\label{fig11}
\end{figure}

\subsection{Results of only assimilating sparse well data}
The history production period is 10 years during which observations are taken from six producers and seven injectors 
every two model timesteps, resulting in 50 time instances. Noisy observations are generated from the model with the “true” permeability field and consisted of bottom-hole 
pressures (BHP) in injectors and fluid rates and WCT in producers, 
both with normally distributed noise. As a result we have 300 fluid rates and 300 WCT values measured in the
producers and 350 bottom-hole pressures measured in the injectors, which gives in total 950 measurement data.

\subsubsection{Study of a base-case}

\begin{figure}[!h]
\centering
\subfloat[]%
  {\includegraphics[width=0.4\linewidth]{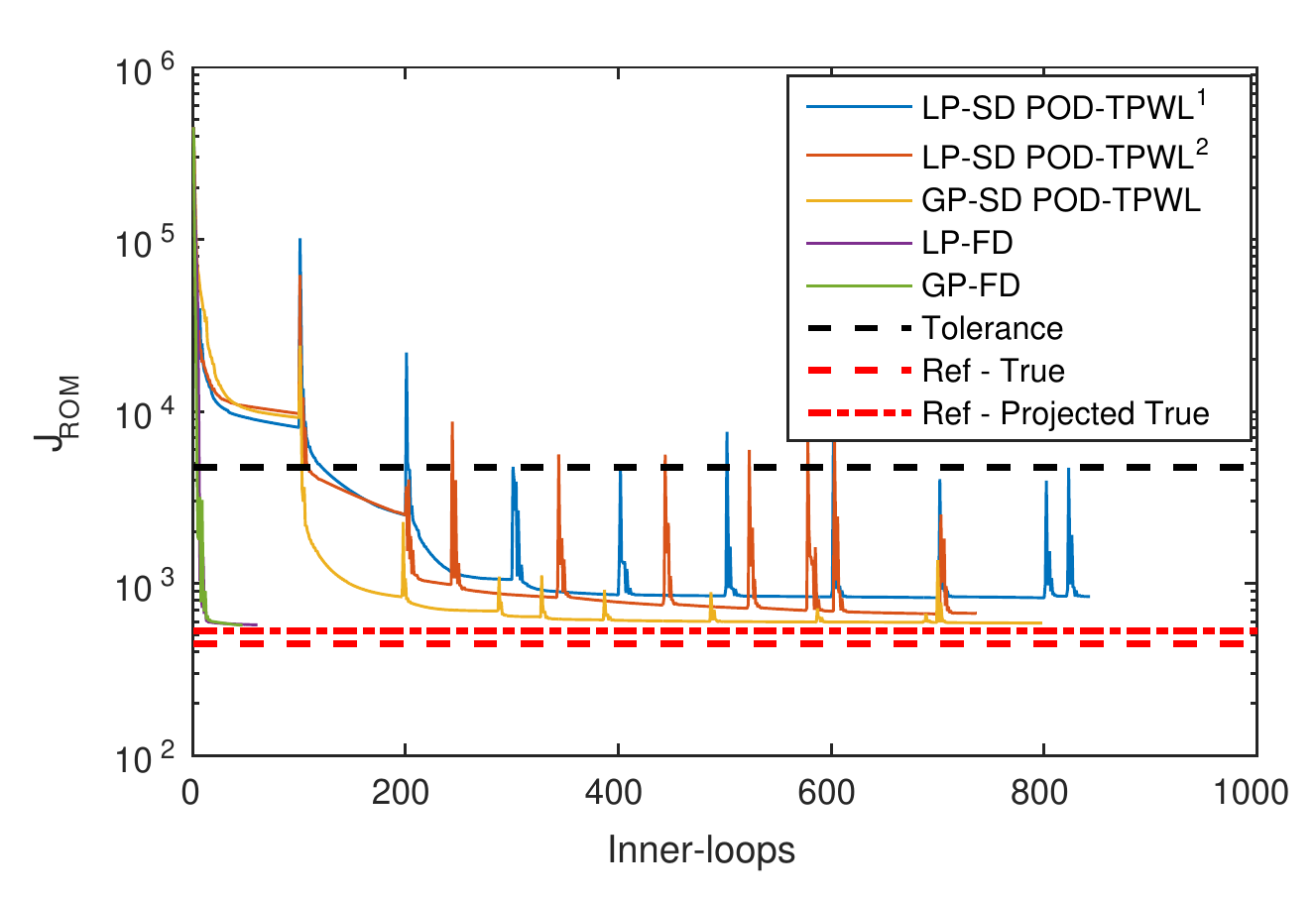}}
\subfloat[]%
  {\includegraphics[width=0.4\linewidth]{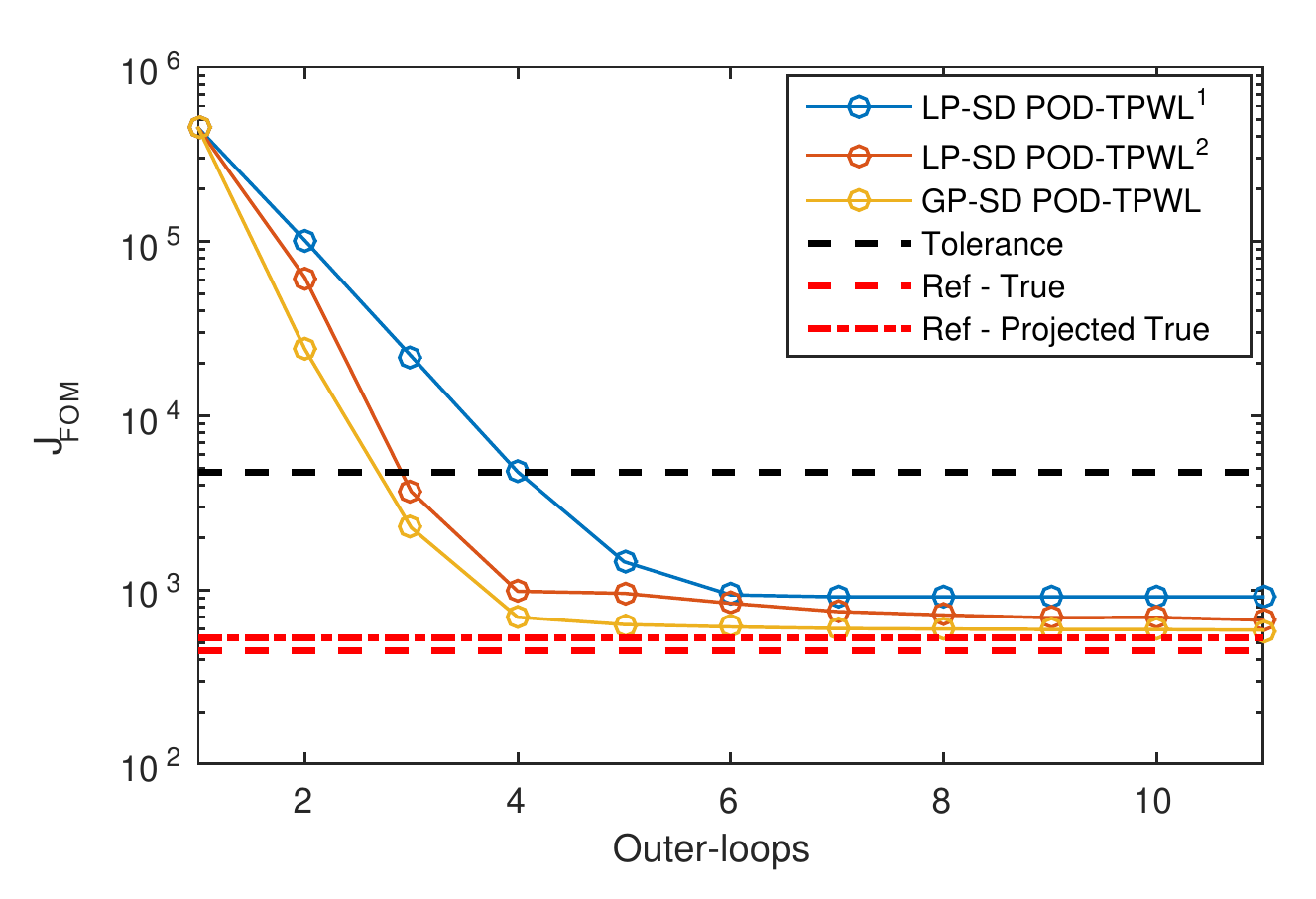}} 
\caption{Cost function values using LP-SD POD-TPWL, GP-SD POD-TPWL, LP-FD and GP-FD method for scenario S1 as a function of outer-loops. The calculation of the cost function for inner-loops
 and outer-loops uses reduced-order linear model and full-order model respectively}\label{fig14}
\end{figure}

\begin{figure*}[!h]
\centering
\subfloat[]%
  {\includegraphics[width=0.3\linewidth]{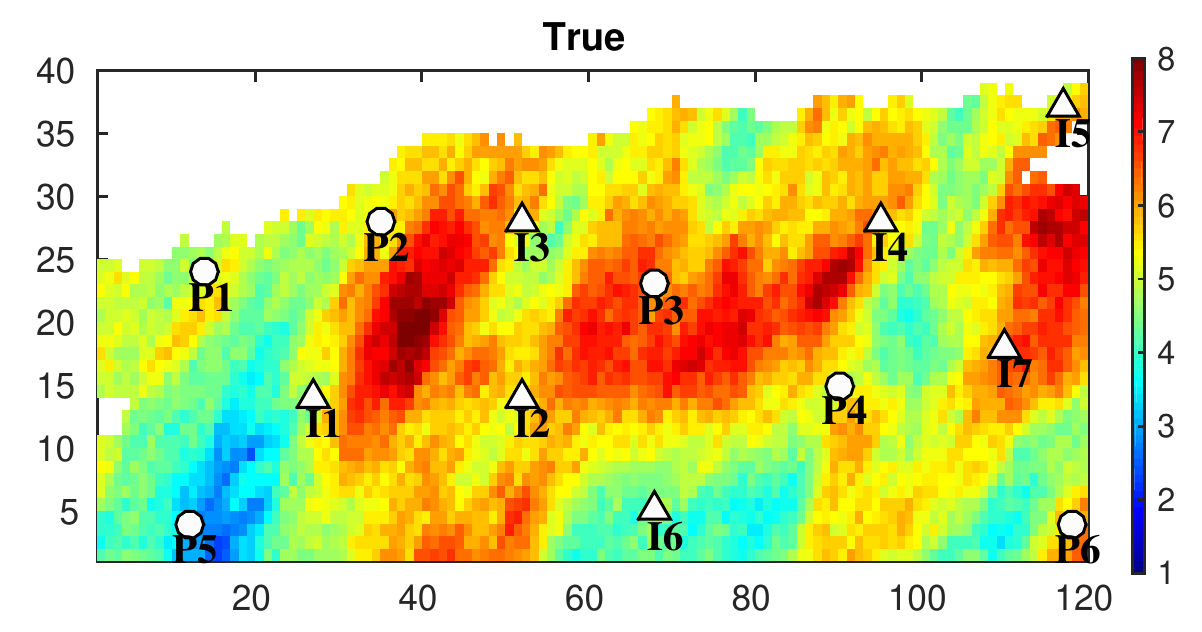}}
\subfloat[]%
  {\includegraphics[width=0.3\linewidth]{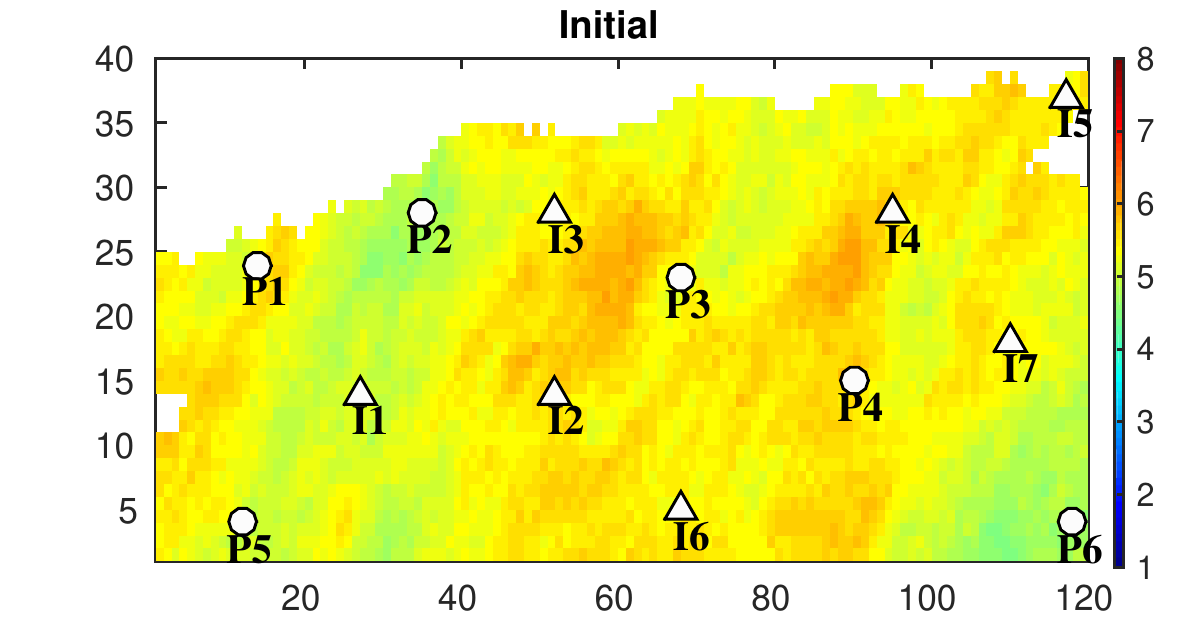}} 
\subfloat[]%
  {\includegraphics[width=0.3\linewidth]{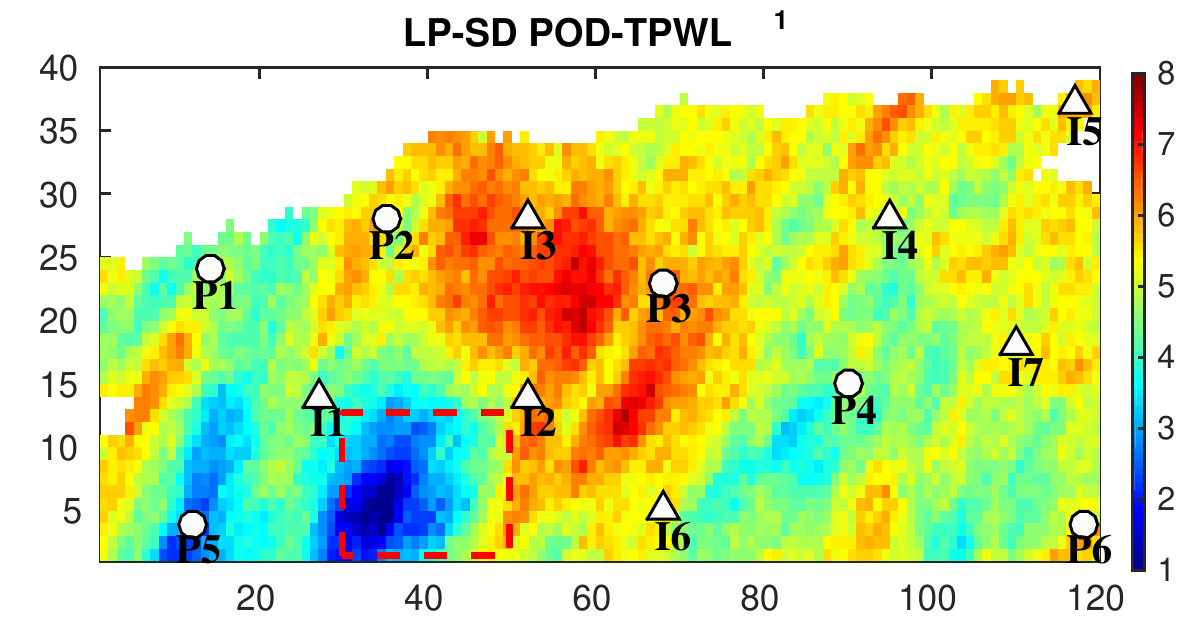}} \\
\subfloat[]%
  {\includegraphics[width=0.3\linewidth]{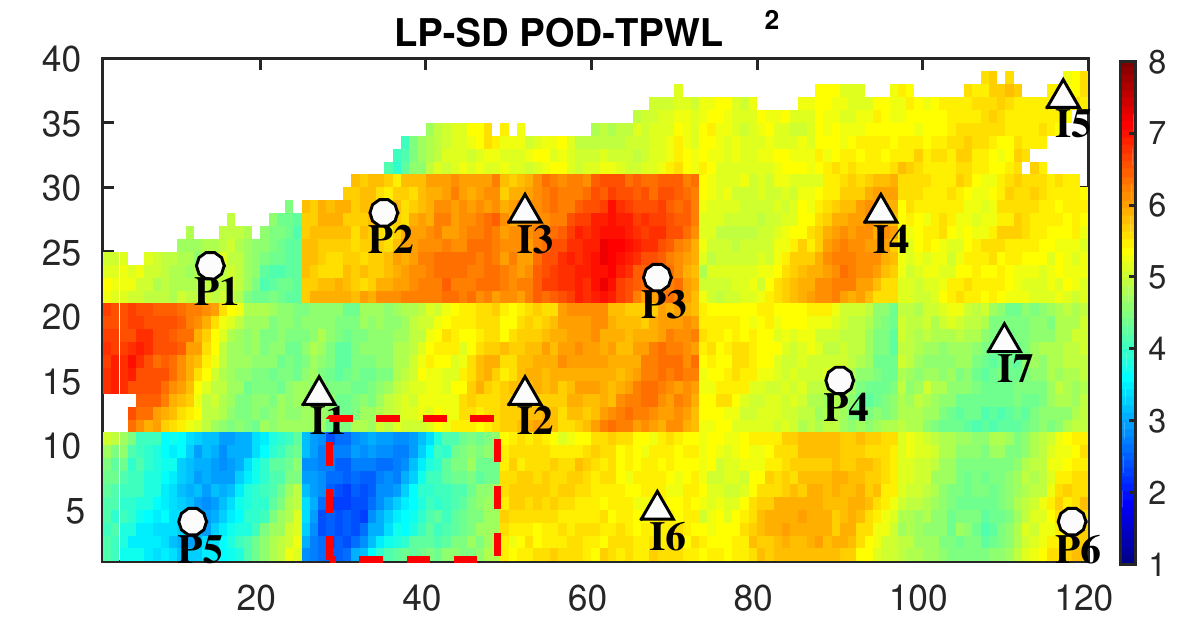}}   
\subfloat[]%
  {\includegraphics[width=0.3\linewidth]{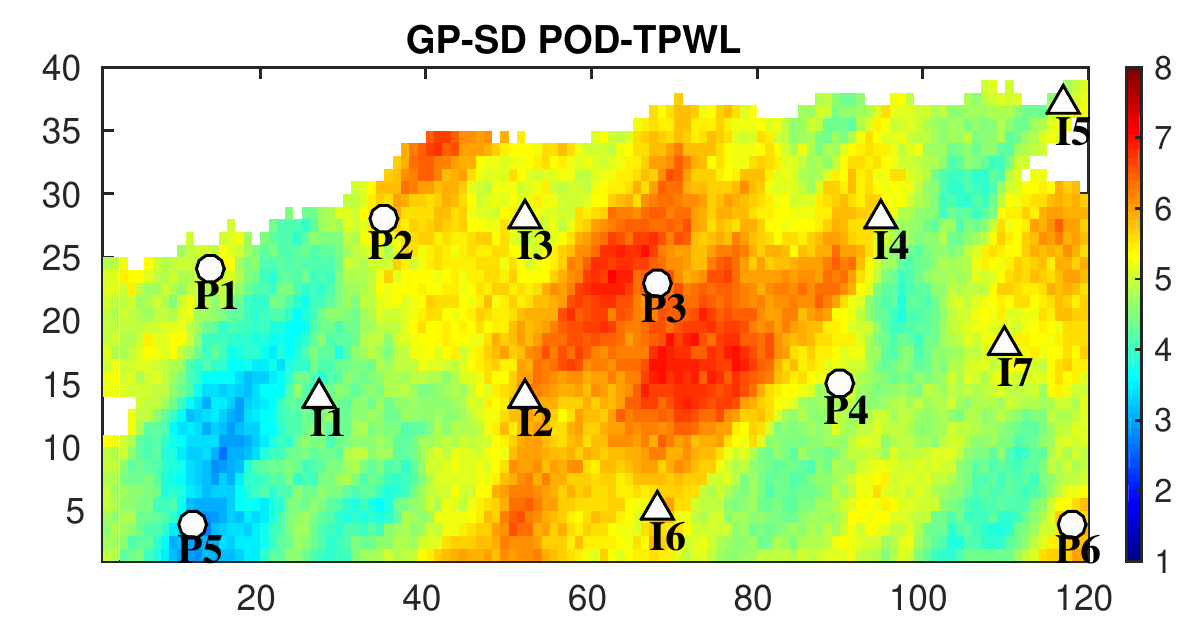}} 
\subfloat[]%
  {\includegraphics[width=0.3\linewidth]{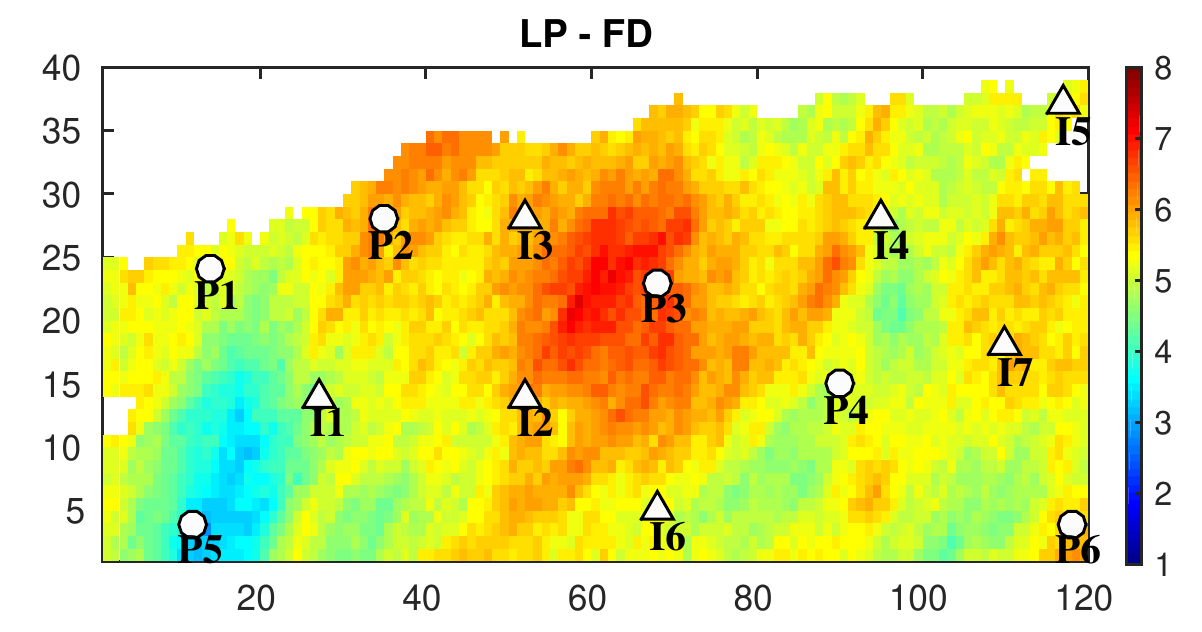}} \\
\subfloat[]%
  {\includegraphics[width=0.3\linewidth]{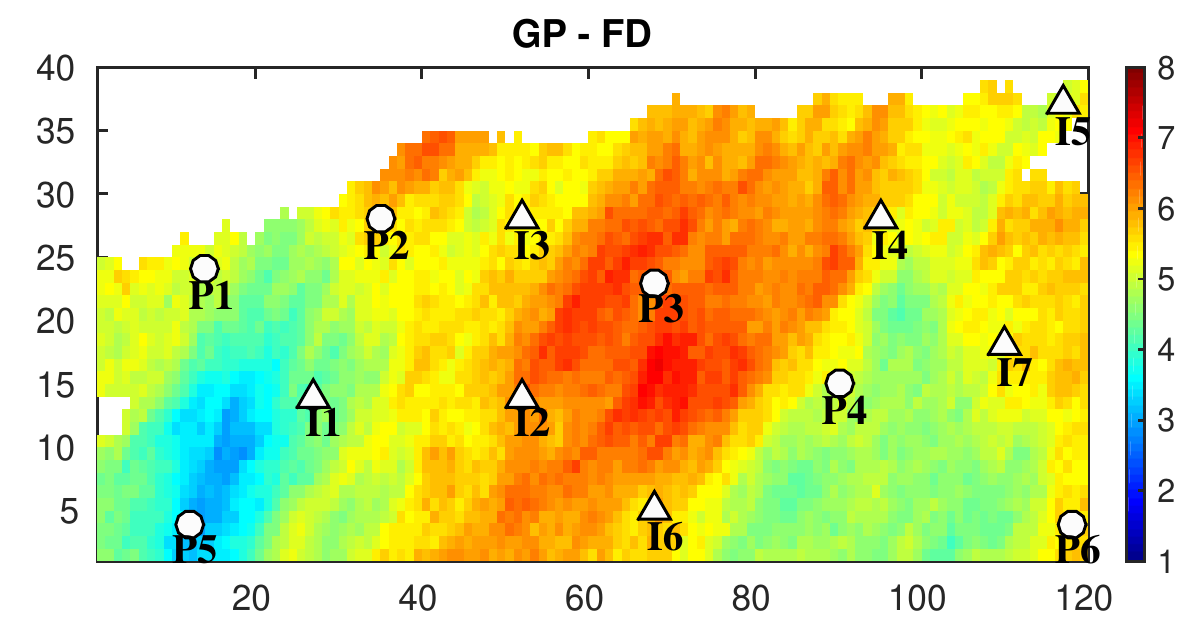}}
\caption{Comparison of the updated permeability fields from the LP-SD POD-TPWL, GP-SD POD-TPWL and FD method for scenario S1}\label{fig15}
\end{figure*}

\begin{figure*}[!h]
\centering
\subfloat[]%
  {\includegraphics[width=0.3\linewidth]{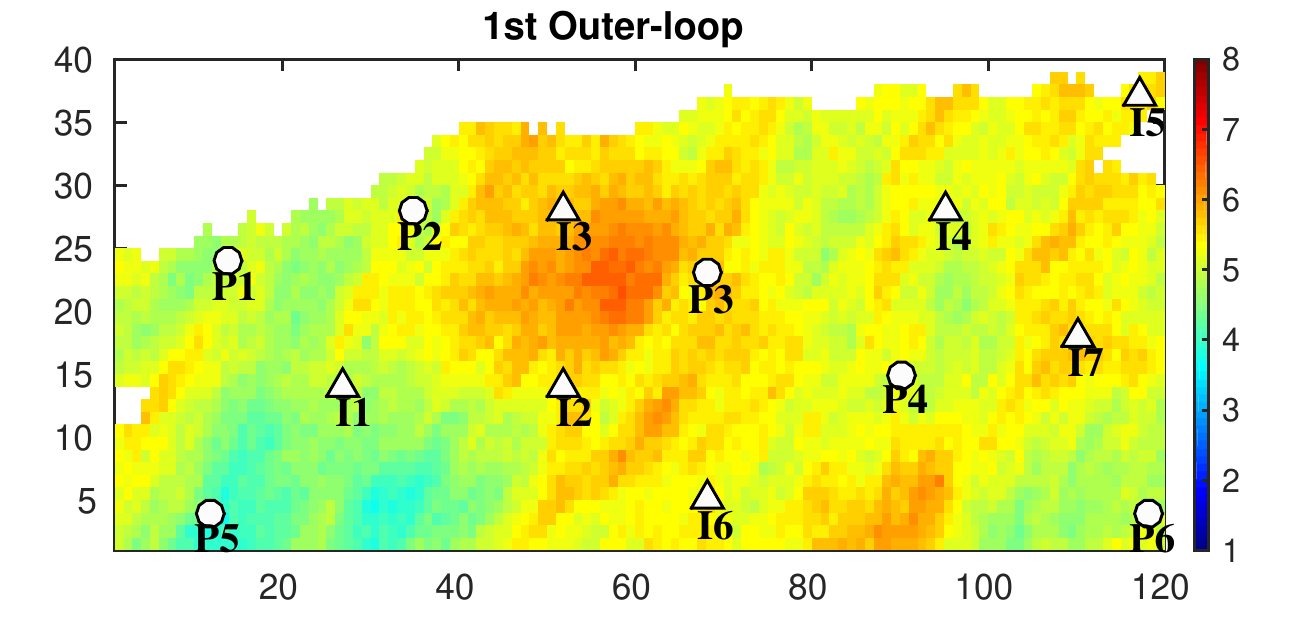}}
\subfloat[]%
  {\includegraphics[width=0.3\linewidth]{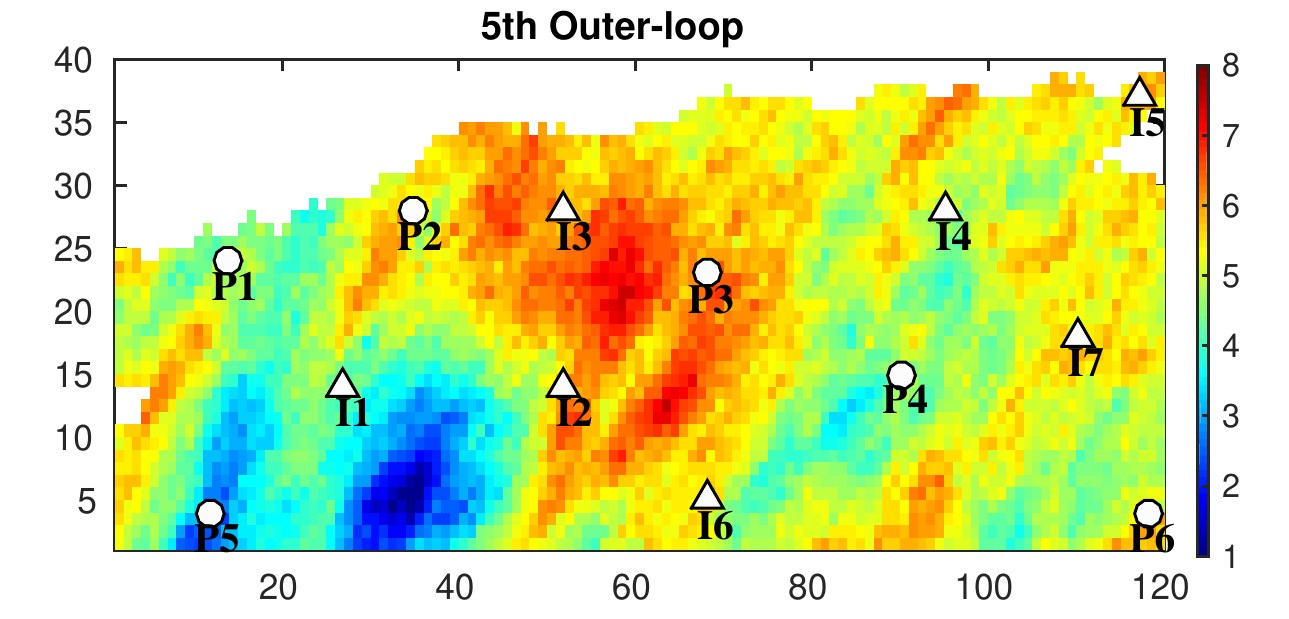}} 
\subfloat[]%
  {\includegraphics[width=0.3\linewidth]{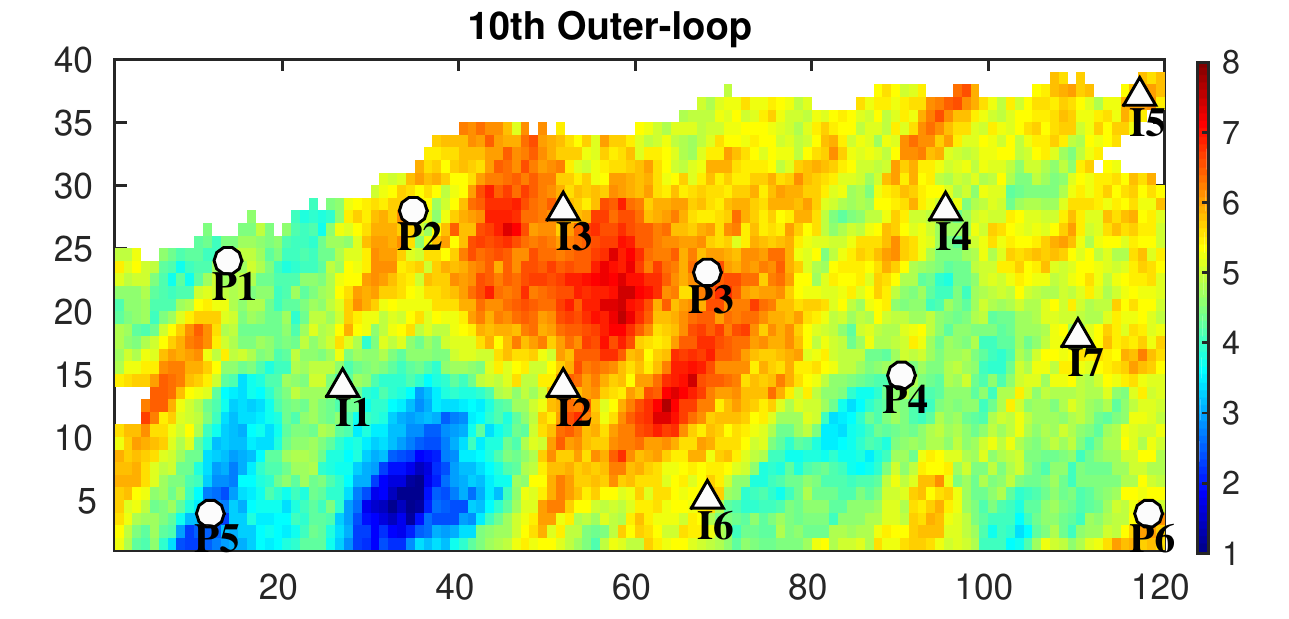}} \\
\subfloat[]%
  {\includegraphics[width=0.3\linewidth]{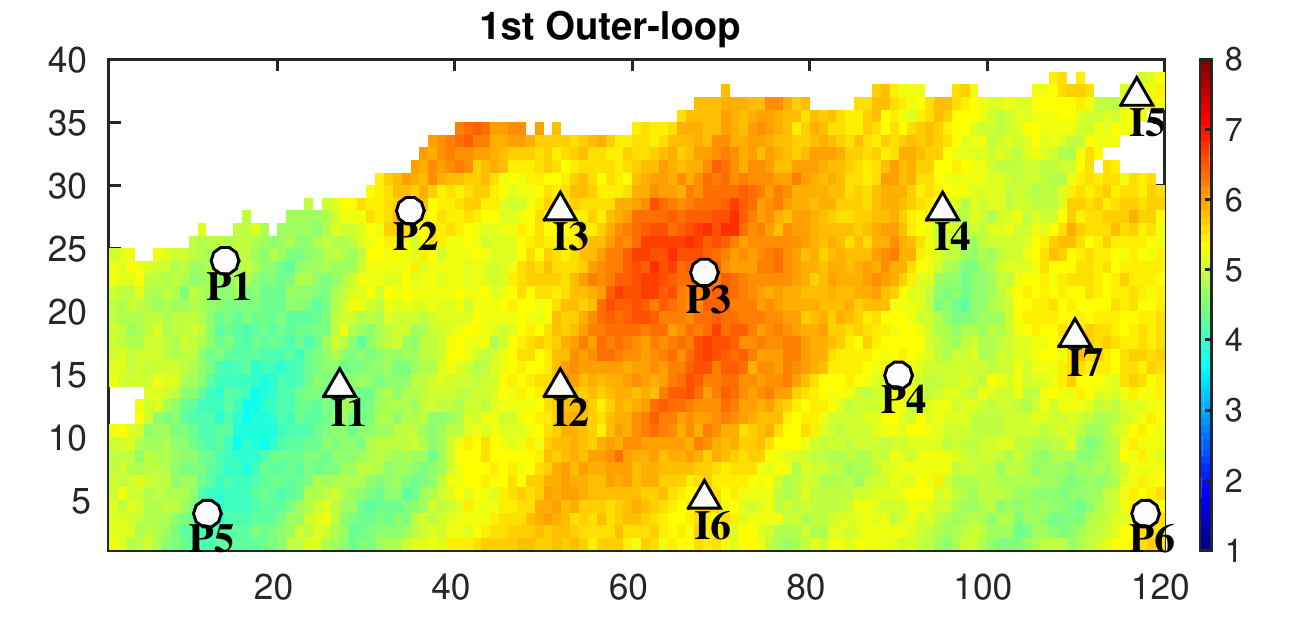}} 
\subfloat[]%
  {\includegraphics[width=0.3\linewidth]{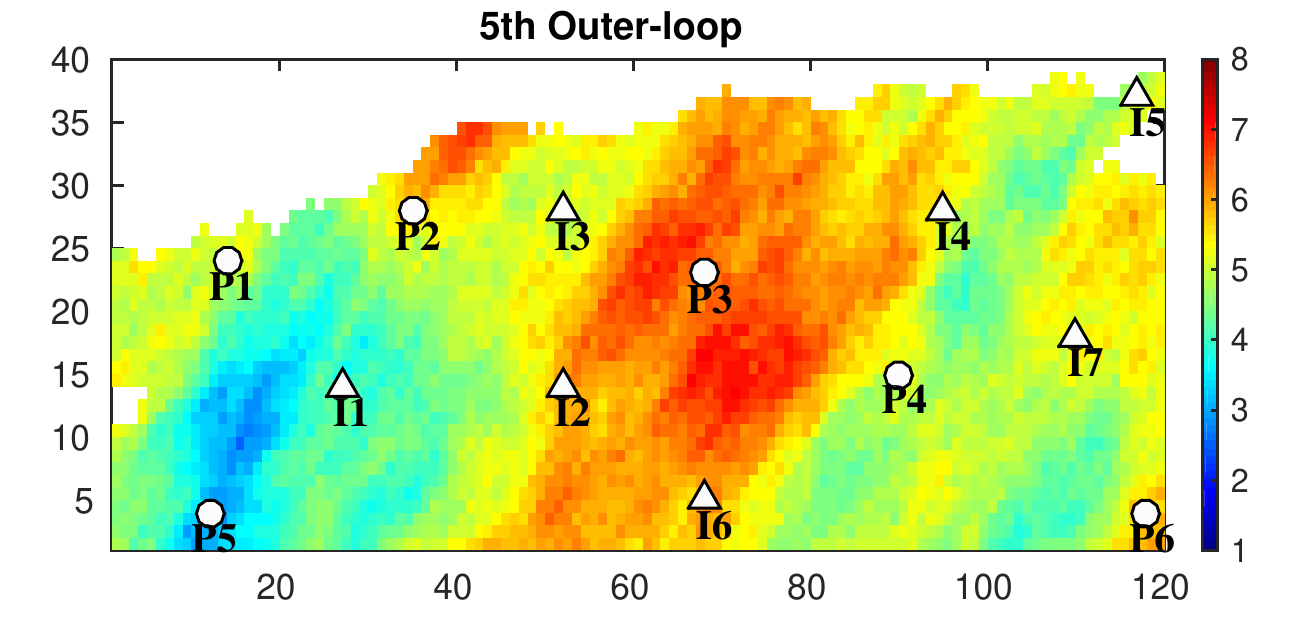}} 
\subfloat[]%
  {\includegraphics[width=0.3\linewidth]{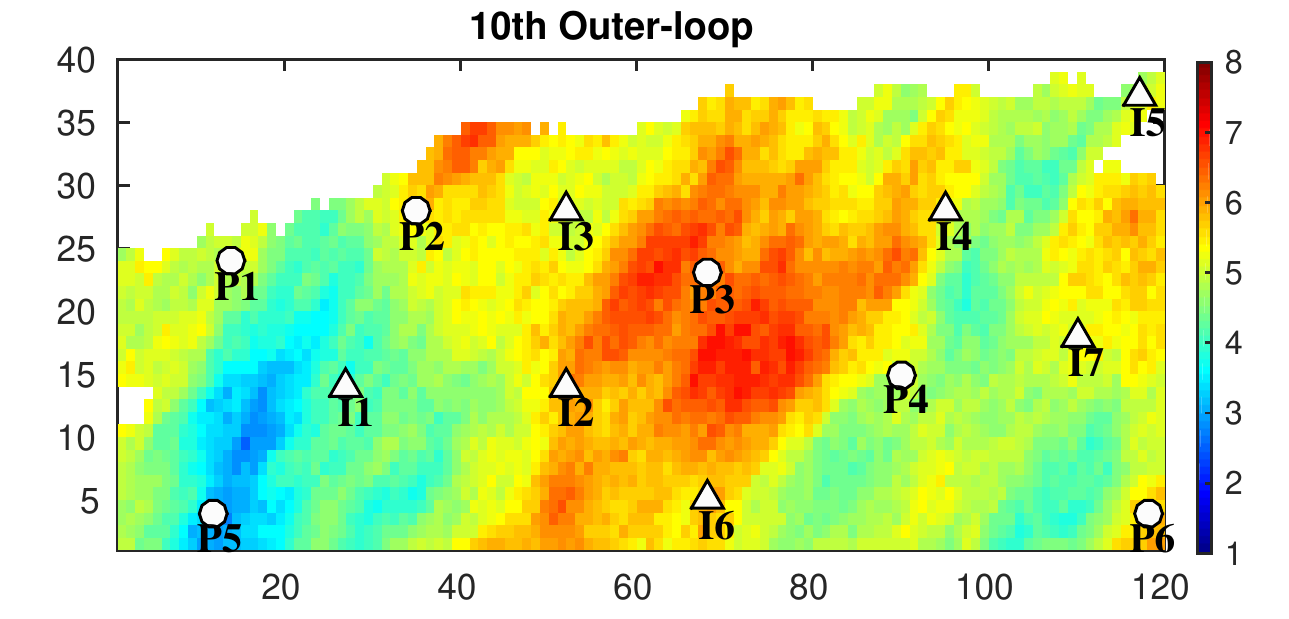}} \\
\caption{Comparison of updated permeability fields from LP-SD POD-TPWL, GP-SD POD-TPWL in three different steps of outer-loop for scenario S1}\label{fig16}
\end{figure*}

\begin{figure*}[!h]
\centering\includegraphics[width=0.8\linewidth]{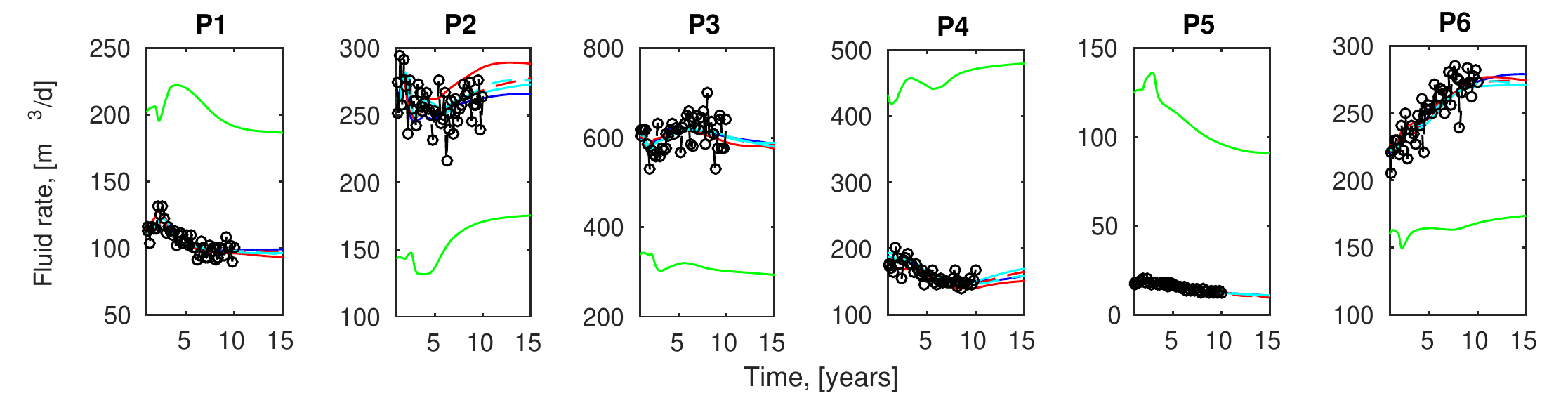}\\
\centering\includegraphics[width=0.8\linewidth]{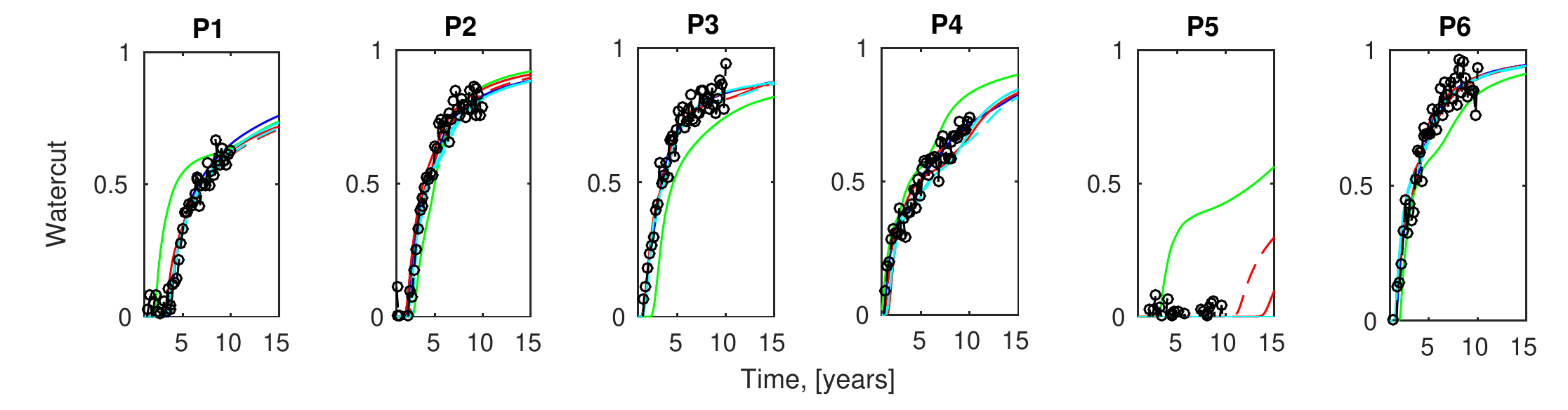}\\
\centering\includegraphics[width=0.9\linewidth]{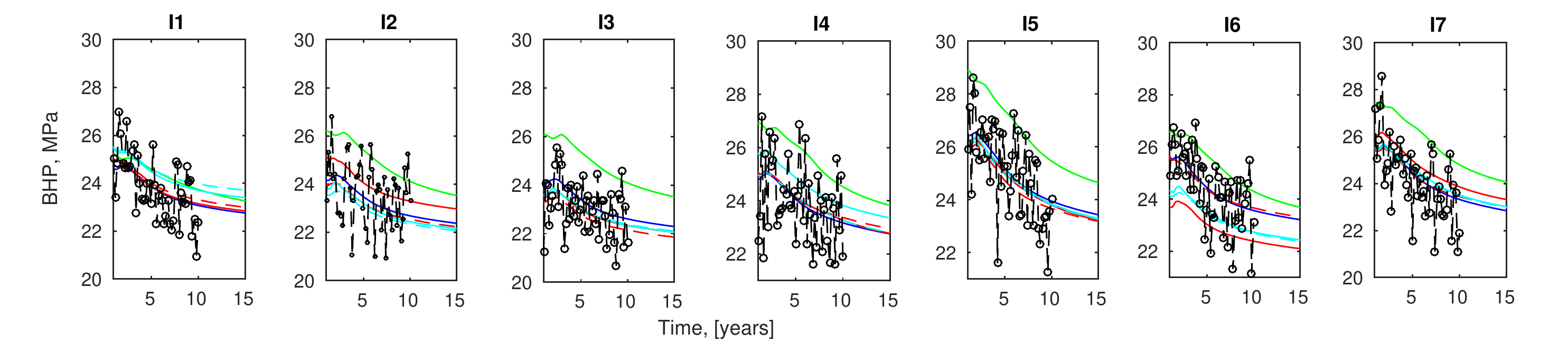}\\
\caption{Forecast of the liquid rate, WCT and BHP for scenario S1: green line-initial model, blue line-'true' model, solid red line-updated model using LP-SD POD-TPWL, dash red line-updated model using GP-SD POD-TPWL, 
solid cyan line-updated model using LP-FD, dash cyan line-updated model using GP-FD}\label{fig17}
\end{figure*}

Fig.\ref{fig14}, Fig.\ref{fig15}, Fig.\ref{fig16}, Fig.\ref{fig17} and Table \ref{tab4} show the results for the base-case, 
including the value of the cost function at each iteration, the mismatch between observed data and predictions, and the updated log-permeability field. 
The subdomain POD-TPWL with O-LS-PCA and LPCA where the LPCA patterns are not mapped to global patterns are referred to as LP-SD POD-TPWL1 and LP-SD POD-TPWL2 respectively. 
The subdomain POD-TPWL with GPCA, referred to as GP-SD POD-TPWL, is also implemented for comparison.
In addition, the one-side finite-difference (FD) with GPCA and O-LS-PCA, referred to to GP-FD and LP-FD, are also implemented, respectively with the cost function 
\begin{align}
\label{eq41}
J_{GP-FD}(\boldsymbol{\xi}_{G}) & = \frac{1}{2}(\boldsymbol{\beta}_{m}+\boldsymbol{\Phi}_{\boldsymbol{\beta}} \boldsymbol{\xi}_{G}-\boldsymbol{\beta}_{b})^{T}{\textbf{R}_{b}}^{-1}(\boldsymbol{\beta}_{m}+\boldsymbol{\Phi}_{\boldsymbol{\beta}} \boldsymbol{\xi}_{G}-\boldsymbol{\beta}_{b}) \notag \\
& +\frac{1}{2}\sum_{m=1}^{N_{0}}(\textbf{d}_{o}^{m}-\textbf{h}^{m}(\textbf{x}^{m},\boldsymbol{\xi}_{G}))^{T}{\textbf{R}_{m}}^{-1}(\textbf{d}_{o}^{m}-\textbf{h}^{m}(\textbf{x}^{m},\boldsymbol{\xi}_{G}))
\end{align}
\begin{align}
\label{eq42}
J_{LP-FD}(\boldsymbol{\xi}_{L}) & = \frac{1}{2}(\boldsymbol{\beta}_{m}+\boldsymbol{\Phi}_{\boldsymbol{\beta}} \textbf{T}_{GL}\boldsymbol{\xi}_{G}-\boldsymbol{\beta}_{b})^{T}{\textbf{R}_{b}}^{-1}(\boldsymbol{\beta}_{m}+\boldsymbol{\Phi}_{\boldsymbol{\beta}} \textbf{T}_{GL}\boldsymbol{\xi}_{G}-\boldsymbol{\beta}_{b}) \notag \\
& +\frac{1}{2}\sum_{m=1}^{N_{0}}(\textbf{d}_{o}^{m}-\textbf{h}^{m}(\textbf{x}^{m},\boldsymbol{\xi}_{L}))^{T}{\textbf{R}_{m}}^{-1}(\textbf{d}_{o}^{m}-\textbf{h}^{m}(\textbf{x}^{m},\boldsymbol{\xi}_{L}))
\end{align}

The stopping criteria are $\eta_{\jmath} =10^{-4} $, $\eta_{\boldsymbol{\xi}} =10^{-3} $, and $N_{max}$=100, the maximum number of outer-loops is 10. 
The initial iteration length $\alpha_{0}$=0.1, once the cost function increases, the iteration length is 
divided by 2 as to ensure a decreasing cost function. 
Fig.\ref{fig13} show that GP-FD, LP-FD and GP-SD POD-TPWL obtain similar cost function values after the minimization, 
while our LP-SD POD-TPWL obtains a slightly less accurate results.
As can be seen from Table \ref{tab4}, both LP-SD POD-TPWL1 and LP-SD POD-TPWL2 need 62 FOM simulations, 
among them, 22 FOM simulations are used to collect the snapshots for the POD, 31=(2$\times$15+1) FOM simulations are used to construct the initial subdomain reduced-order linear model, 
and the remaining 9 FOM simulation are used to
update the reduced-order linear model in the following 9 outer-loops. The GP-SD POD-TPWL needs 224 =22+(4$\times$48+1)+9 FOM simulations.
We also plot reference cost functions which are computed for comparison using the true model and the projected true model.

Fig.\ref{fig14} shows the true, initial and final updated log-permeability field from the FD method, GP-SD POD-TPWL, LP-SD POD-TPWL1 and LP-SD POD-TPWL2. 
Although LP-SD POD-TPWL2 obtains relatively low cost function values, the updated log-permeability filed is non-smooth
which severely violates the geological properties of the original model. Our new LP-SD POD-TPWL1 with O-LS-PCA avoids this problem. 
Global parameterization mitigates the adverse influences of artificial domain decomposition.
The methods with GPCA, e.g, GP-SD POD-TPWL and GP-FD, generate a little better results. In one specific subdomain denoted by a red dash rectangle, the log-permeability fields in this subdomain
are not correctly calibrated due to lack of observations. The choice of domain decomposition has significant influences on the performance of LP-SD POD-TPWL. We will systematically investigate 
this issue in the following sections.

Fig.\ref{fig17} shows that both GP-SD POD-TPWL and LP-SD POD-TPWL obtain acceptable results after 5 outer-loops,
and therefore some additional FOM runs can be avoided. 
Fig.\ref{fig18} illustrates the data match of fluid rate and water-cut up to year 10 and an additional 15-year prediction for all six producers. 
The prediction based on the initial model is far from the true model. After the history matching, the predictions of the updated model match the observations very well. 
Also the prediction of the water breakthrough time is improved for all production wells, including the wells that show water breakthrough only after the history matching period. 

\begin{table}[!h]
\scriptsize
\centering
\caption{The number of required FOM simulations for LP-SD POD-TPWL, GP-SD POD-TPWL and FD method for scenario S1}\label{tab4}
\begin{spacing}{1.25}
\begin{tabular}{|c|c|c|c|}
\hline
- & Iterations & FOM & $J(\boldsymbol{\xi})$  \\
\hline
Initial model & - & - & 4.49$\times$10$^{5}$ \\
LP-SD POD-TPWL1 & 10 & 62 = 22+(2$\times$15+1)+9 & 912.93  \\
LP-SD POD-TPWL2 & 10 & 62 = 22+(2$\times$15+1)+9 & 697.32  \\
GP-SD POD-TPWL & 10 & 224 = 22+(4$\times$48+1)+9 & 587.83  \\
LP-FD& 61 & 4421 & 573.94  \\
GP-FD & 47 & 2773 & 571.73  \\
Tolerance & - & - & 4750 \\
Ref - Projected True & - & - & 528.1 \\
Ref - True & - & - & 447.4 \\
\hline
\end{tabular}
\end{spacing}
\end{table}

\subsubsection{Study of the domain decomposition strategy} 
 
\begin{figure}[!h]
\centering
\includegraphics[width=0.5\linewidth]{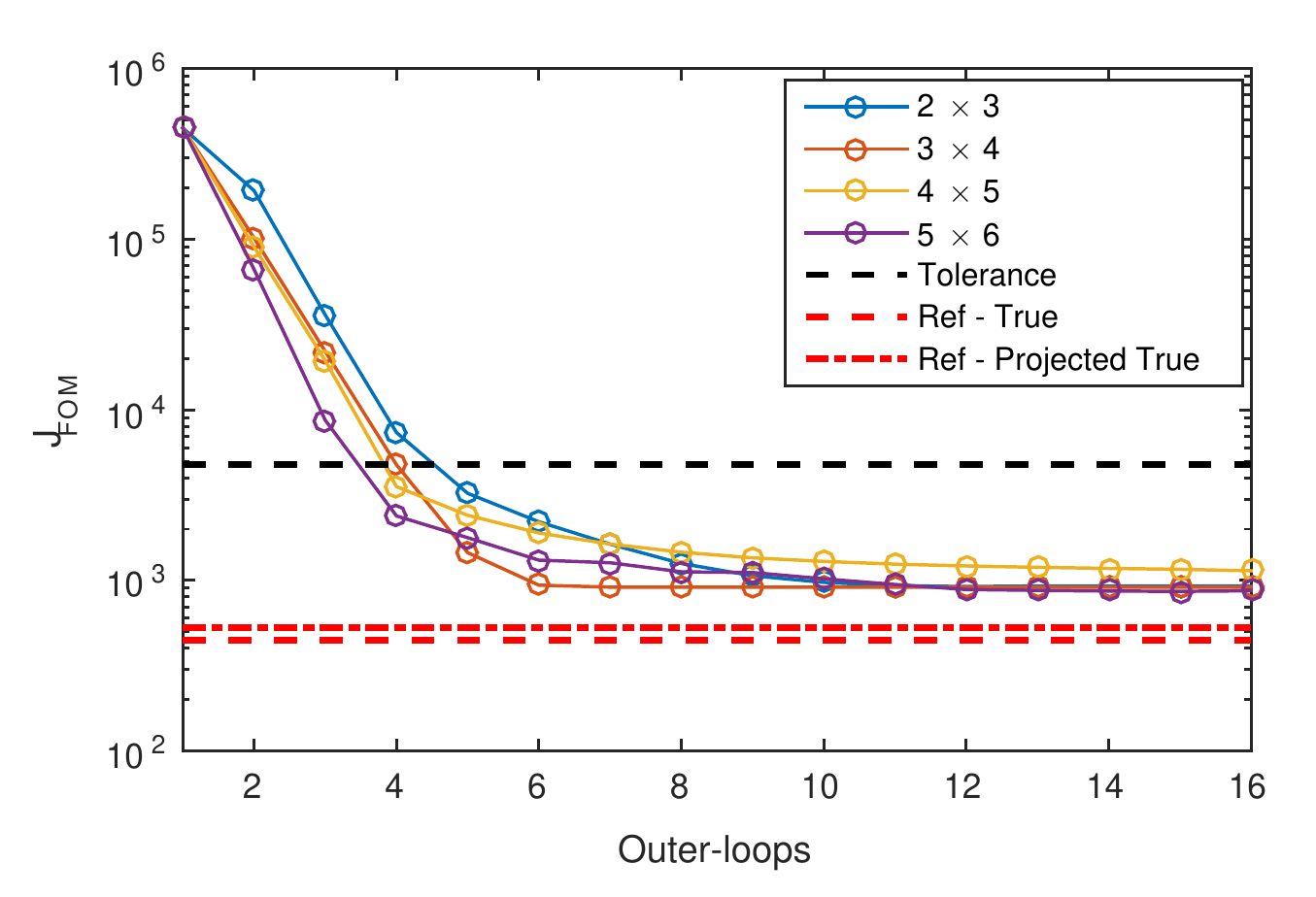} \\
 \caption{Cost function values of LP-SD POD-TPWL by different domain decomposition scheme for scenario S1}\label{fig18}
\end{figure}

\begin{figure*}[!h]
\centering
\subfloat[]%
  {\includegraphics[width=0.3\linewidth]{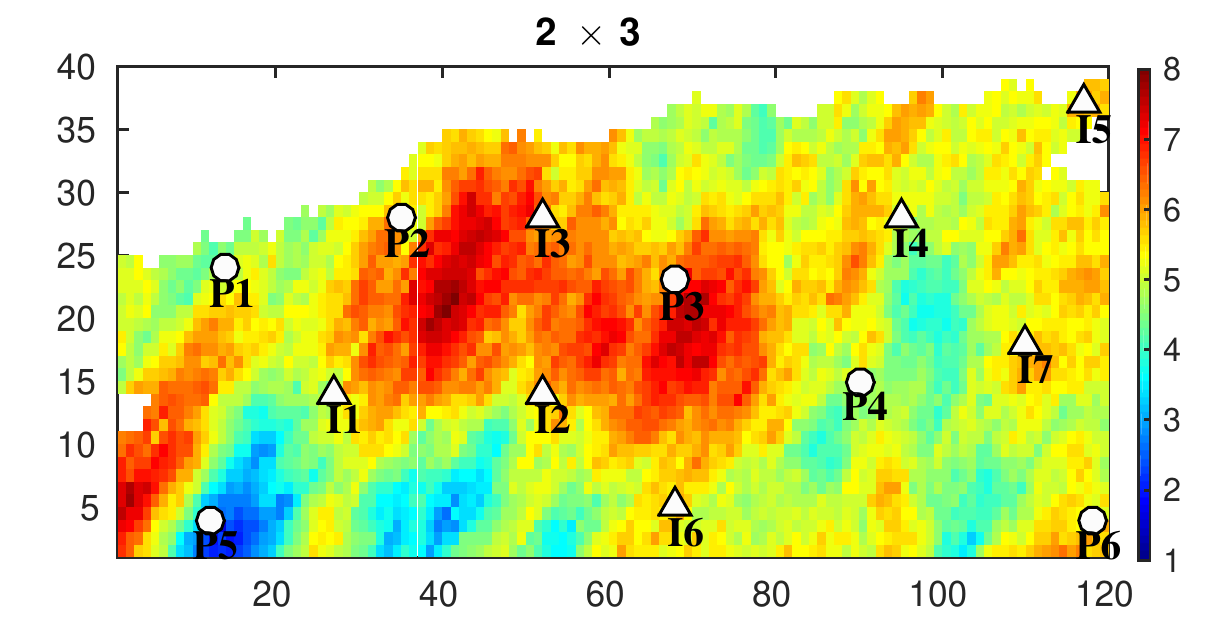}}  
\subfloat[]%
  {\includegraphics[width=0.3\linewidth]{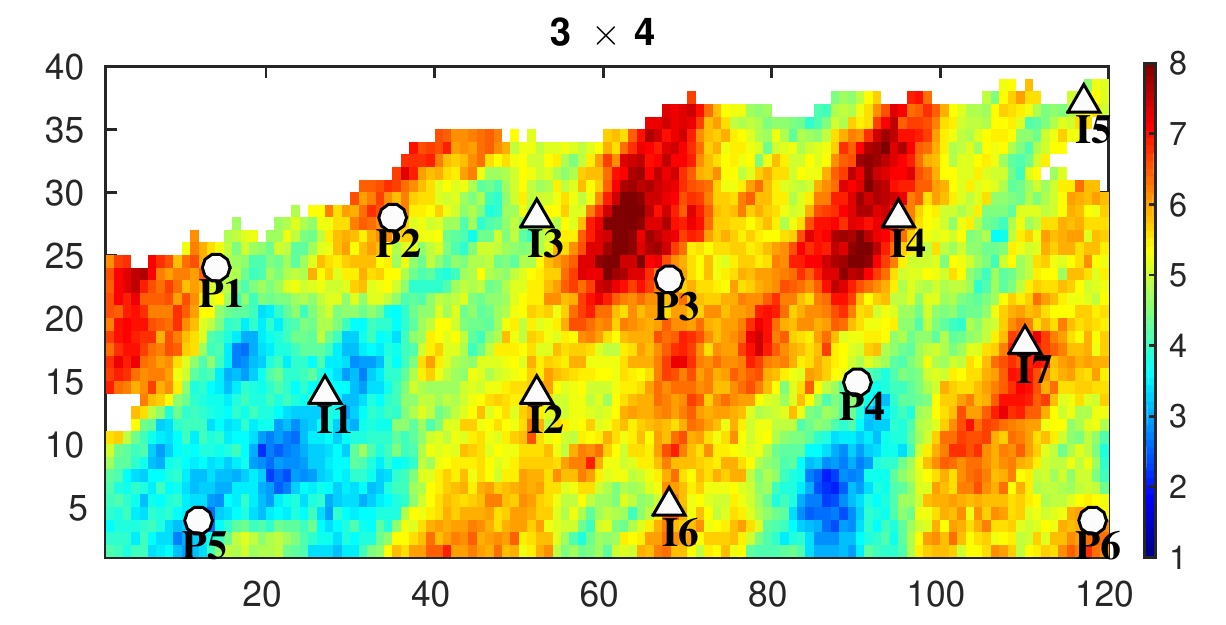}} \\
\subfloat[]%
  {\includegraphics[width=0.3\linewidth]{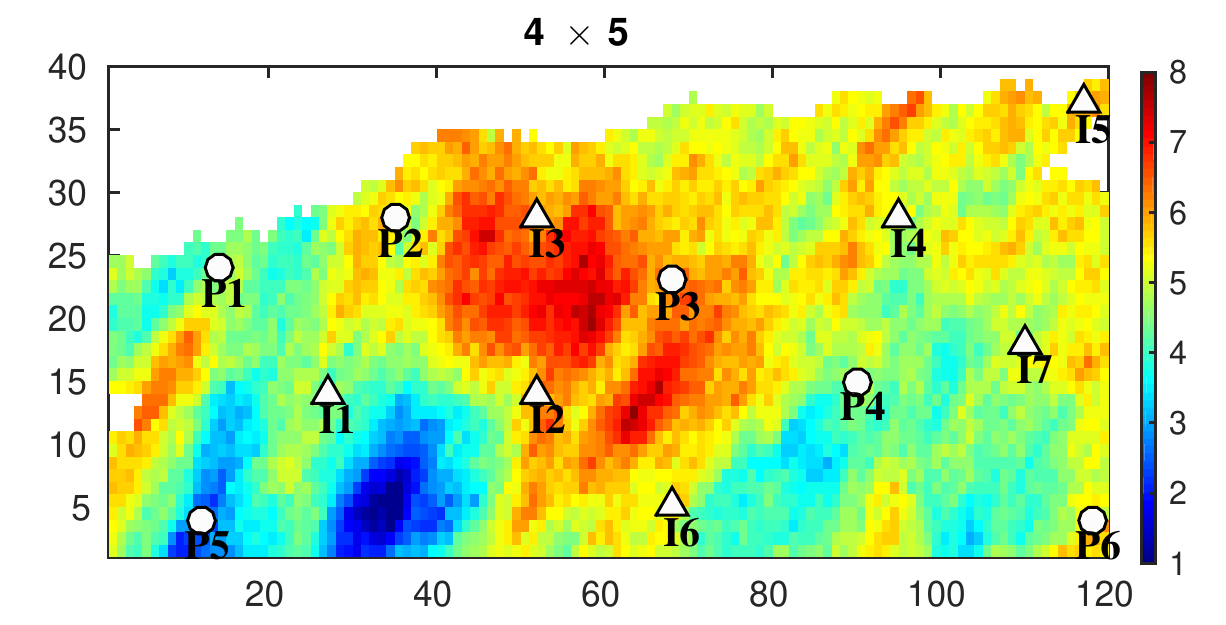}} 
\subfloat[]%
  {\includegraphics[width=0.3\linewidth]{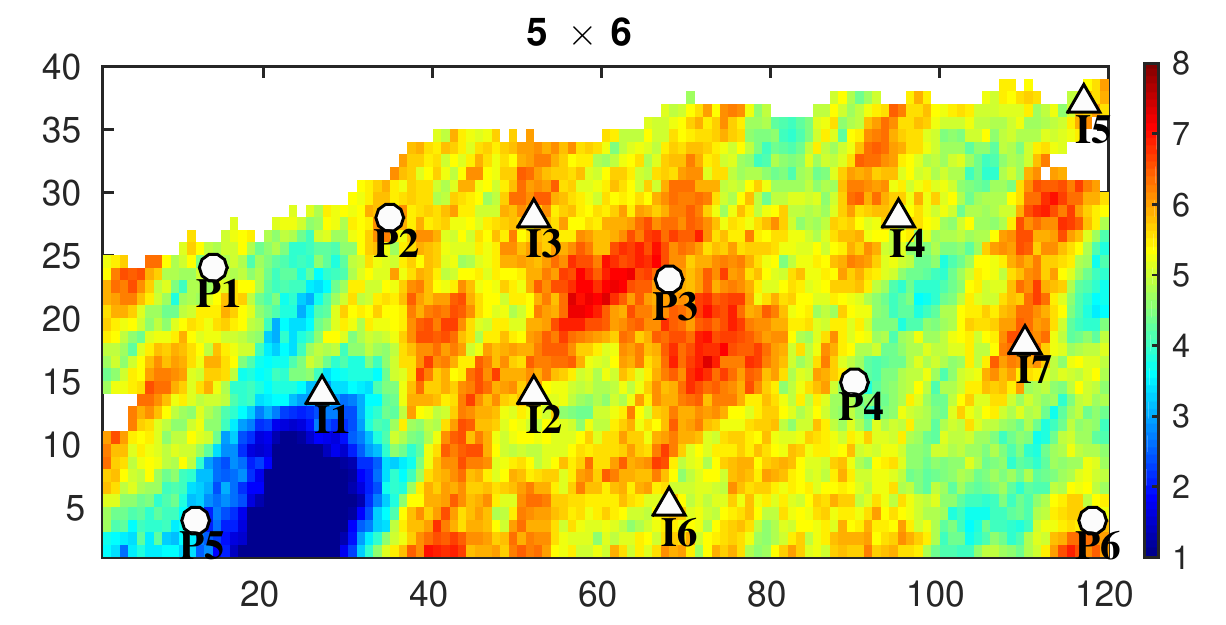}}   
\caption{Updated log-permeability field using different domain decomposition strategy for scenario S1}\label{fig19}
\end{figure*}

Fig.\ref{fig18}, Fig.\ref{fig19} and Table \ref{tab5} show the effects of domain decomposition  
on the minimization process and the final estimate of log-permeability field. 
Four schemes, i.e, 2$\times$3, 3$\times$4, 4$\times$5 and 5$\times$6, are studied. To ensure that all these four schemes are convergent, we specified the maximum number of outer-loops as 15 in this experiment.
The total number of local PCA coefficients and the maximum local PCA coefficients among all subdomains are summarized in Table \ref{tab5}.  

Table \ref{tab5} shows that more subdomains generate fewer local PCA coefficients in each subdomain, 
as a result, fewer FOM simulations are required to implement this subdomain adjoint-based history matching procedure.
Fig.\ref{fig18} and Fig.\ref{fig19} demonstrate that the domain decomposition strategy has significant influence on the performance of LP-SD POD-TPWL.
All four domain decomposition strategies obtain an acceptable cost function value after minimization. 
However, the updated log-permeability fields differ significantly, which implies that different local minimas are generated using
different domain decomposition. 

\begin{table}[!h]
\scriptsize
\centering
\caption{The number of required FOM simulations and cost function values of LP-SD POD-TPWL using different domain decomposition scheme for scenario S1}\label{tab5}
\begin{spacing}{1.25}
\begin{tabular}{|c|c|c|c|c|c|c|}
\hline
- & $N_{L}$ & $l_{d}$ & $N_{G}$ & Iterations & FOM & $J(\boldsymbol{\xi})$ \\
\hline
Initial model & - & - & - & - & - & 4.49$\times$10$^{5}$ \\
2$\times$3 & 112 & 20  &  \multirow{6}{*}{48} & 15 & 77 = 22+(2$\times$20+1)+14 & 901.69 \\
3$\times$4 & 205 & 18  &  \multirow{6}{*}{} & 15 & 73 = 22+(2$\times$18+1)+9 & 878.21 \\
4$\times$5 & 275 & 15  &  \multirow{6}{*}{} & 15 & 67 = 22+(2$\times$15+1)+9 & 912.93 \\
5$\times$6 & 322 & 12  &  \multirow{6}{*}{} & 15 & 61 = 22+(2$\times$12+1)+9 & 869.01 \\
Tolerance & - & - & - & - & - & 4750 \\
Ref - Projected True & - & - & - & - & - & 528.1 \\
Ref - True & - & - & - & - & - & 447.4 \\
\hline
\end{tabular}
\end{spacing}
\end{table}

\subsubsection{Quantification of the different sources of errors}
In the entire approach of subdomain adjoint-based history matching, there are three main sources of errors (SOE) involved, approximation errors of the subdomain POD-TPWL (SOE1), e.g., POD, RBF, and domain decomposition; 
(SOE2) the loss of global PCA patterns due to an insufficient number of local PCA patterns, e.g., retaining 2 local PCA patterns; 
(SOE3) the full energy of global PCA is not preserved, e..g, 95\% energy is retained.  
To distinguish and quantify these three errors, LP-SD POD-TPWL and FD are consecutively implemented. 
After minimizing the cost function using LP-SD POD-TPWL, continuing minimization using FD-LP can quantify the SOE 1, while continuing minimization using FD-GP can quantify the sum of SOE1 and SOE2.
To quantify SOE3, the cost function is minimized through preserving more and more global PCA energy, e.g., 95\%, 98\%, 99\% and 99.5\%. 

\begin{figure*}[!h]
\centering
\subfloat[]%
  {\includegraphics[width=0.3\linewidth]{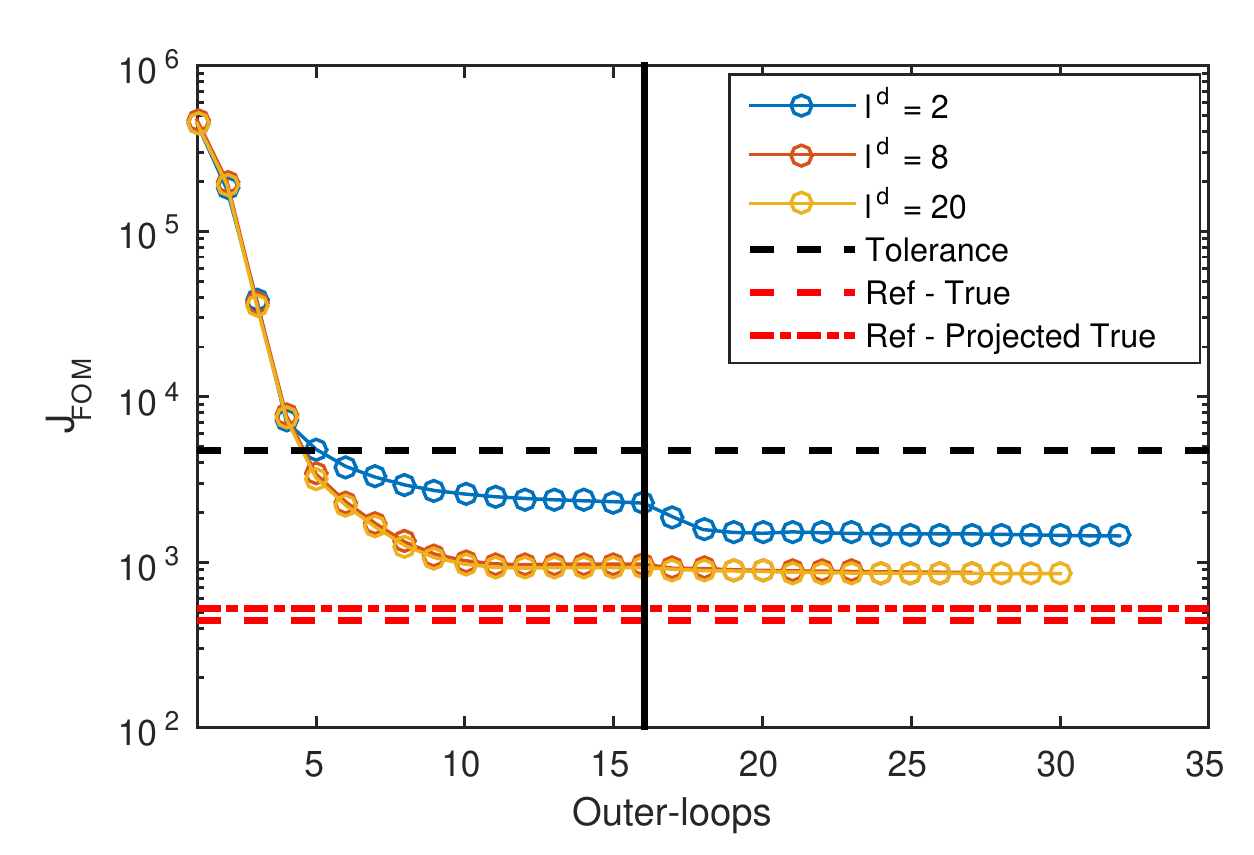}} 
\subfloat[]%
  {\includegraphics[width=0.3\linewidth]{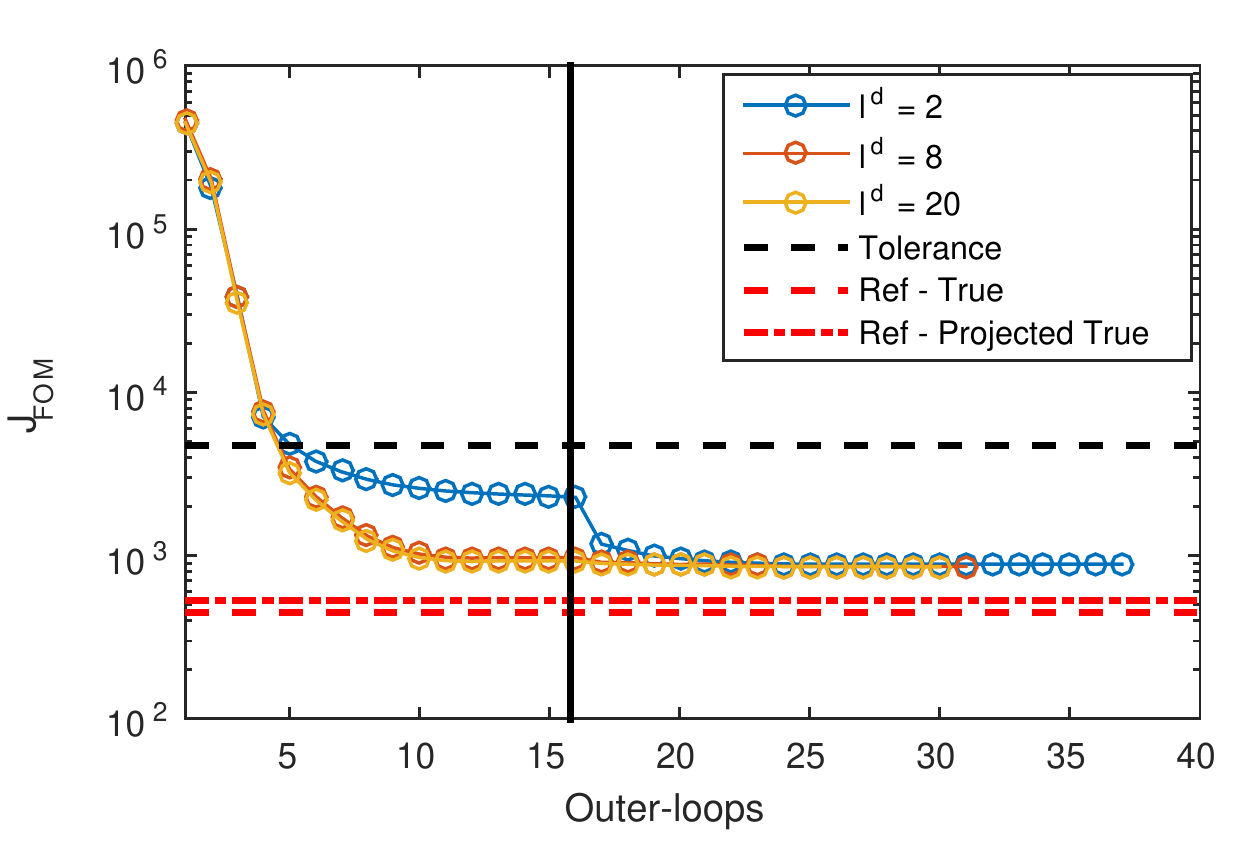}}
\subfloat[]%
  {\includegraphics[width=0.3\linewidth]{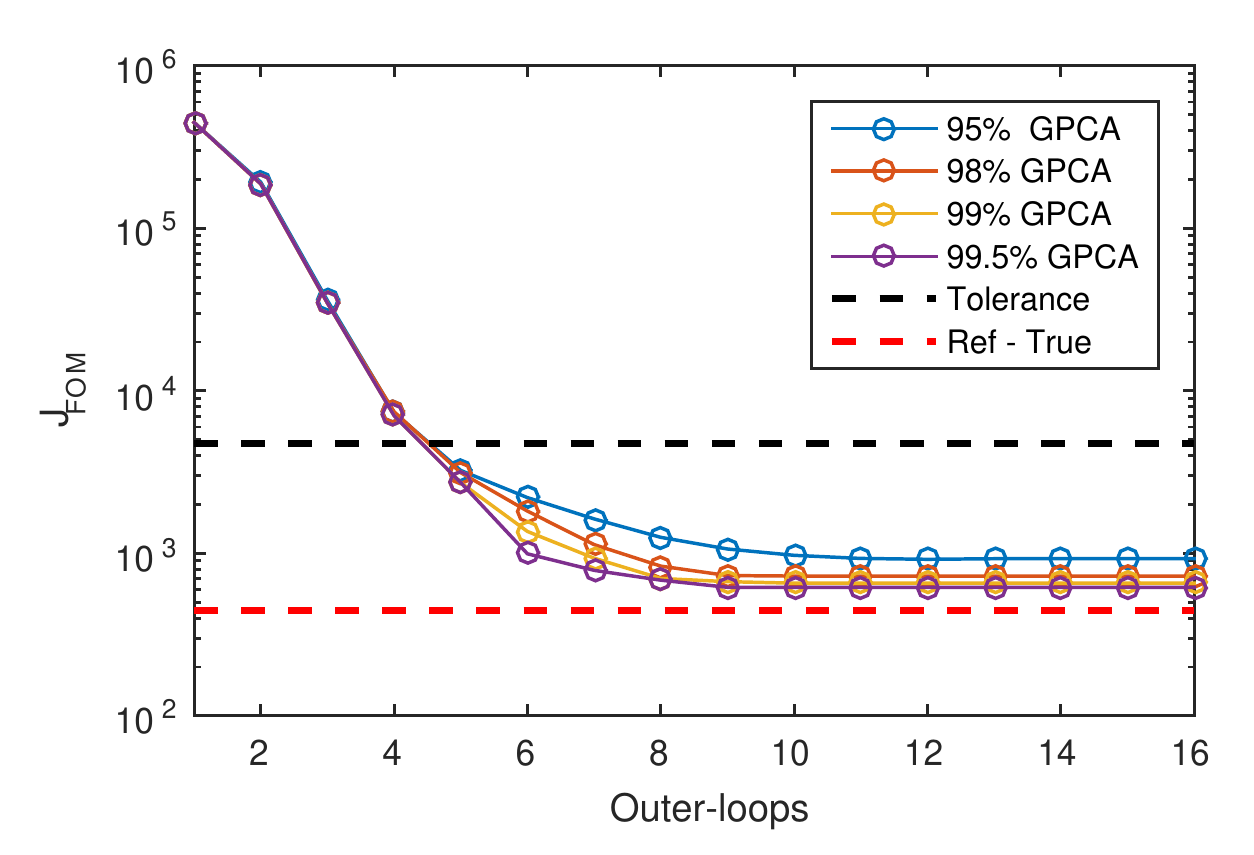}}  
\caption{Changes of the cost function value when the LP-SD POD-TPWL and FD are sequentially implemented. (a) continuing minimization using FD-LP; 
(b) continuing minimization using FD-GP; (c) the cost function is minimized through preserving more and more global PCA energy, e.g., 95\%, 98\%, 99\% and 99.5\%.
The vertical black line represents the starting point of minimization using the FD method}\label{fig20}
\end{figure*}

We have proposed a method to determine the minimum number of local PCA patterns based on whether the global PCA patterns can be fully covered.
2$\times$3 domain decomposition is fixed by retaining 2, 8 and 20 local PCA patterns among all subdomains, respectively.
15 outer-loops are also specified for the minimization process.
Table \ref{tab6} summarizes the initial, final and reference cost function values, the total sum of local PCA patterns, and the required FOM simulations.
8 local PCA patterns are sufficient to fully cover 48 global PCA patterns, therefore, retaining 8 and 20 LPCA patterns obtain similar values of the cost function.
The numerical results imply LP-SD POD-TPWL is very promising for large-scale history matching because retaining only a relative small number of local PCA patterns
can already cover the global PCA patterns fully.
We will further explore this when assimilating a large number of measurements, e.g., 4D seismic data, in the second case-study. 
Fig.\ref{fig20}(a) shows that the minimization using LP-FD does not significantly decrease the cost function no matter 
how many local PCA patterns we retain. This implies that the SOE1 is very small and almost can be ignored. 
Fig.\ref{fig20}(b) quantifies the total sum of SOE1 and SOE2.
Minimization using GP-FD can significantly decrease the cost function particularly in case of only 2 local PCA patterns are preserved.
An insufficient number of local PCA patterns results in a loss of global PCA patterns and therefore in a large SOE2.
SOE2 can be decreased through increasing the number of local PCA patterns, what implies more FOM simulations are required to implement LP-SD POD-TPWL.
Fig.\ref{fig20}(c) and Table.\ref{tab6} indicate that 
SOE3 will be gradually decreased with increasing global PCA energy. Retaining 98\% global PCA patterns can accurately represent the original parameter field in this case.
An additional increase of 60 global PCA patterns from 48 to 108 requires an additional 10 local PCA patterns in each subdomain. This only requires 20 new FOM simulations.

\begin{table}[!h]
\scriptsize
\centering
\caption{The number of FOM simulations and the cost function values when quantifying SOE1 and SOE2 for scenario S1}\label{tab6}
\begin{spacing}{1.25}
\begin{tabular}{|c|c|c|c|c|c|c|}
\hline
\multicolumn{4}{|c|}{-} & \multicolumn{3}{|c|}{$J(\boldsymbol{\xi})$} \\
\hline
\multicolumn{4}{|c|}{Initial model} & \multicolumn{3}{|c|}{1.01$\times$10$^{5}$} \\
\hline
\multicolumn{4}{|c|}{Tolerance} & \multicolumn{3}{|c|}{4750} \\
\hline
\multicolumn{4}{|c|}{Ref - Projected True} & \multicolumn{3}{|c|}{528.1} \\
\hline
\multicolumn{4}{|c|}{Ref - True} & \multicolumn{3}{|c|}{447.4} \\
\hline
$l_{d}$ & $N_{L}$ &  $N_{G}$ & FOM & LP-SD POD-TPWL & LP-FD & GP-FD \\
\hline
2 & 12 &  \multirow{3}{*}{48}  & 53 = 22+(8$\times$2+1)+14 & 2276.15 & 1441 & 886.3 \\
8 & 48  &  \multirow{3}{*}{}  & 53 = 22+(2$\times$8+1)+14 & 902.48 & 869.24 &  860.24\\
20 & 120 &  \multirow{3}{*}{}  & 77 = 22+(2$\times$20+1)+14 & 892.21 & 854.69 & 860.24 \\
\hline
\end{tabular}
\end{spacing}
\end{table}

\begin{table}[!h]
\scriptsize
\centering
\caption{The number of FOM simulations and the cost function values when quantifying SOE3 for scenario S1}\label{tab7}
\begin{spacing}{1.25}
\begin{tabular}{|c|c|c|c|c|c|c|}
\hline
- & $N_{L}$ & $l_{d}$ & $N_{G}$ & Iterations & FOM & $J(\boldsymbol{\xi})$ \\
\hline
Initial model & - & - & - & - & - & 1.01$\times$10$^{5}$ \\
95\% GPCA & 48 & 8  &  72 & 15 & 53 = 22+(2$\times$8+1)+14 & 902.48 \\
98\% GPCA & 72 & 12  &  72 & 15 & 61 = 22+(2$\times$12+1)+14 & 738.25 \\
99\% GPCA & 96 & 16  &  92 & 15 & 69 = 22+(2$\times$16+1)+14 & 694.18 \\
99.5\% GPCA & 108 & 18  & 104 & 15 & 73 = 22+(2$\times$18+1)+14 & 621.52 \\
\hline
\end{tabular}
\end{spacing}
\end{table}

\subsection{Results of estimating a large number of parameters simultaneously assimilating well data and seismic data}
\begin{figure*}[!h]
\centering\includegraphics[width=0.8\linewidth]{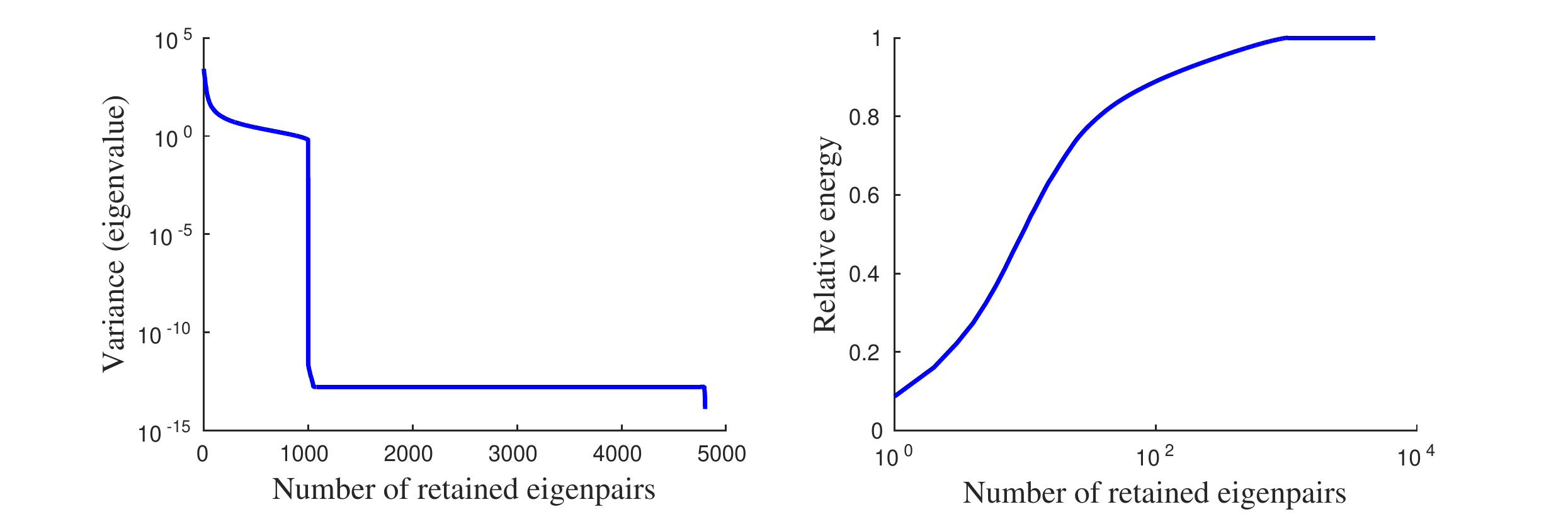}
\caption{The decay of eigen-spectrum of SVDs for scenario S2}\label{fig25}
\end{figure*}

\begin{figure}[!h]
\centering
\subfloat['True' model]%
  {\includegraphics[width=0.35\linewidth]{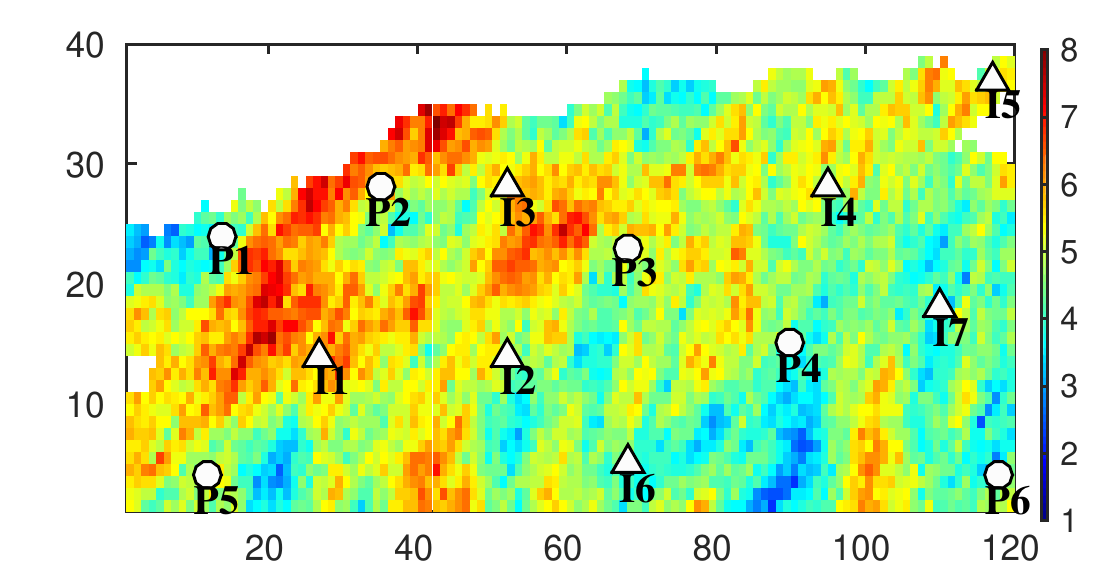}} 
\subfloat[Projected 'true' model using LPCA]%
  {\includegraphics[width=0.35\linewidth]{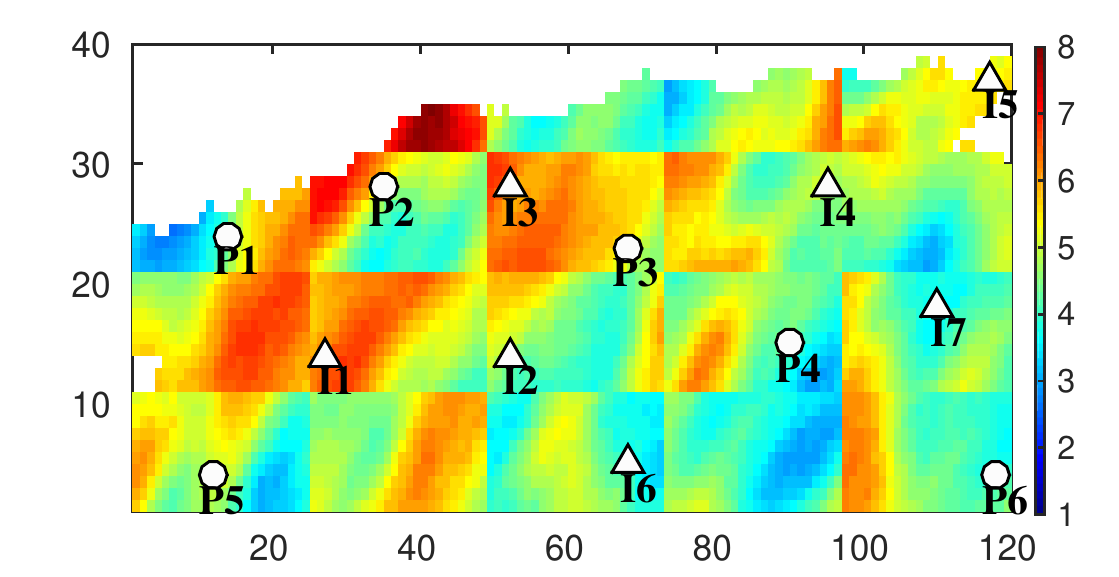}} 
\subfloat[Projected 'true' model where the LPCA results are projected onto the global PCA's]%
  {\includegraphics[width=0.35\linewidth]{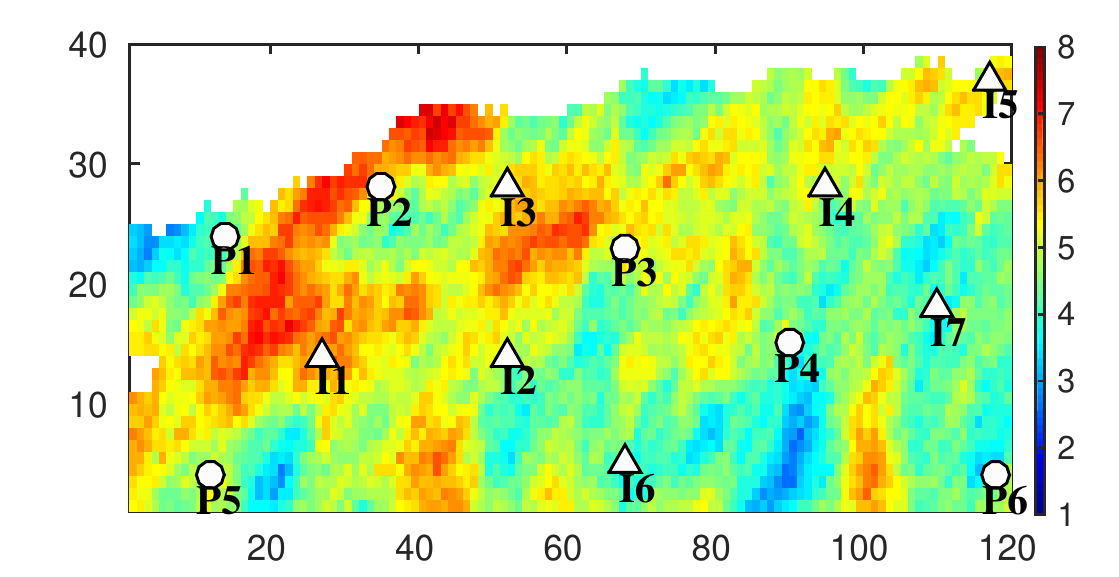}}
\caption{Comparison of the 'true' reservoir model in full-order space and reduced-order space for scenario S2}\label{fig26}
\end{figure}

In this scenario, the ability of LP-SD POD-TPWL to estimate a large number of parameters by simultaneously assimilating well data and seismic data
is investigated for the SAIGUP model with strong spatial variability.
We generate another 1000 Gaussian-distributed log-permeability fields which are used to form the covariance matrix for SVDs.
The decay of Eigen-spectrum of SVDs is illustrated in Fig.\ref{fig25} and retaining 95\% energy results in $N_{G}$ = 282 global PCA patterns.
One of these realizations is considered to be the truth. Fig.\ref{fig26}(a) and Fig.\ref{fig26}(b) separately represents the projected 'true' permeability field using O-LS-PCA and local PCA.

Experiments showed that the changes in singular value spectrum of the snapshot matrix for POD are insignificant 
when the snapshots at every timestep are selected from 32 or more FOM simulations. 
One of domain decomposition schemes, e.g., a total of 20 subdomains with 4 subdomains in $\textit{x}$ direction and 5 subdomains in $\textit{y}$ direction is chosen as a base-case.
For each subdomain, two eigenvalue problems separately for pressure and saturation are solved using POD,
where 95\% of the energy is preserved, and the number of reduced variables for each subdomain is shown in Fig.\ref{fig27}.

\begin{figure}[!h]
\centering\includegraphics[width=0.4\linewidth]{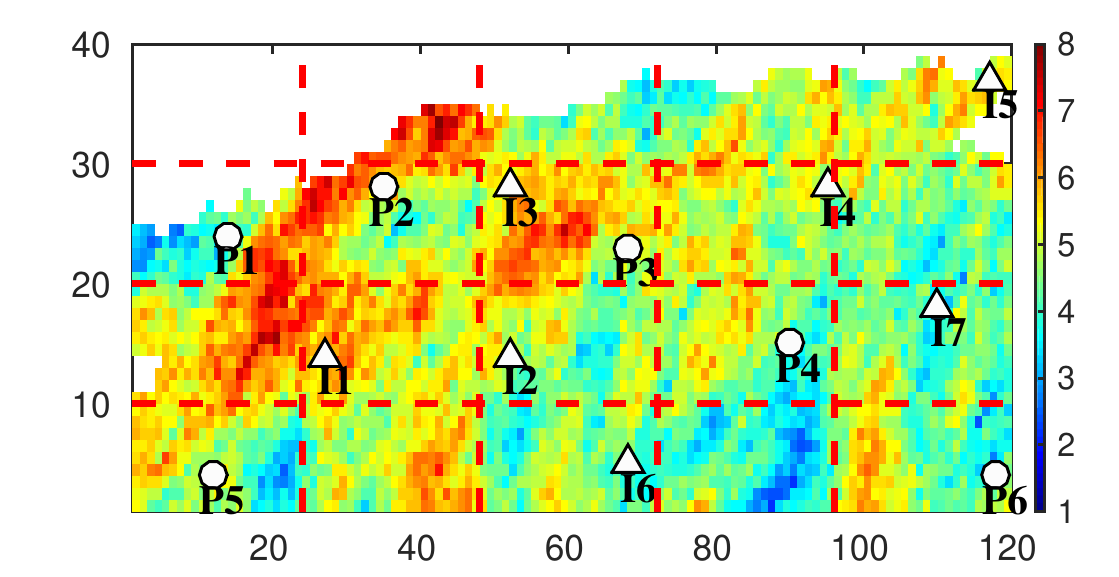}\\
\caption{The illustration of domain decomposition for scenario S2}\label{fig10}
\end{figure}

\begin{figure}[!h]
\centering\includegraphics[width=0.8\linewidth]{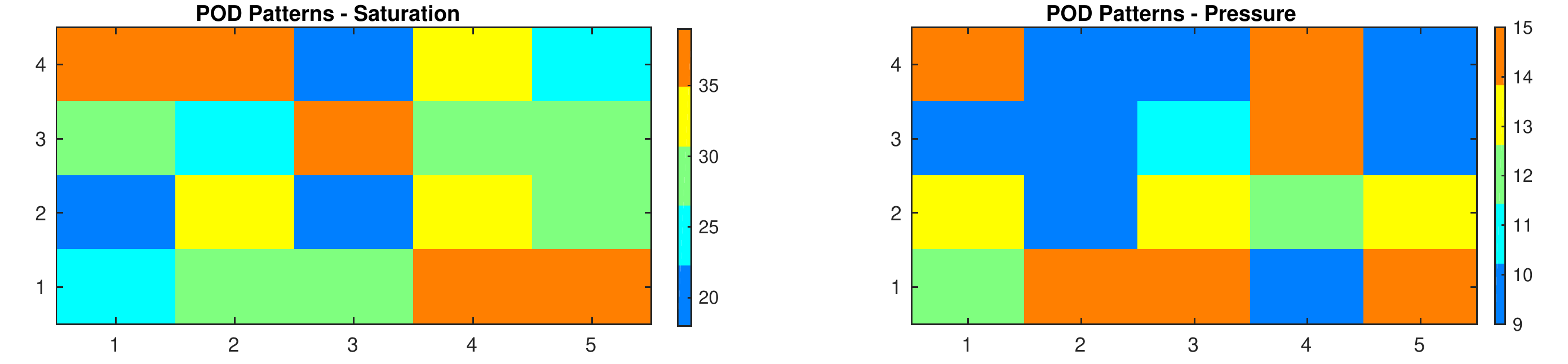}\
\caption{The number of reduced pressure and saturation patterns in each subdomain for scenario S2}\label{fig27}
\end{figure}

\begin{figure}[!h]
\centering
\subfloat[]%
  {\includegraphics[width=0.4\linewidth]{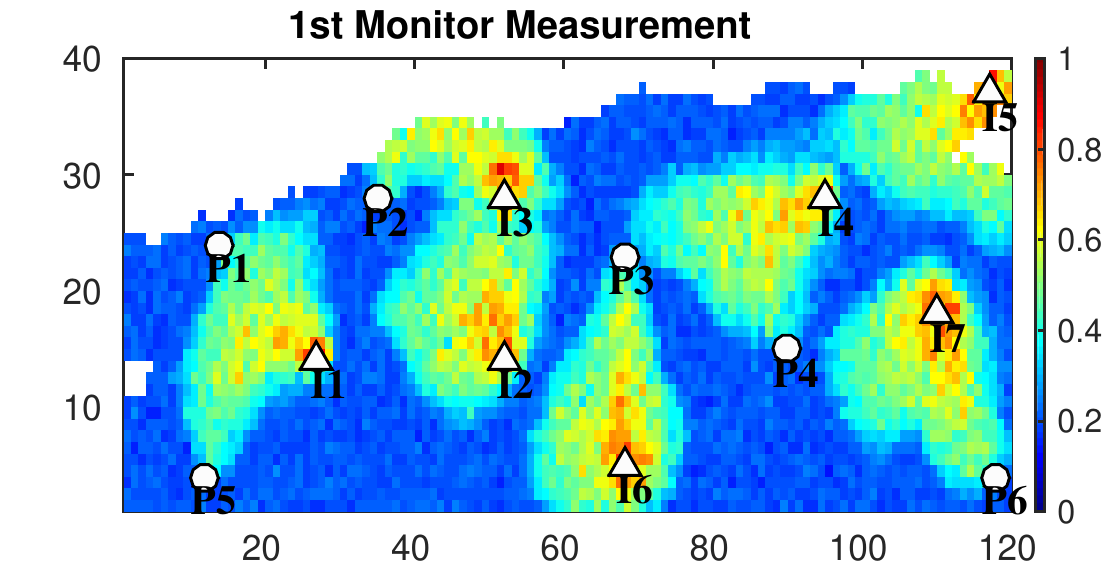}} 
\subfloat[]%
  {\includegraphics[width=0.4\linewidth]{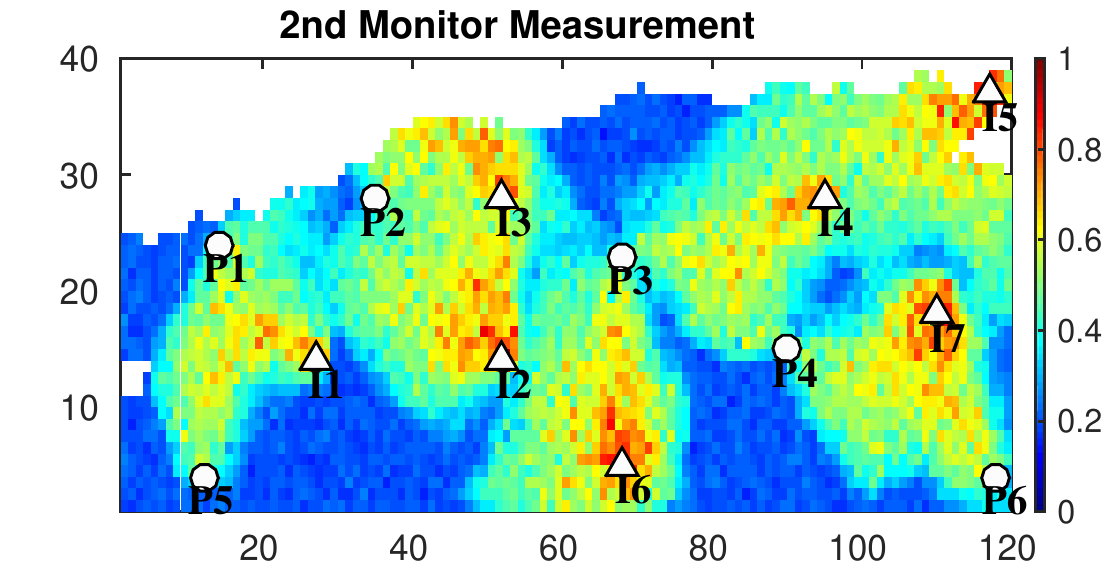}} 
\caption{Noise distribution of water saturation for scenario S2. Normal distributed independent measurement noise with 
a standard deviation equal to 5\% of the 'true' data value, was added to all observations}\label{fig28}
\end{figure}

History matching time-lapse seismic data requires the capability to compute seismic data from a given reservoir
model. This task can be accomplished by introducing
a rock–fluid model to convert the reservoir properties
and the simulator primary variables (pressure and
saturations) into modeled elastic properties. The most
widely used model to predict the seismic response of
a reservoir due to production is the Gassmann model
\cite{gassmann1951elastic}. For a simplification in the paper, we directly assimilate the water saturation in each grid-block 
without the mapping to modeled elastic properties.
The history production period is 10 years during which synthetic seismic data and well data were generated based
on the true reservoir model. The seismic data correspond
to the saturation is collected at two monitor surveys after day 1825 (1st monitor), and day 3650 (2nd monitor) of production.
Well data consisted of bottom-hole pressures in the injector and fluid rates and WCT in the producers are taken from six producers and seven injectors 
in 50 time instances. Thus, there are in total 8920 = 950 + 3985$\times$2 measurements.
The noisy measurements for the two monitors are shown in Fig.\ref{fig28}.

Three different domain decomposition strategies, (1) 3$\times$4 ; 
(2) 4$\times$5 ; and (3) 5$\times$6,  are compared. 
The stopping criteria are $\eta_{\jmath} =10^{-4} $, $\eta_{\boldsymbol{\xi}} =10^{-3} $, and $N_{max}$=100, the number of outer-loops is 10. 
In addition, the finite-difference method GP-FD is also implemented for comparison.
The previous numerical experiments have demonstrated that a minimum number of local PCA patterns is sufficient as long as the total number of local PCA patterns equals to the number of 
global PCA patterns. The number of local PCA patterns corresponding to different domain decomposition strategies is summarized in Table \ref{tab8}.
Fig.\ref{fig30} shows that the true reservoir model can be reconstructed accurately when a large number of measurements is available. 
In this case, the 4$\times$5 domain decomposition strategy leads to a smaller cost function value.
Also, both of the predictions of seismic data and well data are enhanced as well in Fig.\ref{fig31}, Fig.\ref{fig32} and Fig.\ref{fig33}.
Fig.\ref{fig31} and Fig.\ref{fig32} show the predicted saturation and its corresponding RMSE. Compared to the initial model, the model predictions have been significantly improved.
Fig.\ref{fig33} illustrates cross-plots between observed and predicted water saturation at these 2 monitors survey and the dashed lines
correspond to 2 standard deviations of the measurement errors. The uncertainty of the data has been decreased as well.
Compared with the first test case, the number of global PCA patterns
has been increased from 48 to 282. However the required number of FOM simulations has only increased from 53 to 90. 
The degree of freedom for this history matching problem depends on the number of global PCA patterns, while the required FOM simulation depends on the number of local PCA patterns. 
It is therefore very attractive to increase the degree of the freedom by adding local PCA patterns in all subdomains. 
Taking 5 $\times$ 6 domain decomposition as an example, adding one local PCA pattern in each subdomain allows us to retain another 30 global PCA patterns, while only  
2 more FOM simulations are added to the entire history matching procedure.
These numerical results further demonstrate that introducing local parameterization makes subdomain POD-TPWL highly scalable and the required FOM simulations do not grow rapidly with increasing number of uncertain parameters.
Fig.\ref{fig34} illustrates the data match of fluid rate and water-cut up to year 10 and an additional 15-year prediction for all six producers. 
The prediction especially the fluid rate and bottom-hole pressure based on the initial model is far from the true model. 
After the history matching, the predictions of the updated model match the observations very well. 

\begin{figure}[!h]
\centering\includegraphics[width=0.5\linewidth]{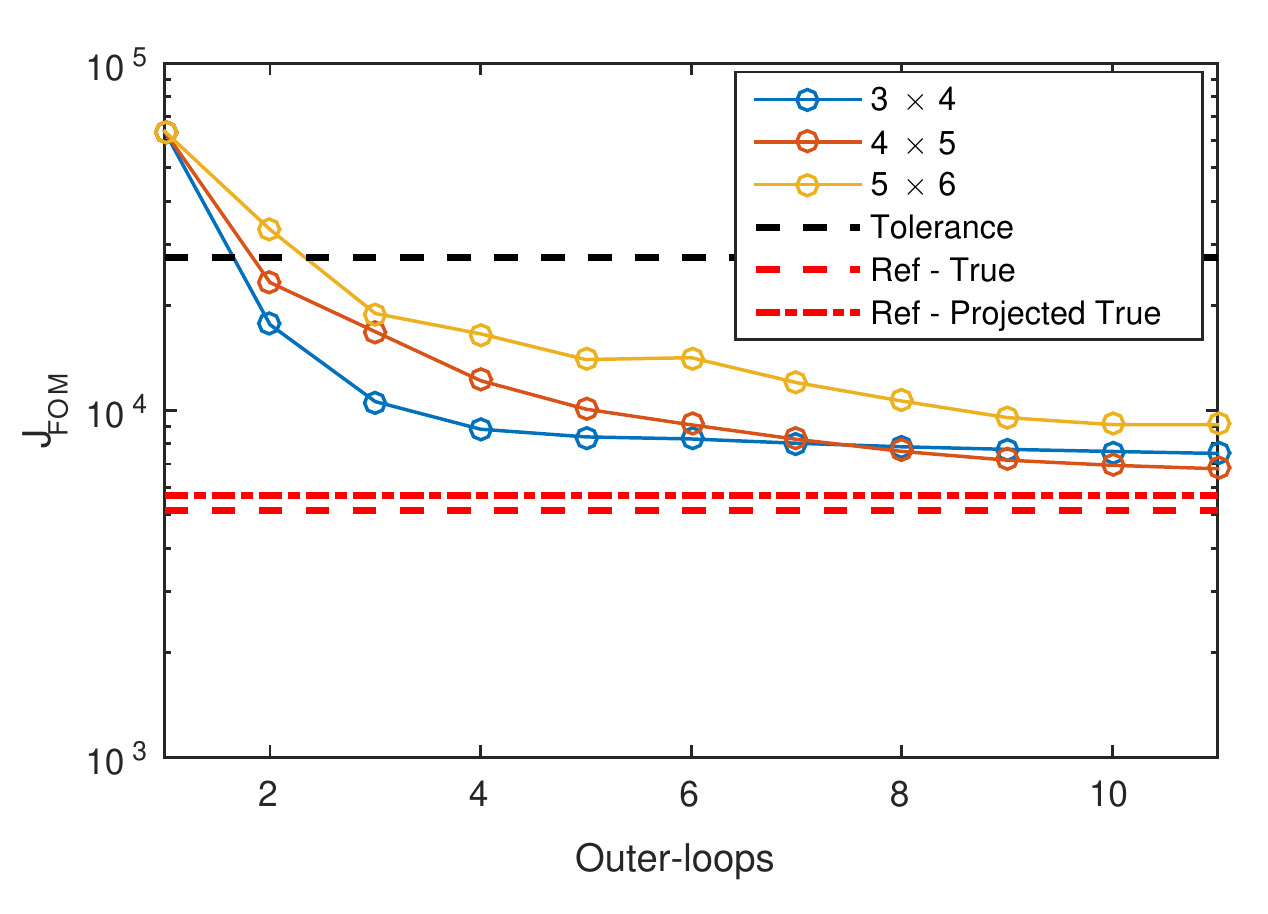}
\caption{The cost function as a function of outer-loops using LP-SD POD-TPWL for scenario S2. The computation of the cost function uses the full-order model as Eq.\ref{eq6}}\label{fig29}
\end{figure}

\begin{table}[!h]
\scriptsize
\centering
\caption{The number of FOM simulations and RMSEs of LP-SD POD-TPWL using different domain decomposition strategies for scenario S2}\label{tab8}
\begin{spacing}{1.25}
\begin{tabular}{|c|c|c|c|c|c|c|}
\hline
- & $l_{d}$ & $N_{L}$ & $N_{G}$ & Iterations & FOM & $J(\boldsymbol{\xi})$  \\
\hline
Initial model & - & - & - & - & - & 6.39$\times$10$^{4}$ \\
3$\times$4 & 24 & 288 & \multirow{3}{*}{282} & 10 & 90 = 32+(2$\times$24+1)+9 & 7508  \\
4$\times$5 & 15 & 300 & \multirow{6}{*}{} & 10 & 72 = 32+(2$\times$15+1)+9 &  6783 \\
5$\times$6 & 10 & 300 & \multirow{6}{*}{} & 10 & 62 = 32+(2$\times$10+1)+9 & 9601   \\
Tolerance & - & - & - & - & - & 2.75$\times$10$^{4}$ \\
GP-FD & - & - & - & - & - & 6416 \\
Projected 'True' model & - & - & - & - & - & 5685 \\
'True' model & - & - & - & - & - & 5149  \\
\hline
\end{tabular}
\end{spacing}
\end{table}

\begin{figure*}[!h]
\centering
\subfloat[]%
  {\includegraphics[width=0.3\linewidth]{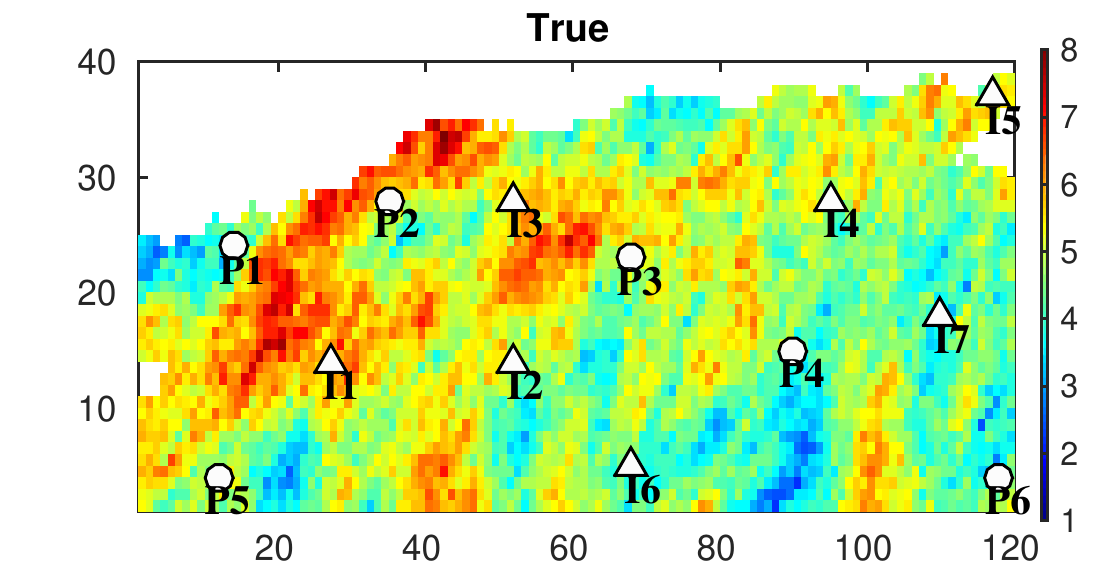}}
\subfloat[]%
  {\includegraphics[width=0.3\linewidth]{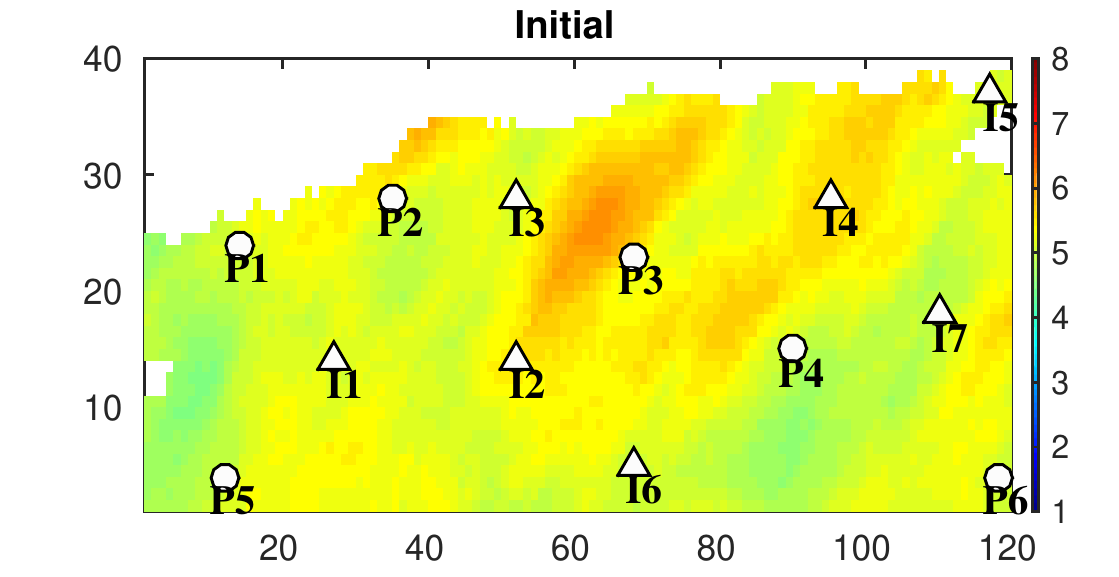}}
  \subfloat[]%
  {\includegraphics[width=0.3\linewidth]{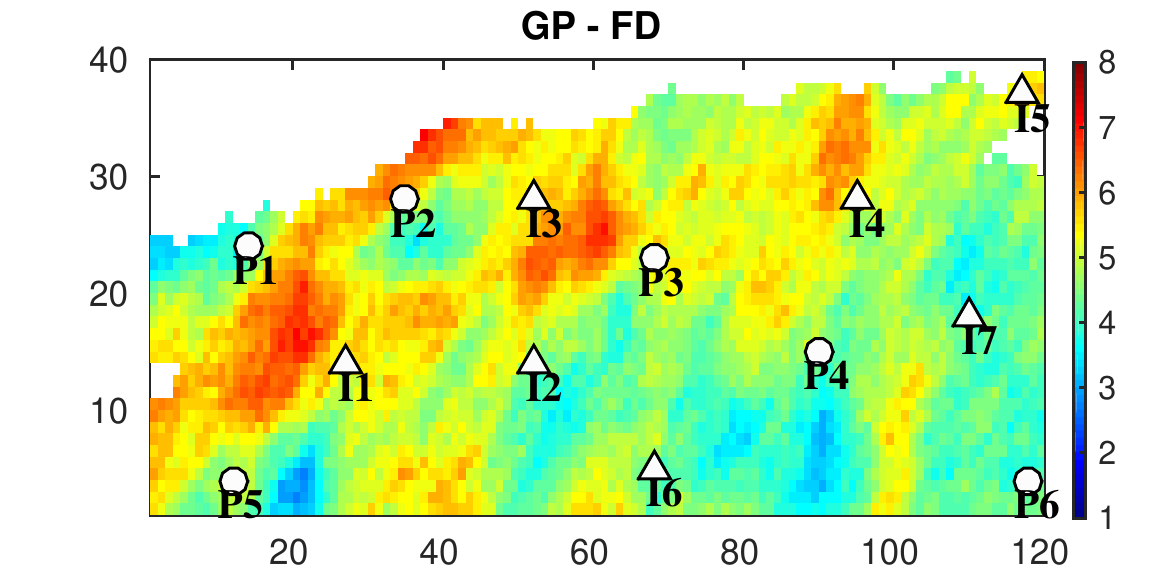}} \\
\subfloat[]%
  {\includegraphics[width=0.3\linewidth]{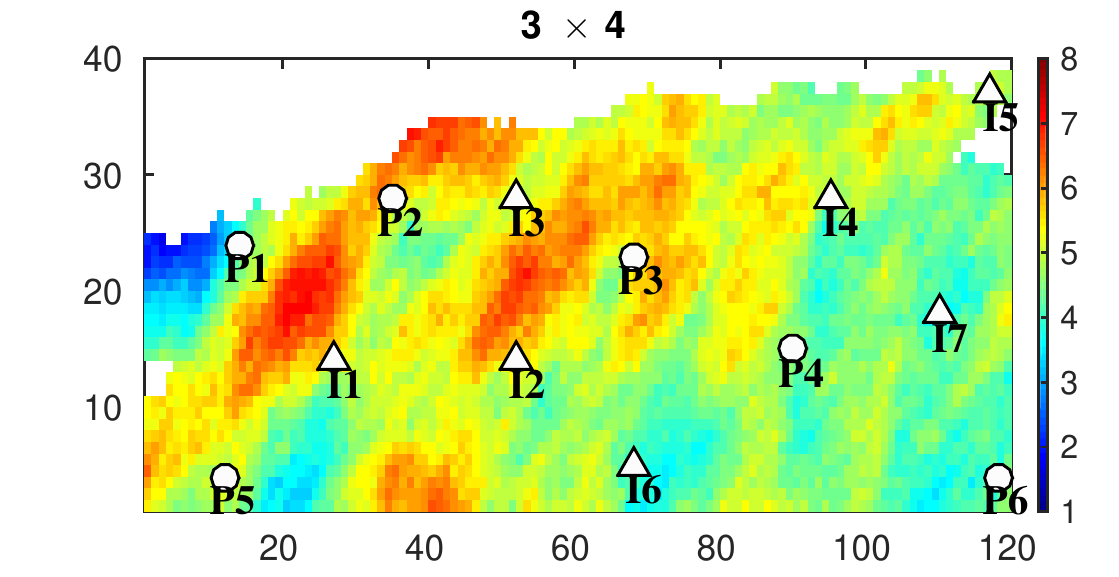}} 
\subfloat[]%
  {\includegraphics[width=0.3\linewidth]{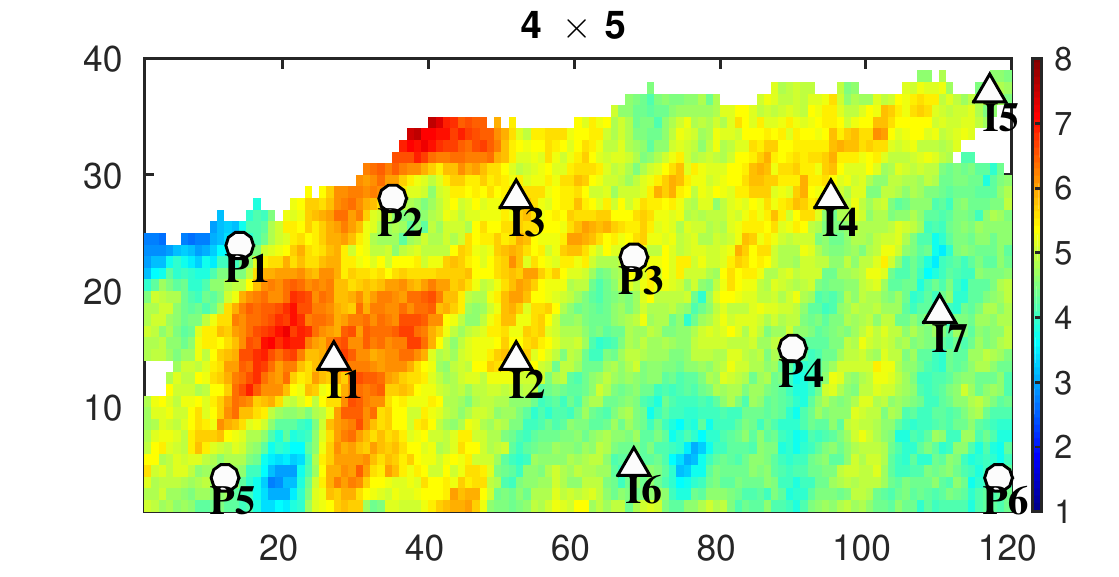}} 
\subfloat[]%
  {\includegraphics[width=0.3\linewidth]{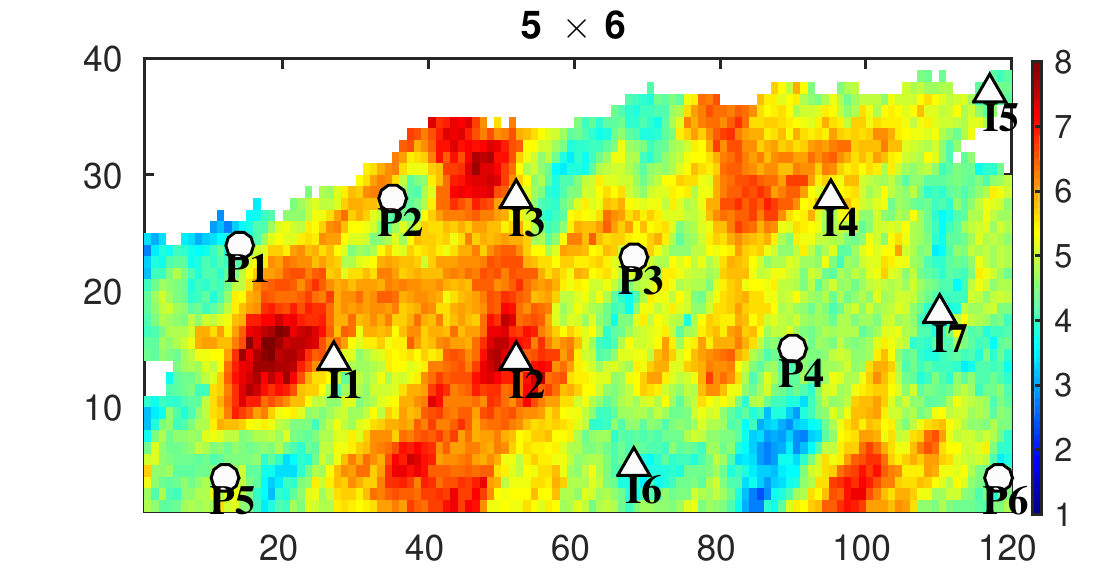}} \\
\caption{Comparison of updated permeability fields using LP-SD POD-TPWL and GP-FD for scenario S2}\label{fig30}
\end{figure*}

\begin{figure}[!h]
\centering
\subfloat[1st monitor]%
  {\includegraphics[width=0.4\linewidth]{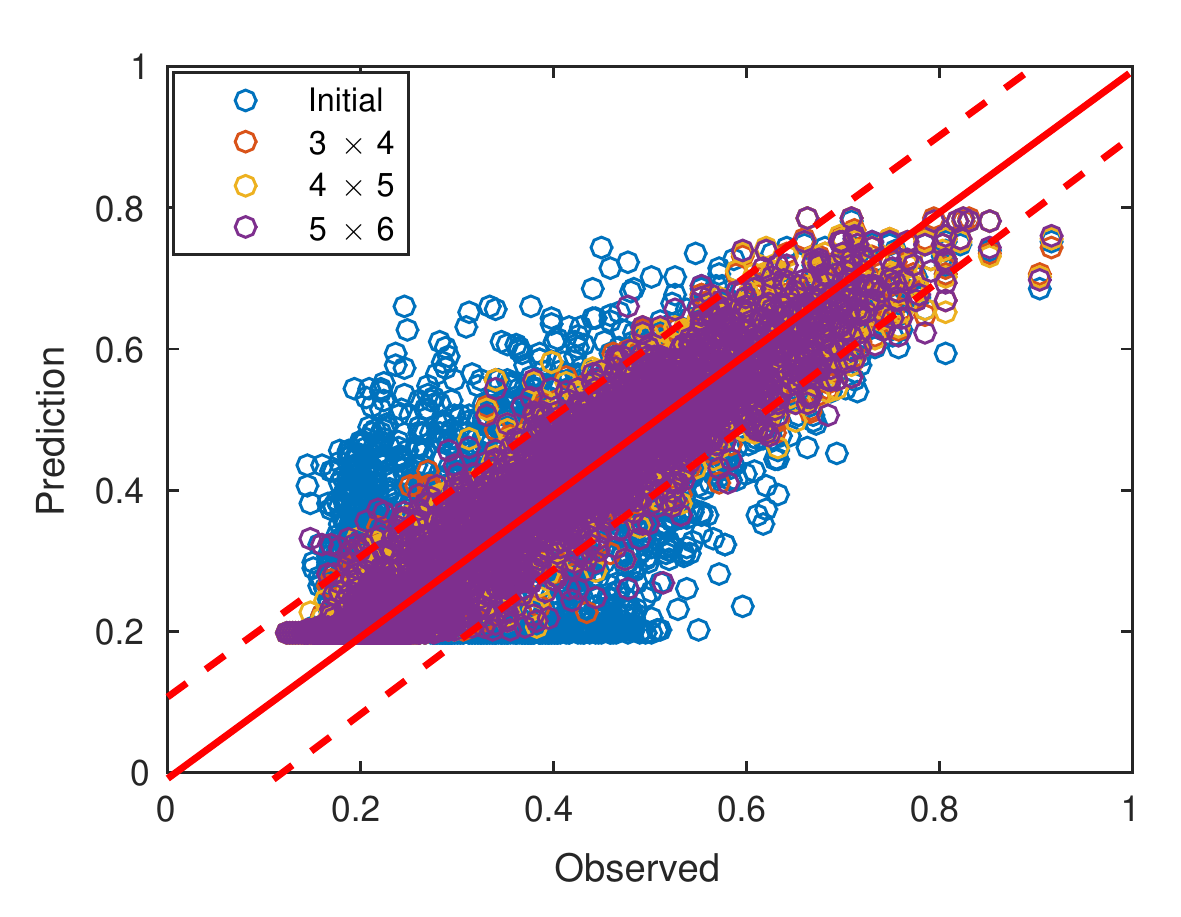}}
\subfloat[2nd monitor]%
  {\includegraphics[width=0.4\linewidth]{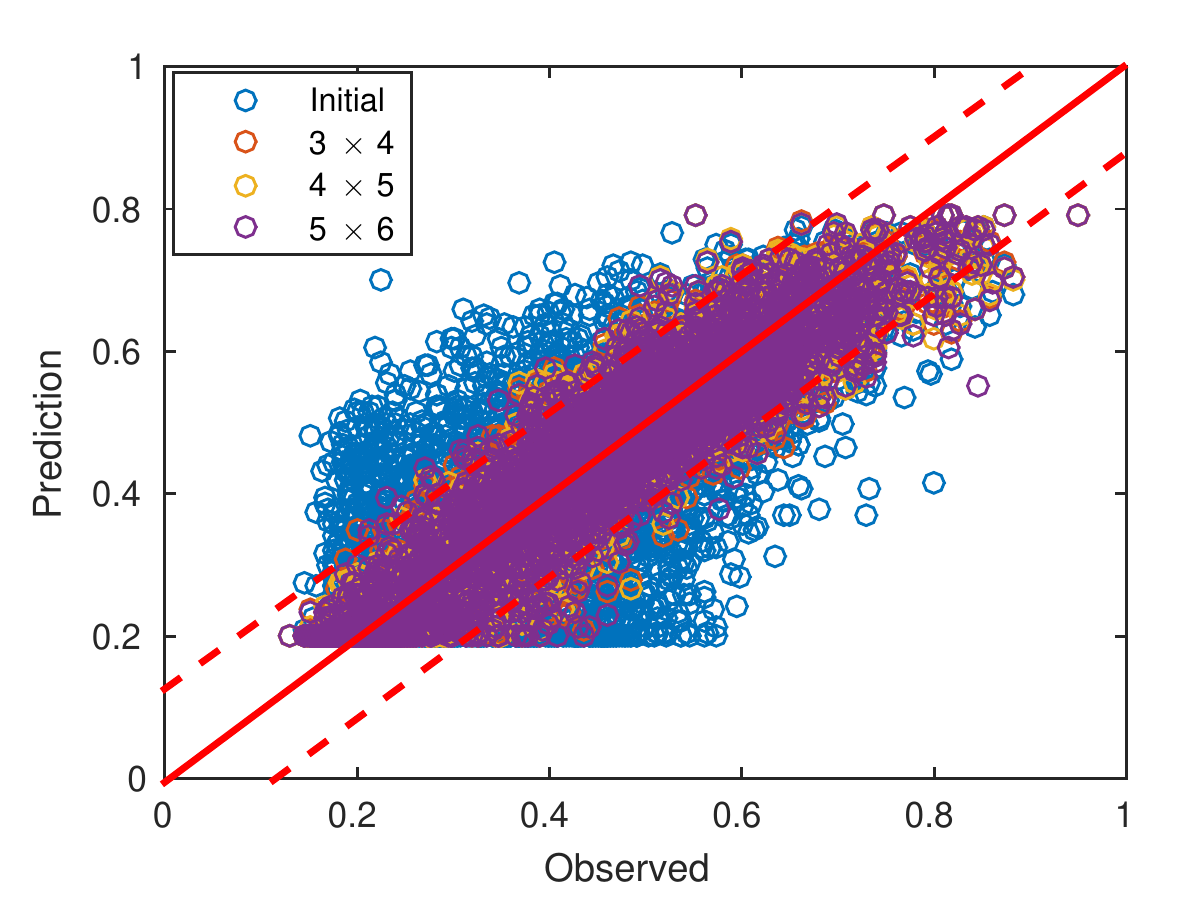}} 
\caption{Cross-plots between observed and predicted water saturation for scenario S2. The dashed lines
correspond to 2 standard deviations of the measurement errors}\label{fig33}
\end{figure}

\begin{figure*}[!h]
\centering\includegraphics[width=0.8\linewidth]{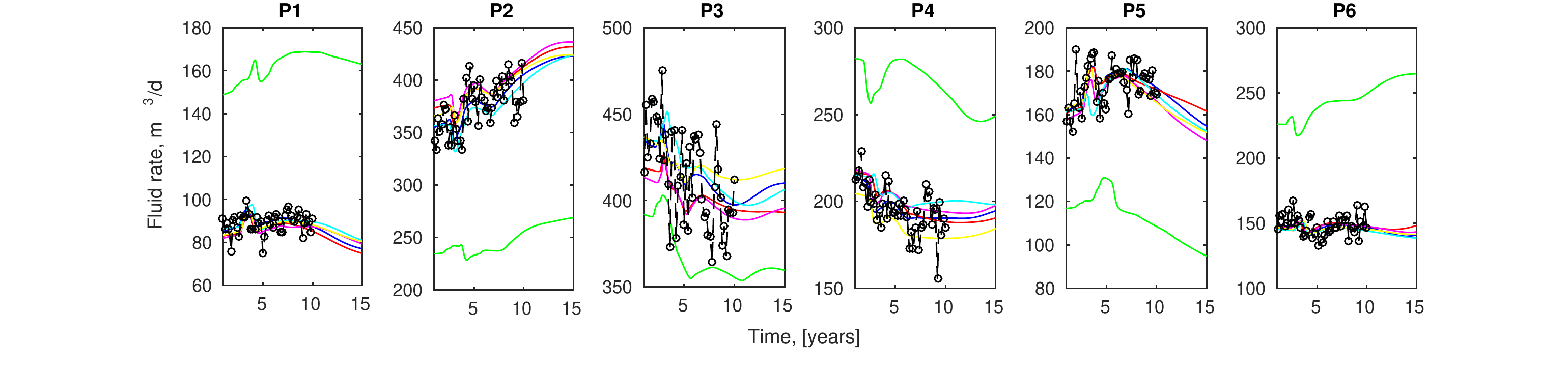}\\
\centering\includegraphics[width=0.8\linewidth]{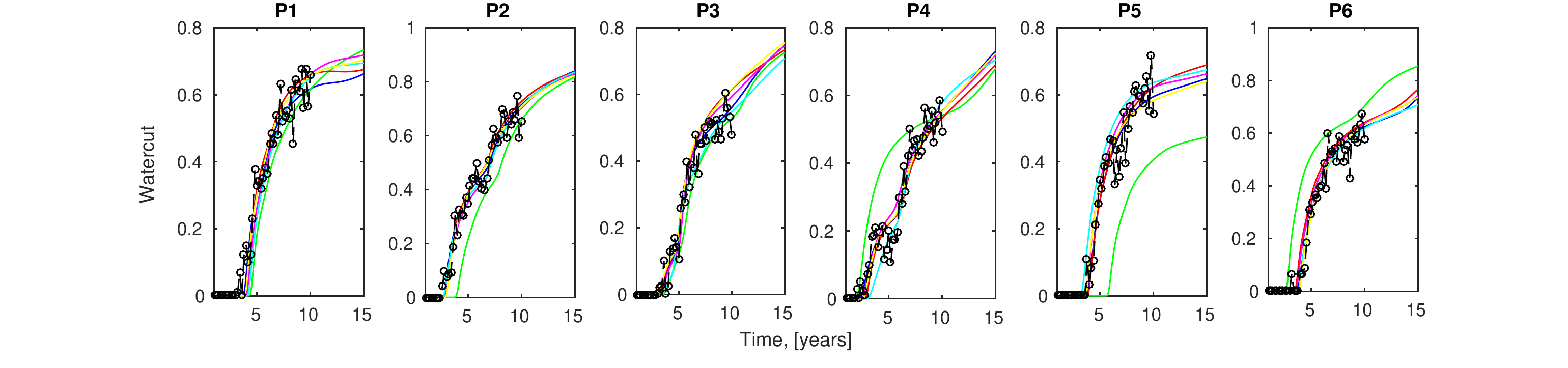}\\
\centering\includegraphics[width=0.8\linewidth]{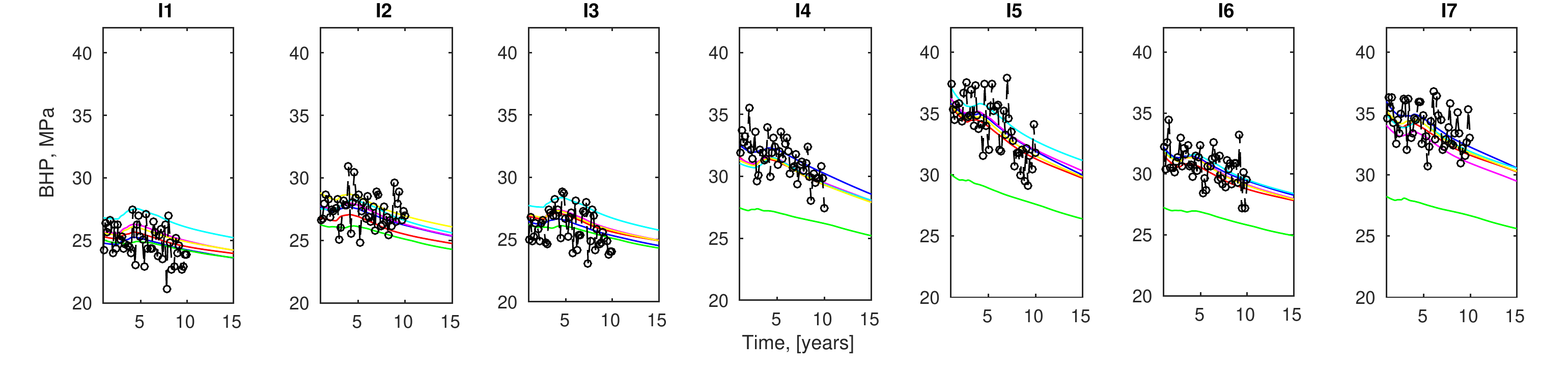}\\
\caption{Forecast of the liquid rate, WCT and BHP for scenario S2: green line - initial model, blue line - 'true' model, red line - updated model using LP-SD POD-TPWL with 3$\times$4 DD,  
magenta line-updated model using LP-SD POD-TPWL with 4$\times$5 DD, 
cyan line - updated model using LP-SD POD-TPWL with 5$\times$6 DD, yellow line - updated model using GP-FD}\label{fig34}
\end{figure*}

\subsection{Computational complexity}
The computational cost of the subdomain model-reduced history matching approach can
be split into two main parts. Offline stage: the cost of constructing the subdomain reduced-order linear model. Online stage:
the cost of solving the reduced system and the parameter estimation problem. \\

\textbf{Offline stage: the cost of constructing subdomain reduced-order linear model}
\begin{itemize}

\item The cost of executing parameterization using eigenvalue decomposition of the global covariance matrix and local covariance matrix
in each subdomain is negligible for small model, while it will become significant for large-scale model, especially for the global covariance matrix. 

\item Generating "good" snapshots is an essential part of the POD method, which
is a data driven method. The actual generation of the snapshots is done by
generating an ensemble of log-permeability around an initial log-permeability. For each member
in the newly generated ensemble the FOM is ran, and the values for
pressure and saturation at each time step are saved. The computational cost of performing this
part of the optimization process, expressed in number of FOM runs is equal
with the number of members in the generated ensemble.

\item The cost of solving the reduced eigenvalue problem is equivalent to the cost of
a Singular Value Decomposition of the snapshot matrix in case of POD method. Since
the number of snapshots can be kept low, this cost is also low.

\item  Cost of approximating derivatives using RBF interpolation represents the most
computationally expensive part of the constructing subdomain reduced-order linear model. 
The computational time expressed in number of FOM runs
is several orders of magnitude higher than the number of local PCA patterns generated.

\end{itemize}

\textbf{Online stage: the cost of solving the model-reduced parameter estimation problem}
\begin{itemize}
\item The cost of solving a system of model-reduced linear equations can be neglected in our case.

\item The cost of the model-reduced optimization procedure, is limited to the number of
times the new subdomain reduced-order linear model constructed for the purpose of improving the reduced
objective function are verified to also improve the original objective function.
This requires one more FOM run for the original objective function calculation.

\end{itemize}
 
\begin{figure}[!h]
\centering
\includegraphics[width=0.5\linewidth]{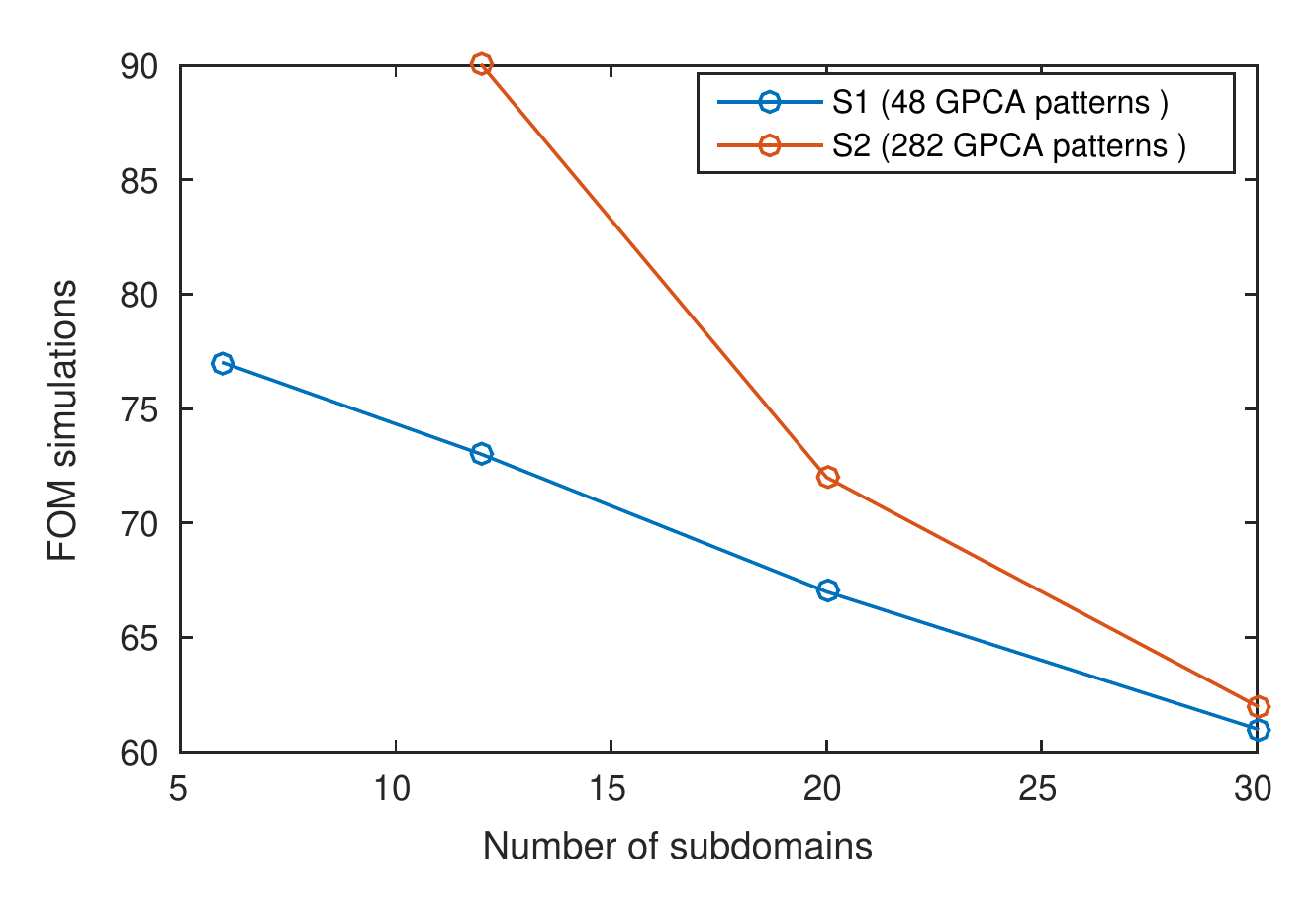} \\
 \caption{Summary of the required FOM simulations for scenario S1 and S2 in this paper}\label{fig35}
\end{figure}

In short, this process is code non-intrusive without any overwhelming programming efforts. 
We should note that the gradient-based reservoir history matching generally requires $O(10^{3}-10^{4})$ FOM simulations, 
thus, an offline cost of $O(10-10^{2})$ FOM simulations in this settings is attractive. For large-scale reservoir history matching problem, 
the main computational cost is dominated by the required number of FOM simulations. 
In our proposed method, most part of the FOM simulations is mainly in offline stage, which makes our method easily implemented.
On the one hand, fig.\ref{fig35} summarizes the required FOM simulations as a function of number of subdomains in this study which indicates that 
we will benefit more computational efficiency with an increasing number of subdomains. On the other hand, previous numerical results also show that
the model updating and minimization of cost function will be deteriorated if too small subdomain is formed. 
It is significant to trade-off efficiency and accuracy by optimizing the domain decomposition strategy.

\section{Conclusions}
In this study we have introduced a subdomain POD-TPWL with local parameterization for large-scale history matching problems. 
The spatial parameter field 
is represented by a low-order parameter subspace in each subdomain through combining principle component analysis (PCA) and domain decomposition (DD).
The optimal local parameters are projected onto a global parameterization based on PCA to eliminate the non-smoothness at the boundaries of neighboring subdomains.
The local parameterization allows us to run a small number of model 
simulations by simultaneously perturbing the parameters in all subdomains.  
The use of local parameterization unlocks the applications of subdomain POD-TPWL to large-scale problems 
since the number of full-order model simulations depends primarily on 
the number of the local parameters in each subdomain.

The approach is tested using a modified version of the SAIGUP benchmark model. 
In the first numerical experiment, subdomain POD-TPWL with global and local parameterization
is able to reconstruct the main geological features, e.g, 
the high permeable zone, of the ‘true’ log-permeability field and show comparable results as finite-difference based history matching. 
Our method also significantly improves the prediction of fluid rate 
and water breakthrough time of production wells.
The results of another example show that also for a more complex history matching problem, very promising 
results could be obtained. 
Activating more subdomains results in much fewer local parameter patterns and enables us to run fewer full-order model simulations.
For the cases studied in this paper, the number of full-order model simulations required for the history matching is roughly 2 times 
the maximum number of local parameter patterns among all subdomains. 

There is a number of aspects of the proposed methodology that could be further improved. 
All examples have shown that domain decomposition strategy has 
significant influences on the model updating and the minimization of the cost function.
The updated model could be far from the 'true' model and shows unsatisfactory long-term predictions, and even the cost function is difficult to be minimized. 
It may be beneficial to choose the subdomains based on information about the main dynamical patterns adaptively. 
We have chosen somewhat arbitrary decompositions of the global domain into subdomains. Although not formally proven, 
the adaptive multi-scale or hierarchical methods is considered to be an option to help 
avoid the local minima or lead to much easier minimization \cite{Chen2012Multiscale}.
The adaptive multi-scale/hierarchical local parameterization is the focus of our ongoing research.
In this study the distribution of both type of data, e.g., well data and seismic data, among all subdomains is almost uniform, thus, an uniform choice of the number of local PCA patterns in each subdomain is also appropriate.
In order to obtain an efficient implementation, a small number of local parameters in each subdomain is to be preferred. Therefore, the domain decomposition should also depend on the amount of available measurement information in 
the subdomain. These aspects also deserve to be explored in the future.

\section*{Acknowledgment}
We thank the research funds by China Scholarship Council (CSC) and Delft University of Technology.


\end{document}